\documentclass[mathpazo]{cicp}
\usepackage{graphicx}
\usepackage{subfigure}
\usepackage{bm}
\usepackage{amsmath}
\numberwithin{equation}{section}
\usepackage{epsfig}
\usepackage{color}

\begin{document}
\title{High-order gas-kinetic scheme with TENO class reconstruction for the Euler and Navier–Stokes equations}


\author[Mu J L et.~al.]{Junlei Mu\affil{1},
     Congshan Zhuo\affil{1,2}\comma\corrauth, Qingdian Zhang\affil{1}, Sha Liu\affil{1,2}, and Chengwen Zhong\affil{1,2}}
\address{\affilnum{1}\ School of Aeronautics,
			 Northwestern Polytechnical University, Xi'an, Shaanxi 710072, China \\
          \affilnum{2}\ National Key Laboratory of Science and Technology on Aerodynamic Design and Research, Northwestern Polytechnical University, Xi'an, Shaanxi 710072, China}
 \emails{{\tt 2021200064@mail.nwpu.edu.cn} (J.~Mu), 
 	{\tt zhuocs@nwpu.edu.cn} (C.~Zhuo), 
 	{\tt zhangqd@mail.nwpu.edu.cn} (Q.~Zhang),
 	{\tt shaliu@nwpu.edu.cn} (S.~Liu), {\tt zhongcw@nwpu.edu.cn} (C.~Zhong)}

\begin{abstract}
The high-order gas-kinetic scheme(HGKS) with WENO spatial reconstruction method has been extensively validated through many numerical experiments, demonstrating its superior accuracy efficiency, and robustness. Compared with WENO class schemes, TENO class schemes exhibit significantly improved robustness, low numerical dissipation and sharp discontinuity capturing. In this paper, two kinds of fifth-order HGKS with TENO class schemes are designed. One involves replacing WENO5 scheme with the TENO5 scheme in the conventional WENO5-GKS. WENO and TENO schemes only provide the non-equilibrium state values at the cell interface. The slopes of the non-equilibrium state along with the equilibrium values and slopes, are obtained by additional linear reconstruction. Another kind of TENO5-D GKS is similar to WENO5-AO GKS. Following a strong scale-separation procedure, a tailored novel ENO-like stencil selection strategy is proposed such that the high-order accuracy is restored in smooth regions by selecting the candidate reconstruction on the large stencil while the ENO property is enforced near discontinuities by adopting the candidate reconstruction from smooth small stencils. The such TENO schemes are TENO-AA and TENO-D scheme. The HGKS scheme based on WENO-AO or TENO-D reconstruction take advantage of the large stencil to provide point values and slopes of the non-equilibrium state. By dynamically merging the reconstructed non-equilibrium slopes, extra reconstruction of the equilibrium state at the beginning of each time step can be avoided. The simplified schemes have better robustness and efficiency than the conventional WENO5-GKS or TENO5-GKS. TENO-D GKS is also as easy to develop as WENO-AO GKS to high-order finite volume method for unstructured mesh.
\end{abstract}


\maketitle

\section{Introduction}
\label{sec1}
	As a high-order finite volume method, the high-order gas-kinetic scheme (HGKS) has witnessed significant development in recent years, including non-compact and compact HGKS scheme~\cite{PAN2016197,PAN2015250,LI20106715,YANG2023105834,JI2018446}. Additionally, HGKS based on Discontinuous Galerkin (DG) and Conservative Projection (CPR) techniques has been progressively developed~\cite{REN2016700,REN2015176,ZHANG2018329,ZHANG2022110830}. In high Mach number flow, especially in hypersonic flow, the high-order gas-kinetic scheme has demonstrated its ability to accurately simulate compressible flow for aircraft applications.\par
	The gas-Kinetic dcheme (GKS), initially introduced by Xu, is a numerical method used for solving the Euler and Navier-Stokes equations~\cite{Xu1998GaskineticSF,XU2001289}. The GKS is a Riemann solver that utilizes the Boltzmann-BGK equation to construct the gas distribution function at the cell interface. It then calculates the flux by considering the relationship between the microscopic gas distribution function and the macroscopic physical quantities. Gas-kinetic scheme is that it incorporates a time-dependent gas distribution function at the cell interface, allowing for the representation of multi-scale flow physics, ranging from kinetic particle transport to hydrodynamic wave propagation. As mentioned previously, a single-step high-order gas-kinetic scheme~\cite{PAN2015250,LI2023112318,LI2020109488} can be achieved by expanding the gas distribution function at the interface with higher-order temporal and spatial terms. However, due to the excessive complexity of expressions beyond the third order, the use of one-stage high-order gas-kinetic scheme is not commonly preferred. Furthermore, the gas-kinetic scheme is computationally demanding due to the intricacies associated with solving the flux. \par
	Later, the two-stage fourth-order generalized Riemann problem (GRP) solver proposed by Li~\cite{DU2018125} greatly contributed to the development of high-order gas-kinetic schemes. Recognizing the time-space coupling characteristics of GRP and GKS, Pan et al. constructed a two-stage fourth-order gas-kinetic scheme based on a two-stage fourth-order temporal discretization~\cite{PAN2016197}. The scheme achieves fourth-order time accuracy with just two steps of time advancement and significantly reduces the computational cost compared to the fourth-order Runge-Kutta (RK4) method. The two-stage fourth-order method was further advanced into high-order multi-stage multi-derivative gas kinetic schemes, such as the two-stage fifth-order and three-stage fifth-order HGKS scheme~\cite{JI2018150}. Additionally, the two-stage fourth-order GKS scheme has emerged as the primary scheme utilized in the advancement of HGKS. Subsequent research endeavors on high-order gas-kinetic scheme has primarily focused on spatial reconstruction. Enhancing the spatial reconstruction method is of paramount importance in attaining high accuracy and resolution in numerical simulation.\par
	The spatial reconstruction process of high-order gas-kinetic scheme primarily relies on WENO class methods. Weighted essentially nonoscillatory (WENO) schemes are widely recognized numerical schemes designed to address problems involving shock waves and fine, smooth structures, including hyperbolic conservation laws. The WENO scheme was initially proposed by Jiang and Shu and named WENO-JS~\cite{SHU1988439,JIANG1996202}. Furthermore, a new scheme called WENO-Z was developed~\cite{BORGES20083191}, which involves obtaining a global higher-order smoothness indicator through a linear combination of the original smoothing indicators. The WENO-Z scheme achieves superior results while requiring almost the same computational effort as the classical WENO method. However, the earliest WENO-GKS, similar to the second-order GKS, reconstructed the non-equilibrium state separately from the equilibrium state. This approach introduced complexity to the algorithm, resulting in numerical oscillation and poor robustness in individual tests. Additionally, constructing a Gaussian point value to solve the flux was not an easy task within this conventional WENO-GKS. Later, Pan and Ji utilized the WENO-AO scheme proposed by Balsara~\cite{BALSARA2016780,BALSARA2020109062} to enhance the performance of the two-stage fourth-order gas-kinetic scheme~\cite{CiCP-28-539,YANG2022110706}. This improvement aimed to simplify the calculation process by avoiding additional reconstruction of the equilibrium state. The WENO-AO scheme is similar to the WENO-ZQ scheme proposed by Qiu and Zhu~\cite{ZHU2016110,ZHAO2019422}, as it introduces flexibility to the weights by modifying the final reconstructed polynomials to prevent negative weights from affecting stability at Gaussian points. Building upon the WENO reconstruction method with arbitrary linear weights, similar to the WENO-AO scheme and WENO-ZQ scheme, the high-order gas-kinetic scheme has gradually evolved into both non-compact and compact three-dimensional high-order gas-kinetic schemes on unstructured mesh~\cite{YANG2023105834,JI2020109367}.\par
	The WENO scheme still exhibits high dissipation in direct numerical simulation for turbulence. In response, Fu proposed a series of higher-order targeted ENO(TENO) scheme~\cite{FU2016333,FU2018724,CiCP-25-1357}. The core concept behind TENO schemes is to either discard candidate stencils that are intersected by discontinuities or reconstruct them with optimal weights. Instead of assembling candidate stencils of the same width as traditional WENO methods, TENO schemes achieve high-order reconstruction by assembling a collection of low-order (higher than third-order) upwind biased candidate stencils with increased width. TENO schemes can adaptively degrade to third order based on local flow characteristics, thereby addressing the issue of multiple discontinuities and restoring the robustness of classical fifth-order WENO-JS scheme~\cite{SHU1988439,JIANG1996202}. Inspired by the work of Hu and Adams~\cite{FU2016333,FU2018724} a more robust scaling separation formula is employed for isolating discontinuities from high wave-number physical waves. Unlike WENO, which assigns smooth weights to candidate stencils, TENO applies either optimal weights or completely eliminates them when true discontinuities with certain intensity are present, thereby utilizing ENO-like stencil selection. This approach effectively limits numerical dissipation to the background linear scheme in the high wave-number region of interest while still retaining strong shock-capturing capability. Following that, the researchers developed the TENO-AA scheme~\cite{FU2021114193}, which achieves a very high order of accuracy. Additionally, they introduced the TENO-D scheme~\cite{JSC} specifically for unstructured mesh by enabling flexible utilization of both large and small stencils.\par
	Indeed, incorporating TENO reconstruction into the gas-kinetic scheme (GKS) provides a means to enhance the performance of the two-stage fourth-order GKS. The algorithm for TENO reconstruction is more simple than the high-order compact GKS scheme~\cite{JI2018446,JI2020109367}. In the case of the classical WENO-Z GKS, the WENO-Z scheme used for constructing non-equilibrium states can be substituted with the standard TENO scheme to construct the TENO GKS. Besides, as mentioned in ~\cite{CiCP-28-539}, the enhanced performance of the WENO-AO GKS compared to the classical WENO GKS can be attributed to two main factors. Firstly, the use of a large stencil in the reconstruction process allows for obtaining high-order non-equilibrium state values and derivatives. This enables more accurate representation of the flow physics and improves the overall accuracy of the scheme. Secondly, in the WENO-AO GKS, the equilibrium state is directly reconstructed by utilizing particle collision dynamics. This method simplifies the process of HGKS spatial reconstruction, and improves the robustness and accuracy of the HGKS scheme. In order to leverage the low-dissipation characteristics of the TENO scheme while retaining the advantages of the WENO-AO GKS scheme, the TENO-D GKS is devised by incorporating the TENO-D scheme with HGKS scheme. This approach allows for the effective handling of discontinuities in the flow. The TENO-D GKS scheme applies the strategy of the TENO scheme when encountering discontinuities. The ENO-like stencil selection of TENO class schemes is used twice. It performs the first ENO-like stencil selection on the large stencil. This enables the determination of point values and their derivatives through the reconstruction of the large stencil in smooth regions, similar to the WENO-AO GKS scheme. Additionally, the TENO-D GKS scheme employs the second ENO-like stencil selection to accurately handle the discontinuities present in all small stencils. By combining these strategies, the TENO-D GKS scheme effectively balances the treatment of smooth regions and discontinuities, resulting in enhanced accuracy and performance.\par
	This paper is organized as follows. Section \ref{sec2} gives a brief review of the review of WENO5-Z GKS and WENO5-AO GKS. Section \ref{sec3} intruduces TENO class schemes and the numerical algorithm for HGKS with TENO class reconstruction. Section \ref{sec4} presents the numerical results from different schemes and their comparison with each other in terms of accuracy, robustness and so on. Finally we end up with concluding remarks.

\section{Brief review of high-order gas-kinetic scheme}
\label{sec2}
\subsection{Gas-kinetic flux solver}
\label{sec2.1}
The 2-D Boltzmann-BGK equation~\cite{PhysRev.94.511} can be written as
\begin{equation}\label{equ1}
{{f}_{t}}+\mathbf{u}\cdot \nabla f=\frac{g-f}{\tau },
\end{equation}
where $\mathbf{u}=\left( u,v \right)$ is the particle velocity and $\tau $ is the collision time. $f$ is the gas distribution function, $g$ is the corresponding two-dimensional Maxwellian distribution for equilibrium state,
\begin{equation}
	g=\rho {{\left( \frac{\lambda }{\pi } \right)}^{\frac{K+2}{2}}}{{e}^{-\lambda \left( {{\left( u-U \right)}			^{2}}+{{\left( v-V \right)}^{2}}+{{\xi }^{2}} \right)}},
\end{equation}
where $\lambda =m/2kT$. $m$ is the molecular mass, $T$ is the temperature and $k$ is the Boltzmann constant. $\rho$ is the density. $U$ and $V$ are the macroscopic velocities in the x-direction and y-direction. $K=\left( 5-3\gamma  \right)/\left( \gamma -1 \right)+1$ is the internal degree of freedom for a 2D flow. $\gamma $ is the specific heat ratio. In the equilibrium state, the internal variable ${{\xi }^{2}}=\xi _{1}^{2}+\xi _{2}^{2}+\cdots +\xi _{K}^{2}$~\cite{PhysRev.94.511,POF}.

  The collision term satisfies the compatibility condition
\begin{equation}\label{equ3}
	\int{\frac{g-f}{\tau }}{\psi }_{\alpha } d\Xi =0,\alpha =1,2,3,4.
\end{equation}
where ${{\psi }_{\alpha }} ={{\left( 1,u,v,\frac{1}{2}\left( {{u}^{2}}+{{v}^{2}}+{{\xi }^{2}} \right) \right)}^{T}}$, $d\Xi =dudvd{{\xi }_{1}}\ldots d{{\xi }_{K}}$. \par
According to the Chapman-Enskog expansion for Boltzmann-BGK equation~\cite{PhysRev.94.511,POF}, the macroscopic governing equations can be derived. The gas distribution function in the continuum regime can be expanded as
$$f=g-\tau {{D}_\mathbf{u}}g+\tau {{D}_\mathbf{u}}\left( \tau {{D}_\mathbf{u}} \right)g-\tau {{D}_\mathbf{u}}\left[ \tau {{D}_\mathbf{u}}\left( \tau {{D}_\mathbf{u}} \right)g \right]+\cdots, $$
where ${{D}_\mathbf{u}}=\partial /\partial t+\mathbf{u}\cdot \nabla $. For the Euler equations, the zeroth order truncation is taken, i.e. $f=g$. For the Navier-Stokes equations, the first order truncation is used and the distribution function is
$$f=g-\tau \left( u{{g}_{x}}+v{{g}_{y}}+{{g}_{t}} \right).$$
With the integral solution of BGK equation, the gas distribution function in Eq. (\ref{equ1}) can be constructed as follows at a cell interface ${{x}_{j+1/2}}$ is
\begin{equation}\label{equ4}
	f\left( {{x}_{i+1/2}},y,t,u,v,\xi  \right)=\frac{1}{\tau }\int_{0}^{t}{g\left( {x}',{y}',{t}',u,v,\xi  \right)}{{e}^{-\left( t-{t}' \right)/\tau }}d{t}'+{{e}^{-t/\tau }}{{f}_{0}}\left( -ut,-vt,u,v,\xi  \right),
\end{equation}
where ${x}'={{x}_{i+1/2}}-u\left( t-{t}' \right)$ and ${y}'=y-v\left( t-{t}' \right)$ are the particle trajectory, ${{x}_{i+1/2}}=0$ is the location of the cell interface. ${{f}_{0}}$ is the initial gas distribution function $f$ at the beginning of each time step $t=0$.\par
With the reconstruction of macroscopic variables, the second-order gas distribution function at the cell interface can be expressed as
\begin{equation}\label{equ5}
	\begin{aligned}
		& f\left( {{x}_{i+1/2}},t,u,v,\xi  \right)=\left( 1-{{e}^{-t/\tau }} \right){{g}_{0}}+\left( \left( t+\tau  \right){{e}^{-t/\tau }}-\tau  \right)\left( {{{\bar{a}}}_{1}}u+{{{\bar{a}}}_{2}}v \right){{g}_{0}} \\ 
		& +\left( t-\tau +\tau {{e}^{-t/\tau }} \right)\bar{A}{{g}_{0}} \\ 
		& +{{e}^{-t/\tau }}{{g}_{r}}\left[ 1-\left( \tau +t \right)\left( a_{1}^{r}u+a_{2}^{r}v \right)-\tau {{A}^{r}} \right]H\left( u \right) \\ 
		& +{{e}^{-t/\tau }}{{g}_{l}}\left[ 1-\left( \tau +t \right)\left( a_{1}^{l}u+a_{2}^{l}v \right)-\tau {{A}^{l}} \right]\left( 1-H\left( u \right) \right),  
	\end{aligned}
\end{equation}
where $H(u)$ is the Heaviside function. Derivations related to gas kinetic scheme can be found in ~\cite{XU2001289}. The equilibrium state ${g}_{0}$ and corresponding conservative variables ${Q}_{0}$ and spatial derivatives in the local coordinate at the quadrature point can be determined by the compatibility condition Eq. (\ref{equ3})
\begin{equation}\label{equ6}
	\begin{aligned}
		& \int{\psi _{\alpha } {{g}_{0}}d\Xi ={{Q}_{0}}=}\int_{u>0}{\psi _{\alpha } {{g}_{l}}d\Xi +}\int_{u<0}{\psi _{\alpha } {{g}_{r}}d\Xi }, \\ 
		& \frac{\partial {{Q}_{0}}}{\partial {{\mathbf{n}}_{x}}}=\int_{u>0}{\psi _{\alpha } a_{1}^{l}{{g}_{l}}d\Xi +}\int_{u<0}{\psi _{\alpha } a_{1}^{r}{{g}_{r}}d\Xi }, \\ 
		& \frac{\partial {{Q}_{0}}}{\partial {{\mathbf{n}}_{y}}}=\int_{u>0}{\psi _{\alpha } a_{2}^{l}{{g}_{l}}d\Xi +}\int_{u<0}{\psi _{\alpha } a_{2}^{r}{{g}_{r}}d\Xi }.\\ 
	\end{aligned}\
\end{equation}
The equilibrium state can also be obtained by extra selection of stencil and reconstruction. The coefficients in Eq. (\ref{equ5}) can be determined by the spatial derivatives of macroscopic flow variables and the compatibility condition as follows
\begin{equation}
\begin{aligned}
	& \left\langle a_{1}^{k} \right\rangle =\frac{\partial {{Q}_{k}}}{\partial {{\mathbf{n}}_{x}}},\left\langle a_{2}^{k} \right\rangle =\frac{\partial {{Q}_{k}}}{\partial {{\mathbf{n}}_{y}}},\left\langle {{{\bar{a}}}_{1}} \right\rangle =\frac{\partial {{Q}_{0}}}{\partial {{\mathbf{n}}_{x}}},\left\langle {{{\bar{a}}}_{2}} \right\rangle =\frac{\partial {{Q}_{0}}}{\partial {{\mathbf{n}}_{y}}}, \\ 
	& \left\langle a_{1}^{k}u+a_{2}^{k}v+{{A}^{k}} \right\rangle =0,\left\langle {{{\bar{a}}}_{1}}u+{{{\bar{a}}}_{2}}v+\bar{A} \right\rangle =0, \\ 
\end{aligned}\
\end{equation}
 where $k=l,r$ and $\left\langle \ldots  \right\rangle $ are the moments of the equilibrium $g$ and defined by
$$\left\langle \ldots  \right\rangle =\int{g\left( \ldots  \right)}\psi _{\alpha } d\Xi. $$
 The relation between the conservative variables $\left( \rho ,\rho U,\rho V,\rho E \right)$ and the distribution function $f$ is
\begin{equation}
\left( \begin{aligned}
  & \rho  \\ 
 & \rho U \\ 
 & \rho V \\ 
 & \rho E \\ 
\end{aligned} 
\right)=\int{{{\psi }_{\alpha }}fd\Xi ,\alpha =1,2,3,4},
\end{equation}
 After obtaining all the required terms, substitute these terms into Eq. (\ref{equ5}) and solve it can obtain the gas distribution function on the interface~\cite{XU2001289,CiCP-28-539}.
 Finally, the gas-kinetic numerical flux in the x-direction on the cell interface can be computed as
\begin{equation}\label{equ9}
	{\mathbf{F}_{i+1/2}}\left( {{\mathbf{W}}^{n}},t \right)=\int{u\left( \begin{aligned}
  & 1 \\ 
 & u \\ 
 & v \\ 
 & \frac{1}{2}\left( {{u}^{2}}+{{v}^{2}}+{{\xi }^{2}} \right) \\ 
\end{aligned} \right)f\left( {{x}_{j+1/2}},t,u,v,\xi  \right)d\Xi .}
\end{equation}
More details of the gas-kinetic scheme can be found in ~\cite{XU2001289}.

\subsection{Two-stage fourth-order temporal discretization}
\label{sec2.2}
The conservation laws can be written as
$${{\mathbf{W}}_{t}}=-\nabla \cdot \mathbf{F}\left( \mathbf{W} \right),$$
$$\mathbf{W}\left( 0,x \right)={{\mathbf{W}}_{0}}\left( x \right),x\in \Omega \subseteq \mathbb{R},$$
where $\mathbf{W}$ is the conservative variables and $\mathbf{F}$ is the corresponding flux. With the spatial discretization ${{\mathbf{W}}^{h}}$ and appropriate evaluation  $-\nabla \cdot \mathbf{F}\left( \mathbf{W} \right)$, the original partial differential equation (PDE) became the ordinary differential equation (ODE).
\begin{equation}
	\mathbf{W}_{t}^{h}=\mathcal{L}\left( {{\mathbf{W}}^{h}} \right),t={{t}_{n}},
\end{equation}
where $\mathcal{L}\left( {{\mathbf{W}}^{h}} \right)$ is the spatial operator of flux. Here we define
$$\mathbf{W}_{t}^{^{\left( m \right)}}\left( {{t}^{n}} \right)=\frac{{{d}^{m}}{{\mathbf{W}}^{n}}}{d{{t}^{m}}}=\frac{{{d}^{m-1}}\mathcal{L}\left( {{\mathbf{W}}^{n}} \right)}{d{{t}^{m-1}}}={{\mathcal{L}}^{^{m-1}}}.$$
The two-stage fourth-order time marching scheme~\cite{DU2018125} is used to solve the initial value problem and is written as
\begin{equation}
\begin{aligned}
  & {{\mathbf{W}}^{*}}={{\mathbf{W}}^{n}}+\frac{1}{2}\Delta t\mathcal{L}\left( {{\mathbf{W}}^{n}} \right)+\frac{1}{8}\Delta {{t}^{2}}\frac{\partial }{\partial t}\mathcal{L}\left( {{\mathbf{W}}^{n}} \right) \\ 
 & {{\mathbf{W}}^{n+1}}={{\mathbf{W}}^{n}}+\Delta t\mathcal{L}\left( {{\mathbf{W}}^{n}} \right)+\frac{1}{6}\Delta {{t}^{2}}\left( \frac{\partial }{\partial t}\mathcal{L}\left( {{\mathbf{W}}^{n}} \right)+2\frac{\partial }{\partial t}\mathcal{L}\left( {{\mathbf{W}}^{*}} \right) \right), \\ 
\end{aligned}
\end{equation}
where $\partial \mathcal{L}\left( \mathbf{W} \right)/\partial t$ is the time derivative of the spatial operator. with the time-dependent gas distribution function Eq. (\ref{equ5}), the flux for the macroscopic flow variables can be calculated by Eq. (\ref{equ9}). For the flux in the time interval $\left[ {{t}_{n}},\ {{t}_{n}}+\Delta t \right]$,  the two-stage gas-kinetic scheme needs the ${\mathbf{F}_{i\pm 1/2}}\left( \mathbf{W} \right)$ and ${{\partial }_{t}}{\mathbf{F}_{i\pm 1/2}}\left( \mathbf{W} \right)$ at both ${{t}_{n}}$ and ${{t}_{*}}={{t}_{n}}+\Delta t/2$. The algorithm of two-stage gas-kinetic scheme is as follows.\par
Firstly, introduce the following notation,
	\begin{equation}\label{equ12}
{{\mathbb{F}}_{i+1/2,j}}\left( {{\mathbf{W}}^{n}},\delta  \right)=\int_{{{t}_{n}}}^{{{t}_{n}}+\delta }{{{\mathbf{F}}_{i+1/2}}\left( {{\mathbf{W}}^{n}},t \right)=\int_{{{t}_{n}}}^{{{t}_{n}}+\delta }{\int{u{{\psi }_{\alpha }}f\left( {{x}_{i+1/2}},t,u,v,\xi  \right)}}}d\Xi dt
\end{equation}
 With the reconstruction at ${{t}_{n}}$, the x-direction flux ${{\mathbb{F}}_{i+1/2,j}}\left( {{\mathbf{W}}^{n}},\Delta t \right)$, ${{\mathbb{F}}_{i+1/2,j}}\left( {{\mathbf{W}}^{n}},\Delta t/2 \right)$ and y-direction flux ${{\mathbb{G}}_{i,j+1/2}}\left( {{\mathbf{W}}^{n}},\Delta t \right)$ and ${{\mathbb{G}}_{i,j+1/2}}\left( {{\mathbf{W}}^{n}},\Delta t/2 \right)$ in the time interval $\left[ {{t}_{n}},\ {{t}_{n}}+\frac{\Delta t}{2} \right]$ can be evaluated by Eq. (\ref{equ12}).
In the time interval $\left[ {{t}_{n}},\ {{t}_{n}}+\Delta t \right]$, the flux is expanded as the linear form as
	\begin{equation}
{\mathbf{F}_{i+1/2,j}}\left( {{\mathbf{W}}^{n}},{{t}_{n}} \right)=\mathbf{F}_{_{i+1/2,j}}^{n}+{{\partial }_{t}}\mathbf{F}_{_{i+1/2,j}}^{n}\left( t-{{t}_{n}} \right).
\end{equation}
The terms $\mathbf{F}_{_{i+1/2,j}}^{{}}\left( {{\mathbf{W}}^{n}},{{t}_{n}} \right)$ and $\partial t\mathbf{F}_{_{i+1/2,j}}^{{}}\left( {{\mathbf{W}}^{n}},{{t}_{n}} \right)$ can be determined as
	\begin{equation}	 
{\mathbf{F}_{i+1/2,j}}\left( {{\mathbf{W}}^{n}},{{t}_{n}} \right)\Delta t+\frac{1}{2}{{\partial }_{t}}{\mathbf{F}_{i+1/2,j}}\left( {{\mathbf{W}}^{n}},{{t}_{n}} \right)\Delta {{t}^{2}}={{\mathbb{F}}_{i+1/2,j}}\left( {{\mathbf{W}}^{n}},\Delta t \right),
\end{equation}
	\begin{equation}	 
\frac{1}{2}{\mathbf{F}_{i+1/2,j}}\left( {{\mathbf{W}}^{n}},{{t}_{n}} \right)\Delta t+\frac{1}{8}{{\partial }_{t}}{\mathbf{F}_{i+1/2,j}}\left( {{\mathbf{W}}^{n}},{{t}_{n}} \right)\Delta {{t}^{2}}={{\mathbb{F}}_{i+1/2,j}}\left( {{\mathbf{W}}^{n}},\Delta t/2 \right).
\end{equation}
The terms $\mathbf{F}_{_{i+1/2,j}}^{{}}\left( {{\mathbf{W}}^{n}},{{t}_{n}} \right)$ and $\partial t\mathbf{F}_{_{i+1/2,j}}^{{}}\left( {{\mathbf{W}}^{n}},{{t}_{n}} \right)$ can be computed by solving this linear system as
\begin{equation}
{\mathbf{F}_{i+1/2,j}}\left( {{\mathbf{W}}^{n}},{{t}_{n}} \right)=\left( 4{{\mathbb{F}}_{i+1/2,j}}\left( {{\mathbf{W}}^{n}},\Delta t/2 \right)-{{\mathbb{F}}_{i+1/2,j}}\left( {{\mathbf{W}}^{n}},\Delta t \right) \right)/\Delta t,
\end{equation}
\begin{equation}
{{\partial }_{t}}{\mathbf{F}_{i+1/2,j}}\left( {{\mathbf{W}}^{n}},{{t}_{n}} \right)=4\left( {{\mathbb{F}}_{i+1/2,j}}\left( {{\mathbf{W}}^{n}},\Delta t \right)-2{{\mathbb{F}}_{i+1/2,j}}\left( {{\mathbf{W}}^{n}},\Delta t/2 \right) \right)/\Delta {{t}^{2}}.
\end{equation}
Similar to $\mathbf{F}_{_{i+1/2,j}}^{{}}\left( {{\mathbf{W}}^{n}},{{t}_{n}} \right)$ and $\partial t\mathbf{F}_{_{i+1/2,j}}^{{}}\left( {{\mathbf{W}}^{n}},{{t}_{n}} \right)$, the $\mathbf{G}_{_{i,j+1/2}}^{{}}\left( {{\mathbf{W}}^{n}},{{t}_{n}} \right)$ and $\partial t\mathbf{G}_{_{i,j+1/2}}^{{}}\left( {{\mathbf{W}}^{n}},{{t}_{n}} \right)$ can be computed from ${{\mathbb{G}}_{i,j+1/2}}\left( {{\mathbf{W}}^{n}},\Delta t \right)$ and ${{\mathbb{G}}_{i,j+1/2}}\left( {{\mathbf{W}}^{n}},\Delta t/2 \right)$.\par
Secondly, update $\mathbf{W}_{ij}^{*}$ at ${{t}_{*}}={{t}_{n}}+\Delta t/2$ by
	 \begin{equation}
\begin{aligned}
  & \mathbf{W}_{ij}^{*}=\mathbf{W}_{ij}^{n}-\frac{1}{\Delta x}\left[ {{\mathbb{F}}_{i+1/2,j}}\left( {{\mathbf{W}}^{n}},\Delta t/2 \right)-{{\mathbb{F}}_{i-1/2,j}}\left( {{\mathbf{W}}^{n}},\Delta t/2 \right) \right] \\ 
 & -\frac{1}{\Delta y}\left[ {{\mathbb{G}}_{i,j+1/2}}\left( {{\mathbf{W}}^{n}},\Delta t/2 \right)-{{\mathbb{G}}_{i,j-1/2}}\left( {{\mathbf{W}}^{n}},\Delta t/2 \right) \right].  
\end{aligned}
\end{equation}
Then we can get $\partial t\mathbf{F}_{_{i+1/2,j}}^{{}}\left( {{\mathbf{W}}^{*}},{{t}_{*}} \right)$ and $\partial t\mathbf{G}_{_{i,j+1/2}}^{{}}\left( {{\mathbf{W}}^{*}},{{t}_{*}} \right)$ in the time interval $\left[ {{t}_{*}},\ {{t}_{*}}+\Delta t \right]$ to compute the middle stage by the same way we did before.\par
Finally, The numerical fluxes $\mathcal{F}_{i+1/2,j}^{n}$ and $\mathcal{G}_{i+1/2,j}^{n}$ can be computed by 
\begin{equation}
\mathcal{F}_{i+1/2,j}^{n}={\mathbf{F}_{i+1/2,j}}\left( {{\mathbf{W}}^{n}},{{t}_{n}} \right)+\frac{\Delta t}{6}\left[ {{\partial }_{t}}{\mathbf{F}_{i+1/2,j}}\left( {{\mathbf{W}}^{n}},{{t}_{n}} \right)+2{{\partial }_{t}}{\mathbf{F}_{i+1/2,j}}\left( {{\mathbf{W}}^{*}},{{t}_{*}} \right) \right],
\end{equation}
\begin{equation}
\mathcal{G}_{i,j+1/2}^{n}={\mathbf{G}_{i,j+1/2}}\left( {{\mathbf{W}}^{n}},{{t}_{n}} \right)+\frac{\Delta t}{6}\left[ {{\partial }_{t}}{\mathbf{G}_{i,j+1/2}}\left( {{\mathbf{W}}^{n}},{{t}_{n}} \right)+2{{\partial }_{t}}{\mathbf{G}_{i,j+1/2}}\left( {{\mathbf{W}}^{*}},{{t}_{*}} \right) \right].
\end{equation}
Update $\mathbf{W}_{ij}^{n+1}$ by
	 \begin{equation}
\mathbf{W}_{ij}^{n+1}=\mathbf{W}_{ij}^{n}-\frac{\Delta t}{\Delta x}\left[ \mathcal{F}_{i+1/2,j}^{n}-\mathcal{F}_{i-1/2,j}^{n} \right]-\frac{\Delta t}{\Delta y}\left[ \mathcal{G}_{i,j+1/2}^{n}-\mathcal{G}_{i,j-1/2}^{n} \right].
\end{equation}\par
More details about two-stage fourth-order gas-kinetic scheme can be found in ~\cite{PAN2016197}.

\section{Previous GKS with fifth-order WENO schemes}
In this section, we present two kinds of high-order gas-kinetic schemes using fifth-order WENO  schemes: WENO5-Z GKS and WENO5-AO GKS. Apart from employing different WENO spatial discretization techniques, these two HGKS schemes also utilize slightly different reconstruction methods for values and their gradients of non-equilibrium state and equilibrium state to solving the GKS flux. In ~\cite{CiCP-28-539}, it is extensively discussed that WENO5-AO GKS exhibits significant performance improvements compared to WENO5-Z GKS.\par
\label{sec3}
\subsection{WENO(Z) reconstruction for GKS}
\label{sec3.1}
For the classical fifth-order WENO-Z scheme~\cite{PAN2016197,CiCP-28-539}, three sub-stencils
$${{S}_{0}}=\left\{ {{I}_{i-2}},{{I}_{i-1}},{{I}_{i}} \right\},{{S}_{1}}=\left\{ {{I}_{i-1}},{{I}_{i}},{{I}_{i+1}} \right\},{{S}_{2}}=\left\{ {{I}_{i}},{{I}_{i+1}},{{I}_{i+2}} \right\}$$
are selected to reconstruct the value $Q_{i+1/2}^{l}$ and  $Q_{i-1/2}^{r}$ at both sides in a cell. Take $Q_{i+1/2}^{l}$ for example. The three quadratic polynomials $p_{k}^{r3}\left( x \right)$ corresponding to the sub-stencils ${{S}_{k}},k=0,1,2$ are constructed by requiring
\begin{equation}
	\frac{1}{\Delta x}\int_{{{I}_{i-j-k-1}}}{p_{k}^{r3}\left( x \right)dx}={{\bar{Q}}_{i-j-k-1}},j=-1,0,1,
\end{equation}
where $\bar{Q}$ represents the cell-averaged value. Each sub-stencil can achieve a third-order spatial accuracy in $r=3$ smooth case. For the reconstructed polynomials, the point values at the cell interface  ${{x}_{i+1/2}}$ is given in terms of the cell-averaged value as follows
\begin{equation}
	\begin{aligned}
		& p_{0}^{r3}\left( {{x}_{i+1/2}} \right)=\frac{1}{3}{{{\bar{Q}}}_{i-2}}-\frac{7}{6}{{{\bar{Q}}}_{i-1}}+\frac{11}{6}{{{\bar{Q}}}_{i}}, \\ 
		& p_{1}^{r3}\left( {{x}_{i+1/2}} \right)=-\frac{1}{6}{{{\bar{Q}}}_{i-1}}+\frac{5}{6}{{{\bar{Q}}}_{i}}+\frac{1}{3}{{{\bar{Q}}}_{i+1}},\\ 
		& p_{2}^{r3}\left( {{x}_{i+1/2}} \right)=\frac{1}{3}{{{\bar{Q}}}_{i}}+\frac{5}{6}{{{\bar{Q}}}_{i+1}}-\frac{1}{3}{{{\bar{Q}}}_{i+2}},.\\ 
	\end{aligned}
\end{equation}\par
On the large stencil ${{S}_{3}}=\left\{ {{S}_{0}},{{S}_{1}},{{S}_{2}} \right\}$, the point value at cell interface can be written as 
\begin{equation}\label{equ3.3}
Q_{i+1/2}^{l}=\frac{1}{60}\left( 2{{{\bar{Q}}}_{i-2}}-13{{{\bar{Q}}}_{i-1}}+47{{{\bar{Q}}}_{i}}+27{{{\bar{Q}}}_{1}}-3{{{\bar{Q}}}_{i+2}} \right)
\end{equation}
The linear weights ${{d}_{k}},k=0,1,2$ can be obtained such that
\begin{equation}
p_{3}^{5}\left( {{x}_{i+1/2}} \right)=\sum\limits_{k=0}^{2}{{{d}_{k}}p_{k}^{r3}\left( {{x}_{i+1/2}} \right)},
\end{equation}
where ${{d}_{0}}=0.1, {{d}_{1}}=0.6, {{d}_{2}}=0.3$. These three unique weights are called optimal weights. It lifts the reconstructed low order value from the small stencils to a higher-order one from the large stencil. To deal with discontinuities, the non-normalized WENO-Z type nonlinear weight is introduced as follows
\begin{equation}\label{equ3.5}
{{\omega }_{k}}={{d}_{k}}\left( 1+\frac{\tau }{{{\beta }_{k}}+\varepsilon } \right),
\end{equation}
where the global smooth indicator $\tau $ is designed as 
\begin{equation}
\tau =\left| {{\beta }_{0}}-{{\beta }_{2}} \right|.
\end{equation}
The smoothness indicators ${{\beta }_{k}}$ are defined as 
\begin{equation}\label{equ3.7}
	{{\beta }_{k}}={{\sum\limits_{q=1}^{{{q}_{k}}}{\Delta {{x}^{2q-1}}\int_{{{x}_{i-1/2}}}^{{{x}_{i+1/2}}}{\left( \frac{{{d}^{q}}}{d{{x}^{q}}}{{p}_{k}}\left( x \right) \right)}}}^{2}}dx=O\left( \Delta {{x}^{2}} \right),
\end{equation}
where ${{q}_{k}}$ is the order of ${{p}_{k}}\left( x \right)$. For $p_{k}^{r3},k=0,1,2,{{q}_{k}}=2$; for $p_{3}^{r5},{{q}_{3}}=4$. The small parameter $\varepsilon ={{10}^{-8}}$ is taken for WENO schemes in current work.
The normalized weights ${{\bar{\omega }}_{k}}$ is defined as follows~\cite{BORGES20083191}
\begin{equation}\label{equ3.8}
{{\bar{\omega }}_{k}}=\frac{{{\omega }_{k}}}{\sum\nolimits_{0}^{2}{{{\omega }_{k}}}}.
\end{equation}
Thus, the reconstructed left interface value $Q_{i+1/2}^{l}$ can be written as
\begin{equation}\label{equ3.9}
Q_{i+1/2}^{l}=\sum\limits_{k=0}^{2}{{{{\bar{\omega }}}_{k}}p_{k}^{r3}\left( {{x}_{i+1/2}} \right)}.
\end{equation}
Finally, $ Q $ should be changed to the corresponding conservative variables $ W$.\par
After ${{W}^{l,r}}$ are obtained, the $W_{x}^{l,r}$ of none-equilibrium state are obtained by constructing a third order polynomial by requiring
\begin{equation}
		\begin{aligned}
\frac{1}{\Delta x}\int_{{{I}_{i}}}{p\left( x \right)}dV={{\bar{W}}_{i}},p\left( {{x}_{i-1/2}} \right)=W_{i-1/2}^{r},p\left( {{x}_{i+1/2}} \right)=W_{i+1/2}^{l}.
	\end{aligned}
\end{equation}
Therefore, the $W_{x}^{l,r}$ are solved by 
\begin{equation}\label{equ3.11}
{{\left( W_{x}^{r} \right)}_{i-1/2}}=-\frac{2\left( 2W_{i-1/2}^{r}+W_{i+1/2}^{l}-3{{W}_{i}} \right)}{\Delta x},{{\left( W_{x}^{l} \right)}_{i+1/2}}=\frac{2\left( W_{i-1/2}^{r}+2W_{i+1/2}^{l}-3{{W}_{i}} \right)}{\Delta x}.
\end{equation}
Only third-order accuracy is achieved for the slopes on the targeted
Locations.\par
With the reconstructed $W_{i+1/2}^{l}$ and $W_{i+1/2}^{r}$ at both sides of a cell interface ${{x}_{i+1/2}}$, the macroscopic variables $W_{i+1/2}^{0}$ and the corresponding equilibrium state can be determined according to compatibility condition Eq. (\ref{equ6}).\par
On the large stencil ${{S}_{3}}=\left\{ {{S}_{0}},{{S}_{1}},{{S}_{2}} \right\}$, the ${{\left( W_{x}^{0} \right)}_{i+1/2}}$ of the equilibrium state ${{g}_{0}}$ at cell interface can be got by fifth-order linear reconstruction as
\begin{equation}\label{equ3.12}
{{\left( W_{x}^{0} \right)}_{i+1/2}}=\left[ -\frac{1}{12}\left( {{{\bar{W}}}_{i+2}}-{{{\bar{W}}}_{i-1}} \right)+\frac{5}{4}\left( {{{\bar{W}}}_{i+1}}-{{{\bar{W}}}_{i}} \right) \right]/\Delta x
\end{equation}\par
The direction-by-direction reconstruction strategy is typically utilized in WENO5-Z GKS for two-dimensional reconstruction on rectangular meshes. In the case of a fourth-order scheme, numerical flux integration necessitates the use of two Gaussian points on each interface.The two-dimensional reconstruction process of WENO5-Z GKS is as follows.\par
{\bfseries Step 1.} According to the one-dimensional WENO-Z reconstruction in Section \ref{sec3.1} and  Eq. (\ref{equ3.11}), the line averaged reconstructed values and slopes
$${{\left( {{Q}^{l}} \right)}_{i+1/2,j}},{{\left( Q_{x}^{l} \right)}_{i+1/2,j}},{{\left( {{Q}^{r}} \right)}_{i+1/2,j}},{{\left( Q_{x}^{r} \right)}_{i+1/2,j}}$$
can be obtained along the normal direction by using the cell averaged values ${{\left( {\bar{Q}} \right)}_{i+l,j}}$ and ${{\left( {\bar{Q}} \right)}_{i+l+1,j}}$,$l=-2,\ldots ,2$.\\
{\bfseries Step 2.} Next, according to Eq. (\ref{equ6}) and Eq. (\ref{equ3.11}), the line averaged reconstructed values and slopes
$${{\left( {{Q}^{0}} \right)}_{i+1/2,j}},{{\left( Q_{x}^{0} \right)}_{i+1/2,j}}$$
can be obtained along the normal direction by using the cell averaged values ${{\left( {\bar{Q}} \right)}_{i+l,j}}$ and ${{\left( {\bar{Q}} \right)}_{i+l+1,j}}$,$l=-2,\ldots ,2$.\\
{\bfseries Step 3.} Again with the one-dimensional WENO-Z reconstruction in Section \ref{sec3.1} and  Eq. (\ref{equ3.11}), the values at each Gaussian point
$${{\left( {{Q}^{l}} \right)}_{i+1/2,jm}},{{\left( Q_{y}^{l} \right)}_{i+1/2,jm}},{{\left( {{Q}^{r}} \right)}_{i+1/2,jm}},{{\left( Q_{y}^{r} \right)}_{i+1/2,jm}}$$
with $y={{y}_{jm}},m=0,1$ can be obtained by using the line averaged values ${{\left( {{Q}^{l}} \right)}_{i+1/2,j+l}},{{\left( {{Q}^{r}} \right)}_{i+1/2,j+l}},l=-2,\ldots ,2$ constructed above. In the same way, the point-wise derivatives at Gaussian point $\left( {{x}_{i+1/2}},{{y}_{jm}} \right),m=1,2$ ${{\left( Q_{x}^{l} \right)}_{i+1/2,jm}},{{\left( Q_{x}^{r} \right)}_{i+1/2,jm}}$ can be constructed by using the above line averaged derivatives ${{\left( Q_{x}^{l} \right)}_{i+1/2,j+l}},{{\left( Q_{x}^{r} \right)}_{i+1/2,j+l}},l=-2,\ldots ,2$ with the WENO-Z method in Section\ref{sec3.1}.\\
{\bfseries Step 4.} With linear fourth-order polynomial Eq. (\ref{equ3.11}), the values at each Gaussian point
$${{\left( {{Q}^{0}} \right)}_{i+1/2,jm}},{{\left( Q_{y}^{0} \right)}_{i+1/2,jm}}$$
with $y={{y}_{jm}},m=0,1$ can be obtained by using the line averaged values ${{\left( {{Q}^{0}} \right)}_{i+1/2,j+l}},l=-2,\ldots ,2$ constructed above. In the same way, the point-wise derivatives at Gaussian point ${{\left( {Q_{x}^{0}} \right)}_{i+1/2,jm}},m=1,2$ can be constructed by ${{\left( Q_{x}^{0} \right)}_{i+1/2,j+l}},l=-2,\ldots ,2$. \par
The details about tangential reconstruction for HGKS are described in ~\cite{CiCP-28-539}.\par

\subsection{WENO-AO reconstruction for GKS}
\label{sec3.2}
WENO-AO scheme proposed by Balsara~\cite{BALSARA2016780,BALSARA2020109062} can attain fifth-order accuracy when the solution's smoothness within the fifth-order stencil justifies it. It also possesses the flexibility to adaptively lower its accuracy to third order if the solution on the mesh does not necessitate higher-order accuracy. This capability to dynamically adjust the accuracy level of finite-volume WENO schemes for hyperbolic conservation laws, especially on unstructured meshes, can be of great value.\par
For the fifth-order WENO-AO scheme, the details are as follows~\cite{CiCP-28-539}. On the large stencil ${{S}_{3}}=\left\{ {{S}_{0}},{{S}_{1}},{{S}_{2}} \right\}$, a fourth-order polynomial $p_{3}^{r5}\left( x \right)$ can be constructed by requiring
\begin{equation}
	\frac{1}{\Delta x}\int_{{{I}_{i+j}}}{p_{3}^{r5}\left( x \right)dx}={{\bar{Q}}_{i+j}},j=-2,-1,0,1,2.
\end{equation}
The $p_{3}^{r5}\left( x \right)$ can be written by Balsara as
\begin{equation}
	p_{3}^{r5}\left( x \right)={{d}_{3}}\left( \frac{1}{{{d}_{3}}}p_{3}^{r5}\left( x \right)-\sum\limits_{0}^{2}{\frac{{{d }_{k}}}{{{d}_{3}}}p_{k}^{r3}\left( x \right)} \right)+\sum\limits_{0}^{2}{{{d}_{k}}p_{k}^{r3}\left( x \right)},{{r}_{1}}\ne 0,
\end{equation}
where ${{r}_{k}},l=1,2,3$ are defined linear weights. The linear weights for the large stencil ${{S}_{3}}$ and the sub-stencils ${{S}_{0}}$,${{S}_{1}}$ and ${{S}_{2}}$ are given by
\begin{equation}
	{{d}_{3}}={{d}_{Hi}},{{d}_{1}}={{d}_{3}}=\left( 1-{{d }_{Hi}} \right)\left( 1-{{d }_{Lo}} \right)/2,{{d}_{2}}=\left( 1-{{d }_{Hi}} \right){{d}_{Lo}},
\end{equation}
which satisfy $\sum\limits_{0}^{3}{{{d}_{k}}}=1$. Typically, ${{d}_{Hi}}\in \left[ 0.85,\ 0.95 \right]$ and ${{d}_{Lo}}\in \left[ 0.85,\ 0.95 \right]$. These linear weights offer more flexibility compared to the linear weights used in WENO-Z schemes. Moreover, when the smoothness indicators indicate that the larger stencil ${{S}_{3}}$ is smooth, the WENO-AO scheme can ensure fifth-order accuracy. Interestingly, even when the smoothness indicators do not indicate smoothness, the scheme can still guarantee third-order accuracy.\par
To deal with discontinuities and avoid the loss of order of accuracy at inflection points, the non-normalized WENO-Z type nonlinear weights ~\cite{BORGES20083191} are obtained by Eq. (\ref{equ3.5}).
But the global smooth indicator $\tau $ is designed as
\begin{equation}
	\tau =\frac{1}{3}\left( \left| \beta _{3}^{r5}-\beta _{0}^{r3} \right|+\left| \beta _{3}^{r5}-\beta _{1}^{r3} \right|+\left| \beta _{3}^{r5}-\beta _{3}^{r3} \right| \right)=O\left( \Delta {{x}^{4}} \right).
\end{equation}
The global smoothness indicators $\beta _{3}^{r5}$ are defined as 
\begin{equation}
	\begin{aligned}
		& \beta _{3}^{r5}={{\left( {{\left( p_{3}^{r5} \right)}_{x}}+{{\left( p_{3}^{r5} \right)}_{xxx}}/10 \right)}^{2}}+13/3{{\left( {{\left( p_{3}^{r5} \right)}_{xx}}+123/455{{\left( p_{3}^{r5} \right)}_{xxxx}} \right)}^{2}}+ \\ 
		& +781/20\left( p_{3}^{r5} \right)_{xxx}^{2}+1421461/2275\left( p_{3}^{r5} \right)_{xxxx}^{2},  
	\end{aligned}
\end{equation}
where 
\begin{equation}
	\begin{aligned}
		& {{\left( p_{3}^{r5} \right)}_{x}}=\left( -82{{{\bar{Q}}}_{-1}}+11{{{\bar{Q}}}_{-1}}+82{{{\bar{Q}}}_{-1}}-11{{{\bar{Q}}}_{-1}} \right)/120, \\ 
		& {{\left( p_{3}^{r5} \right)}_{xx}}=\left( 40{{{\bar{Q}}}_{-1}}-3{{{\bar{Q}}}_{-2}}-74{{{\bar{Q}}}_{0}}+40{{{\bar{Q}}}_{1}}-3{{{\bar{Q}}}_{2}} \right)/56, \\ 
		& {{\left( p_{3}^{r5} \right)}_{xxx}}=\left( 2{{{\bar{Q}}}_{-1}}-{{{\bar{Q}}}_{-2}}-2{{{\bar{Q}}}_{1}}+{{{\bar{Q}}}_{2}} \right)/12, \\ 
		& {{\left( p_{3}^{r5} \right)}_{xxxx}}=\left( -4{{{\bar{Q}}}_{-1}}+{{{\bar{Q}}}_{-2}}+6{{{\bar{Q}}}_{0}}-4{{{\bar{Q}}}_{1}}+{{{\bar{Q}}}_{2}} \right)/24, \\ 
	\end{aligned}
\end{equation}\par
The normalized weights are given by
\begin{equation}
	{{\bar{\omega }}_{k}}=\frac{{{\omega }_{k}}}{\sum\limits_{0}^{3}{{{\omega }_{k}}}}.
\end{equation}
Then the final form of the reconstructed polynomial is
\begin{equation}
	{{p}^{AO\left( 5,3 \right)}}\left( x \right)={{\bar{\omega }}_{3}}\left( \frac{1}{{{\gamma }_{3}}}p_{3}^{r5}\left( x \right)-\sum\limits_{0}^{2}{\frac{{{\gamma }_{k}}}{{{\gamma }_{3}}}p_{k}^{r3}\left( x \right)} \right)+\sum\limits_{0}^{2}{{{{\bar{\omega }}}_{k}}p_{k}^{r3}\left( x \right)}.
\end{equation}
The desired values at the cell interface can be fully written as 
\begin{equation}
	Q_{i-1/2}^{r}={{P}^{AO\left( 5,3 \right)}}\left( {{x}_{i-1/2}} \right),Q_{i+1/2}^{l}={{P}^{AO\left( 5,3 \right)}}\left( {{x}_{i+1/2}} \right).
\end{equation}\par
The WENO-AO reconstruction procedure is over because it is applied to schemes with Riemann solvers where only point-wise values are needed. In order to calculate the GKS flux, we supplement the derivatives at the cell interfaces on the large stencil and sub-stencils as follows
\begin{equation}
	\begin{aligned}
		& \left( {{p}_{x}} \right)_{0}^{r3}\left( {{x}_{i+1/2}} \right)=\left( {{{\bar{Q}}}_{i-2}}-3{{{\bar{Q}}}_{i-1}}+2{{{\bar{Q}}}_{i}} \right)/\Delta x,\left( {{p}_{x}} \right)_{0}^{r3}\left( {{x}_{i-1/2}} \right)=\left( -{{{\bar{Q}}}_{i-1}}+{{{\bar{Q}}}_{i}} \right)/\Delta x, \\ 
		& \left( {{p}_{x}} \right)_{1}^{r3}\left( {{x}_{i+1/2}} \right)=\left( -{{{\bar{Q}}}_{i}}+{{{\bar{Q}}}_{i+1}} \right)/\Delta x,\left( {{p}_{x}} \right)_{1}^{r3}\left( {{x}_{i-1/2}} \right)=\left( -{{{\bar{Q}}}_{i-1}}+{{{\bar{Q}}}_{i}} \right)/\Delta x, \\ 
		& \left( {{p}_{x}} \right)_{2}^{r3}\left( {{x}_{i+1/2}} \right)=\left( -{{{\bar{Q}}}_{i}}+{{{\bar{Q}}}_{i+1}} \right)/\Delta x,\left( {{p}_{x}} \right)_{2}^{r3}\left( {{x}_{i-1/2}} \right)=\left( -2{{{\bar{Q}}}_{i}}+3{{{\bar{Q}}}_{i+1}}-{{{\bar{Q}}}_{i+2}} \right)/\Delta x. \\ 
	\end{aligned}
\end{equation}
\begin{equation}
	\begin{aligned}
		& \left( {{p}_{x}} \right)_{3}^{r5}\left( {{x}_{i+1/2}} \right)=\frac{1}{12\Delta x}\left( {{{\bar{Q}}}_{i-1}}-15{{{\bar{Q}}}_{i}}+15{{{\bar{Q}}}_{i+1}}-{{{\bar{Q}}}_{i+2}} \right), \\ 
		& \left( {{p}_{x}} \right)_{3}^{r5}\left( {{x}_{i-1/2}} \right)=\frac{1}{12\Delta x}\left( {{{\bar{Q}}}_{i-2}}-15{{{\bar{Q}}}_{i-1}}+15{{{\bar{Q}}}_{i}}-{{{\bar{Q}}}_{i+1}} \right). \\ 
	\end{aligned}
\end{equation}
The desired derivatives at the cell interfaces can be fully determined as
\begin{equation}
	{{\left( Q_{x}^{r} \right)}_{i-1/2}}=P_{x}^{^{AO\left( 5,3 \right)}}\left( {{x}_{i-1/2}} \right),{{\left( Q_{x}^{l} \right)}_{i+1/2}}=P_{x}^{^{AO\left( 5,3 \right)}}\left( {{x}_{i+1/2}} \right).
\end{equation}\par
The non-equilibrium states are obtained by either the upwind linear reconstruction or the WENO-AO reconstruction, and the equilibrium states can be obtained by the following simple method. 
\begin{equation}\label{equ3.25}
	\int{\psi }{{g}^{0}}d\Xi ={{\mathbf{W}}^{0}}=\int_{u>0}{\psi {{g}^{l}}d\Xi }+\int_{u>0}{\psi {{g}^{r}}d\Xi },
\end{equation}
\begin{equation}\label{equ3.26}
	\int{\psi }g_{x}^{0}d\Xi =\mathbf{W}_{x}^{0}=\int_{u>0}{\psi g_{x}^{l}d\Xi }+\int_{u>0}{\psi g_{x}^{r}d\Xi },
\end{equation}
where ${{g}^{0}},g_{x}^{0},g_{xx}^{0}\ldots $ are the equilibrium states and ${{g}^{l,r}},g_{x}^{l,r},g_{xx}^{l,r}\ldots $ are the non-equilibrium states. This is a kinetic-based weighting of the values and derivatives on the left and right sides of the cell interface, while introducing the upwind mechanics. Arithmetic averaging can also be used for smooth flow. In classic WENO5-GKS~\cite{PAN2016197}, an extra linear polynomial reconstruction for the equilibrium states is required. This method requires no additional process and achieves a 5th-order spatial accuracy for the equilibrium states. In the above way, all components of the microscopic slopes across the interface have been obtained.\par
The HGKS with WENO-AO for 2-D reconstruction in ~\cite{CiCP-28-539} are written as follows.\par
In the two-dimensional case, the values of each Gaussian point $\left( {{x}_{i+1/2}},{{y}_{jm}} \right),m=1,2$ which need to be obtained by reconstruction are
$${{W}^{l}},W_{x}^{l},W_{y}^{l},{{W}^{r}},W_{x}^{r},W_{y}^{r}.$$
The scheme for obtaining the values of Gaussian points is as follows through dimension-by-dimensional reconstruction as follows. The time level is omitted here.\\
{\bfseries Step 1.} According to the one-dimensional WENO-AO reconstruction, the line averaged reconstructed values and slopes
$${{\left( {{Q}^{l}} \right)}_{i+1/2,j}},{{\left( Q_{x}^{l} \right)}_{i+1/2,j}},{{\left( {{Q}^{r}} \right)}_{i+1/2,j}},{{\left( Q_{x}^{r} \right)}_{i+1/2,j}}$$
can be obtained along the normal direction by using the cell averaged values ${{\left( {\bar{Q}} \right)}_{i+l,j}}$ and ${{\left( {\bar{Q}} \right)}_{i+l+1,j}}$,$l=-2,\ldots ,2$.\\
{\bfseries Step 2.} Again with the one-dimensional WENO-AO reconstruction, the values at each Gaussian point
$${{\left( {{Q}^{l}} \right)}_{i+1/2,jm}},{{\left( Q_{y}^{l} \right)}_{i+1/2,jm}},{{\left( {{Q}^{r}} \right)}_{i+1/2,jm}},{{\left( Q_{y}^{r} \right)}_{i+1/2,jm}}$$
with $y={{y}_{jm}},m=0,1$ can be obtained by using the line averaged values ${{\left( {{Q}^{l}} \right)}_{i+1/2,j+l}},{{\left( {{Q}^{r}} \right)}_{i+1/2,j+l}},l=-2,\ldots ,2$ constructed above. In the same way, the point-wise derivatives at Gaussian point $\left( {{x}_{i+1/2}},{{y}_{jm}} \right),m=1,2$ ${{\left( Q_{x}^{l} \right)}_{i+1/2,jm}},{{\left( Q_{x}^{r} \right)}_{i+1/2,jm}}$ can be constructed by using the above line averaged derivatives ${{\left( Q_{x}^{l} \right)}_{i+1/2,j+l}},{{\left( Q_{x}^{r} \right)}_{i+1/2,j+l}},l=-2,\ldots ,2$ with the  WENO-AO method. The details about tangential reconstruction are described in ~\cite{CiCP-28-539}.\\
{\bfseries Step 3.}. The quantities related to the non-equilibrium states at each Gaussian point are all obtained. And then the quantities related to the equilibrium states can be obtained by the unified weighting Eq. (\ref{equ3.25}) and Eq. (\ref{equ3.26}).\\
\section{HGKS with fifth-order TENO schemes}
\label{sec4}
\subsection{TENO reconstruction for GKS}
\label{sec4.1}
The TENO scheme has been systematically introduced by Fu et al.~\cite{FU2016333,FU2018724}. This framework allows for the achievement of arbitrary high-order spatial accuracy by utilizing a set of low-order stencils with incrementally increasing width. In this section, we introduce the fifth-order TENO scheme, which employs a simple yet effective procedure.\par
Inspired by Hu et al.~\cite{FU2016333} and Borges et al. ~\cite{BORGES20083191}, the smoothness measurement of fifth order TENO scheme is given as
\begin{equation}\label{equ4.1}
{{\gamma }_{k}}={{\left( C+\frac{\tau }{{{\beta }_{k}}+\varepsilon } \right)}^{q}},k=0,1,2.
\end{equation}
It can be found the parameters $\tau $ and ${{\beta }_{k}}$ of WENO-JS scheme or WENO-Z scheme have been reused, and the small threshold remains the value of WENO-Z scheme, i.e. $\varepsilon ={{10}^{-40}}$. $C$ is set as 1, and the integer power $q$ is set as 6. It should be mentioned that, for fifth-order TENO scheme, the local smooth indicator of WENO-JS scheme, i.e., ${{\beta }_{k}}$, is completely reused. 
However, for higher-order TENO schemes, the application of incremented-width stencils leads to a slightly different unified formulation compared to classical WENO schemes.\par
In order to recover the optimal weight in smooth region, TENO scheme does not
directly use the weights in Eq. (\ref{equ4.1}). The measurement in Eq. (\ref{equ4.1}) is normalised at first, i.e.
\begin{equation}\label{equ4.2}
{{\chi }_{k}}=\frac{{{\gamma }_{k}}}{\sum\nolimits_{k=0}^{2}{{{\gamma }_{k}}}},
\end{equation}
and then a cut-off function is defined as
\begin{equation}\label{equ4.3}
{{\delta }_{k}}=\left\{ \begin{aligned}
	& 0,if{{\chi }_{k}}<{{C}_{T}}, \\ 
	& 1,otherwise. \\ 
\end{aligned} \right.
\end{equation}
Finally, the weights of TENO scheme for Eq. (\ref{equ3.9}) are defined by a normalizing procedure
\begin{equation}\label{equ4.4}
{{\bar{\omega }}_{k}}=\frac{{{d}_{k}}{{\delta }_{k}}}{\sum\nolimits_{k=0}^{2}{{{d}_{k}}{{\delta }_{k}}}},
\end{equation}
where the optimal weights are utilised without rescaling, and only the stencil containing discontinuity is removed from the final reconstruction completely. As a result, the TENO scheme ensures numerical robustness and fully recovers the optimal weight, denoted as ${{d}_{k}}$, as well as accuracy and spectral properties in smooth regions, including at smooth critical points.\par
It can be found that parameter ${{C}_{T}}$ is also an effective and a direct mean to control the spectral properties of TENO scheme for a specific problem, e.g. compressible turbulence simulation in which embedded shocklets need to be captured without increasing overall dissipation. Haimovich and Frankel~\cite{HAIMOVICH2017105} has conducted a series of numerical cases, in which the TENO solution with ${{C}_{T}}={{10}^{-3}}$ is still superior in comparison to the WENO-Z solution. In this paper, the parameter is simply set as ${{C}_{T}}={{10}^{-7}}$ for all the simulations without detailed discussion.\par
For TENO5 GKS, the TENO scheme, like the WENO-Z scheme in Section \ref{sec3.1}, is only responsible for providing the values of the non-equilibrium state ${{W}^{l}}$ and ${{W}^{r}}$ at the cell interface. The rest of the variables for non-equilibrium state and equilibrium state $W_{x}^{l},W_{y}^{l},{{W}^{r}},W_{x}^{r},W_{y}^{r}$ construction process required to solve the GKS flux is the same as WENOZ-GKS in Section \ref{sec3.1}.
\subsection{TENO-D reconstruction for GKS}
\label{sec4.2}
In ~\cite{JSC,FU2021114193}, the candidates include a large central-biased stencil and a set of small directional stencils. The targeted high-order reconstruction is built upon the large stencil to effectively resolve smooth scales in the low-wave number range. On the other hand, several low-order reconstructions are designed based on the small directional stencils. The compactness of these small stencils is beneficial for capturing discontinuities, as they are less likely to be crossed by such discontinuities compared to the large central-biased stencil. The nonlinear adaptation among these small directional stencils, facilitated by the new TENO weighting strategy, ensures the preservation of the ENO property.\par
 Building upon the aforementioned ideas, we construct TENO5-D GKS , which shares similarities with WENO5-AO GKS. For the fifth-order TENO-D GKS scheme, the large central-biased stencil takes ${{S}_{3}}=\left( {{I}_{i-2}},{{I}_{i-1}},{{I}_{i}},{{I}_{i+1}},{{I}_{i+2}} \right)$. And three substencils are ${{S}_{0}}=\left( {{I}_{i-2}},{{I}_{i-1}},{{I}_{i}} \right),{{S}_{1}}=\left( {{I}_{i-1}},{{I}_{i}},{{I}_{i+1}} \right),{{S}_{2}}=\left( {{I}_{i}},{{I}_{i+1}},{{I}_{i+2}} \right)$. For the TENO concept, an efficient scale separation procedure that effectively distinguishes discontinuities from smooth flow scales is crucial for accurate shock wave capture~\cite{FU2021114193}. This paper defines the scale-separation formula as follows:
\begin{equation}\label{equ4.5}
{{\gamma }_{k}}=\frac{1}{{{\left( {{\beta }_{k}}+\varepsilon  \right)}^{q}}}
\end{equation}
where $\varepsilon ={{10}^{-40}}$ is introduced to avoid the zero denominator and $q$ denotes the total candidate stencil number. Unlike that in the original WENO-JS schemes~\cite{SHU1988439}, $q$ is set as 7 to achieve sufficient scale separation rather than 2. ${{\beta }_{k}}$, which measures the smoothness of each candidate stencil, can be evaluated by Eq. (\ref{equ3.7}).\par
Secondly, the smoothness indicators are normalized as
\begin{equation}\label{equ4.6}
{{\chi }_{k,1}}=\frac{{{\gamma }_{k}}}{\sum\nolimits_{k=0}^{3}{{{\gamma }_{k}}}}
\end{equation}
and then filtered by a sharp cutoff function\par
\begin{equation}\label{equ4.7}
	{{\delta }_{k,1}}=\left\{ \begin{aligned}
		& 0,if{{\chi }_{k,1}}<{{C}_{T,1}}, \\ 
		& 1,otherwise, \\ 
	\end{aligned} \right.
\end{equation}
where the cut-off parameter ${{C}_{T,1}}$ determines the nonlinear adaptation and can be set as ${{C}_{T,1}}={{10}^{-7}}$. If the large candidate stencil is judged to be smooth, i.e., ${{\delta }_{3,1}}=1$, the final high-order reconstruction on the cell interface can be given by
\begin{equation}\label{equ4.8}
	Q_{i+1/2}^{l}=\frac{1}{60}\left( 2{{{\bar{Q}}}_{i-2}}-13{{{\bar{Q}}}_{i-1}}+47{{{\bar{Q}}}_{i}}+27{{{\bar{Q}}}_{i+1}}-3{{{\bar{Q}}}_{i+2}} \right)
\end{equation}
\begin{equation}\label{equ4.9}
{{\left( Q_{x}^{l} \right)}_{i+1/2}}=\frac{1}{12\Delta x}\left( {{{\bar{Q}}}_{i-1}}-15{{{\bar{Q}}}_{i}}+15{{{\bar{Q}}}_{i+1}}-{{{\bar{Q}}}_{i+2}} \right)
\end{equation}\par
Under this circumstance, the desirable high-order accuracy is restored exactly without any compromise for resolving smooth flow scales. Otherwise, if ${{\delta }_{3,1}}=0$, it indicates the presence of a discontinuity within the large candidate stencil. To ensure the ENO property for capturing such discontinuities, a second ENO-like stencil selection is applied to the remaining small upwind stencils. Similarly, the smoothness indicators are first normalized as
\begin{equation}\label{equ4.11}
{{\chi }_{k,2}}=\frac{{{\gamma }_{k}}}{\sum\nolimits_{k=0}^{3-1}{{{\gamma }_{k}}}}
\end{equation}
and then filtered by a sharp cutoff function
\begin{equation}\label{equ4.12}
{{\delta }_{k,2}}=\left\{ \begin{aligned}
	& 0,if{{\chi }_{k,2}}<{{C}_{T,2}}, \\ 
	& 1,otherwise, \\ 
\end{aligned} \right.
\end{equation}
where the cut-off parameter ${{C}_{T,2}}$ determines the nonlinear adaptation and ${{C}_{T,2}}={{10}^{-5}}$. Here, ${{C}_{T,2}}$ is larger than ${{C}_{T,1}}$ for a stronger separation of the discontinuities. In other words, sufficient numerical dissipation is generated for capturing discontinuities stably and sharply.\par
Consequently, the final reconstructed data at the cell interface is given by the nonlinear combination of small candidate stencils as
\begin{equation}\label{equ4.13}
Q_{_{i+1/2}}^{l}=\sum\limits_{0}^{2}{{{{\bar{\omega }}}_{k}}p_{k}^{r3}}\left( {{x}_{i+1/2}} \right),
\end{equation}
\begin{equation}\label{equ4.14}
{{\left( Q_{x}^{l} \right)}_{i+1/2}}=\sum\limits_{0}^{2}{{{{\bar{\omega }}}_{k}}\left( {{p}_{x}} \right)_{k}^{r3}}\left( {{x}_{i+1/2}} \right),
\end{equation}
where
$${{\bar{\omega }}_{k}}=\frac{{{d}_{k}}{{\delta }_{k}}}{\sum\nolimits_{k=0}^{3-1}{{{d}_{k}}{{\delta }_{k}}}},$$
and the optimal linear weights ${{d}_{k}}$ can be simply determined as ${{d}_{k}}=1/3$. \par
The TENO-D GKS and WENO-AO GKS methods have already computed the point value and its derivative during the reconstruction process. Moreover, both methods can readily obtain the Gaussian point value in the tangential reconstruction. Hence, for two-dimensional reconstruction using TENO-D GKS, the remaining reconstruction process is identical to that of WENO-AO GKS in Section \ref{sec3.2}.

\section{Numerical tests}
\label{sec5}
In this section, 1-D and 2-D numerical tests will be presented to validate the WENO5-Z GKS, WENO5-AO GKS, TENO5-GKS and TENO5-D GKS. For the parameters of the HGKS in the follow tests, the ratio of specific heats takes $\gamma =1.4$. For the inviscid flow, the collision time $\tau $ is
$$\tau ={{c}_{1}}\Delta t+{{c}_{2}}\left| \frac{{{p}_{l}}-{{p}_{r}}}{{{p}_{l}}+{{p}_{r}}} \right|\Delta t,$$
where ${{p}_{l}}$ and ${{p}_{r}}$ denote the pressure on the left and right cell interface. Usually ${{c}_{1}}=0.01$ and ${{c}_{2}}=1$ are chosen in the classic HGKS. The pressure jump term in $\tau $ can add artificial dissipation to enlarge the shock thickness to the scale of numerical cell size in the discontinuous region. Besides, it can keep the non-equilibrium dynamics in the shock layer through the kinetic particle transport to mimic the real physical mechanism inside the shock structure. \par
 For the viscous flow~\cite{CiCP-28-539,YANG2022110706}, the collision time term related to the viscosity coefficient is defined as 
$$\tau =\frac{\mu }{p}+{{c}_{2}}\left| \frac{{{p}_{l}}-{{p}_{r}}}{{{p}_{l}}+{{p}_{r}}} \right|\Delta t,$$
where $\mu $ is the dynamic viscous coefficient and $p$ is the pressure at the cell interface. In smooth viscous flow region, it reduces to $\tau =\mu /p$. The time step is determined by
$$\Delta t={{C}_{CFL}}Min\left( \frac{\Delta x}{\left\| \mathbf{U} \right\|+{{a}_{s}}},\frac{{{\left( \Delta x \right)}^{2}}}{4\nu } \right),$$
where $\left\| \mathbf{U} \right\|$ is the magnitude of velocities, ${{C}_{CFL}}$ is the CFL number, ${{a}_{s}}$ is the sound speed and $\nu =\mu /\rho$ is the kinematic viscosity coefficient.

\subsection{Accuracy test in 1-D}
\label{sec5.1}
The advection of density perturbation is tested whose initial condition is set as follows
$$\rho \left( x \right)=1+0.2sin\left( \pi x \right),U\left( x \right)=1,p\left( x \right)=1,x\in \left[ 0,2 \right].$$
Both the left and right sides of the test case are periodic boundary conditions. The analytic solution of the advection of density perturbation is
$$\rho \left( x,t \right)=1+0.2sin\left( \pi \left( x-t \right) \right),U\left( x,t \right)=1,p\left( x,t \right)=1,x\in \left[ 0,2 \right].$$
In the computation, a uniform mesh with $N$  points are used. The time step $\Delta t=0.2\Delta x$ is fixed. The collision time $\tau =0$ is set since the flow is smooth and inviscid. Based on the two-stage fourth-order time-marching method, the HGKS with WENO-Z, WENO-AO, TENO and TENO-D method is expected to achieve the same fifth-order spatial accuracy and fourth-order temporal accuracy as analyzed in ~\cite{PAN2016197}. The ${{L}^{1}},{{L}^{2}}$ and ${{L}^{\infty }}$ errors and corresponding orders at $t=2.0$ are given in the follow tables. With the mesh refinement in Table \ref{1Daccuracy01}-Table \ref{1Daccuracy04}, the expected orders of accuracy are obtained and the numerical errors are identical.\\
\begin{table}
\begin{tabular}{c|cc|cc|cc}
	\hline
	mesh length & ${{L}^{1}}$error & Order & ${{L}^{2}}$error & Order & ${{L}^{\infty }}$error & Order\\
	\hline
	1/5 & 1.1127301e-03 &  & 1.2229163e-03 &  & 1.7483320e-03 & \\ 
	1/10 & 3.0430860e-05 & 5.19 & 3.4930110e-05 & 5.13 & 5.2960032e-05 & 5.04 \\
	1/20 & 9.0136622e-07 & 5.08 & 1.0151350e-06 & 5.10 & 1.5138367e-06 & 5.13\\
	1/40 & 2.8089445e-08 & 5.00 & 3.1209611e-08 & 5.02 & 4.6531500e-08 & 5.02\\
	1/80 & 8.7828277e-10 & 5.00 & 9.7371574e-10 & 5.00 & 1.4489871e-09 & 5.01\\
	\hline
\end{tabular}
\centering 
	\caption{\label{1Daccuracy01}Accuracy test in 1-D for the advection of density perturbation by the WENO-Z reconstruction. $\Delta t=0.2\Delta x.$}
\end{table}
\begin{table}
\begin{tabular}{c|cc|cc|cc}
	\hline
	mesh length & ${{L}^{1}}$error & Order & ${{L}^{2}}$error & Order & ${{L}^{\infty }}$error & Order\\
	\hline
	1/5 & 8.8980542e-04 &  & 9.9719282e-04 &  & 1.3990624e-03 & \\ 
	1/10 & 2.8462550e-05 & 4.97 & 3.1616280e-05 & 4.98 & 4.6506743e-05 & 4.91 \\
	1/20 & 8.9766689e-07 & 4.99 & 9.9424894e-07 & 4.99 & 1.4737229e-06 & 4.98\\
	1/40 & 2.8078509e-08 & 5.00 & 3.1116562e-08 & 5.00 & 4.6205231e-08 & 5.00\\
	1/80 & 8.7827033e-10 & 5.00 & 9.7334592e-10 & 5.00 & 1.4455303e-09 & 5.00\\
	\hline
\end{tabular}
\centering 
	\caption{\label{1Daccuracy02}Accuracy test in 1-D for the advection of density perturbation by the WENO-AO reconstruction. $\Delta t=0.2\Delta x.$}
\end{table}
\begin{table}
	\begin{tabular}{c|cc|cc|cc}
		\hline
		mesh length & ${{L}^{1}}$error & Order & ${{L}^{2}}$error & Order & ${{L}^{\infty }}$error & Order\\
		\hline
		1/5 & 8.5618115e-04 &  & 9.6826196e-04 &  & 1.3768690e-03 & \\ 
		1/10 & 2.8377750e-05 & 4.92 & 3.1550478e-05 & 4.94 & 4.6413792e-05 & 4.89 \\
		1/20 & 8.9738457e-07 & 4.98 & 9.9404369e-07 & 4.99 & 1.4727714e-06 & 4.98\\
		1/40 & 2.8075789e-08 & 5.00 & 3.1115052e-08 & 5.00 & 4.6193336e-08 & 5.00\\
		1/80 & 8.7823923e-10 & 5.00 & 9.7332582e-10 & 5.00 & 1.4453295e-09 & 5.00\\
		\hline
	\end{tabular}
	\centering 
	\caption{\label{1Daccuracy03}Accuracy test in 1-D for the advection of density perturbation by the TENO reconstruction. $\Delta t=0.2\Delta x.$}
\end{table}
\begin{table}
	\begin{tabular}{c|cc|cc|cc}
		\hline
		mesh length & ${{L}^{1}}$error & Order & ${{L}^{2}}$error & Order & ${{L}^{\infty }}$error & Order\\
		\hline
		1/5 & 8.5627987e-04 &  & 9.6840762e-04 &  & 1.3781734e-03 & \\ 
		1/10 & 2.8397550e-05 & 4.91 & 3.1558226e-05 & 4.94 & 4.6497543e-05 & 4.89 \\
		1/20 & 8.9755629e-07 & 4.98 & 9.9413704e-07 & 4.99 & 1.4736869e-06 & 4.98\\
		1/40 & 2.8078309e-08 & 5.00 & 3.1116332e-08 & 5.00 & 4.6205146e-08 & 5.00\\
		1/80 & 8.7827023e-10 & 5.00 & 9.7334572e-10 & 5.00 & 1.4455303e-09 & 5.00\\
		\hline
	\end{tabular}
	\centering 
	\caption{\label{1Daccuracy04}Accuracy test in 1-D for the advection of density perturbation by the TENO-D reconstruction. $\Delta t=0.2\Delta x.$}
\end{table}\par
The reference solutions for the following one dimensional Riemann problems are obtained using classic WENO5-GKS with 10,000 uniform mesh points.
\subsection{Shock-tube problem}
\label{sec5.2} 
\emph{(a) Lax problem}\par
\begin{figure}[htbp]
	\centering
	\subfigure{\includegraphics[width=0.4\textwidth]{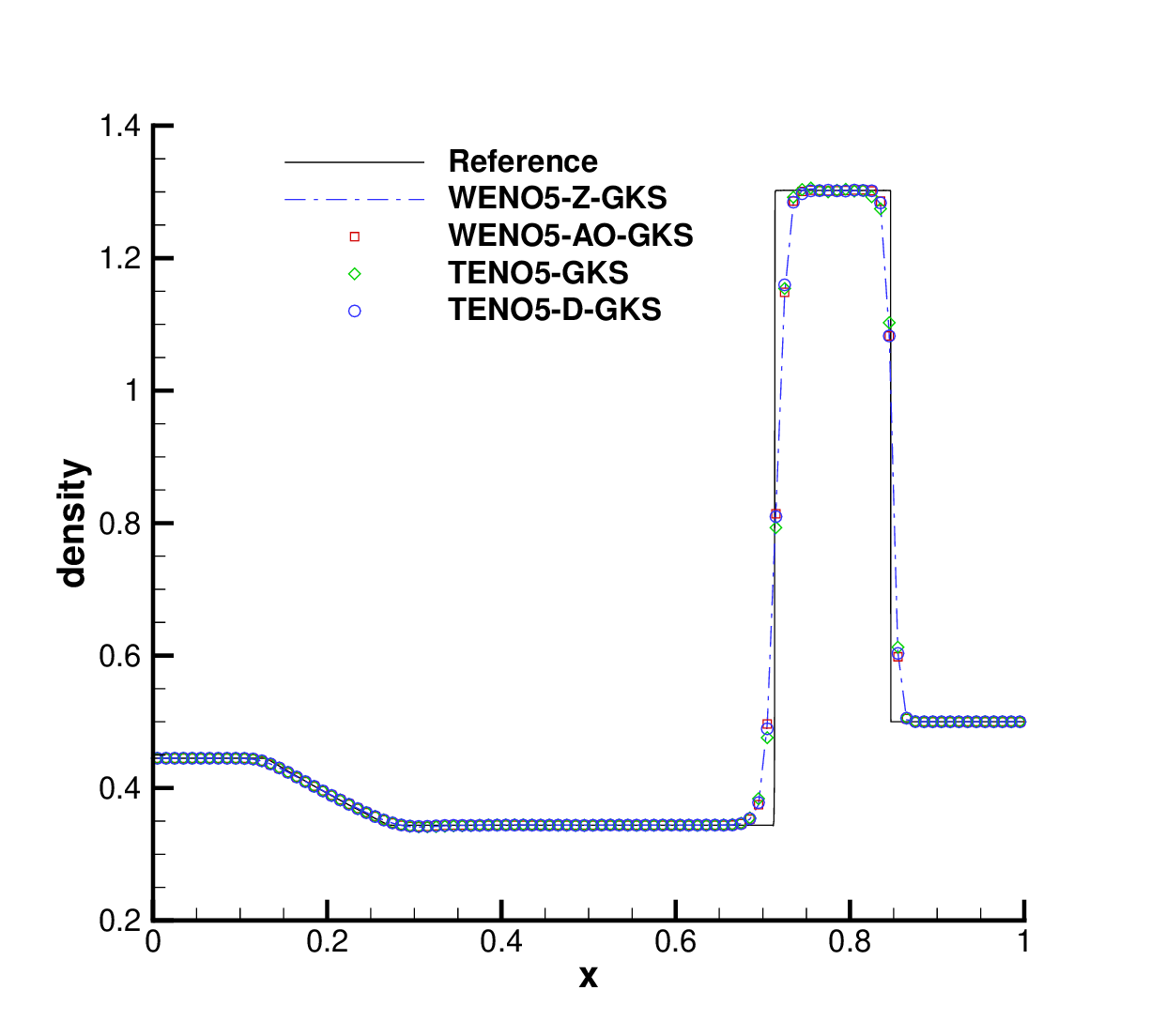}}
	\subfigure{\includegraphics[width=0.4\textwidth]{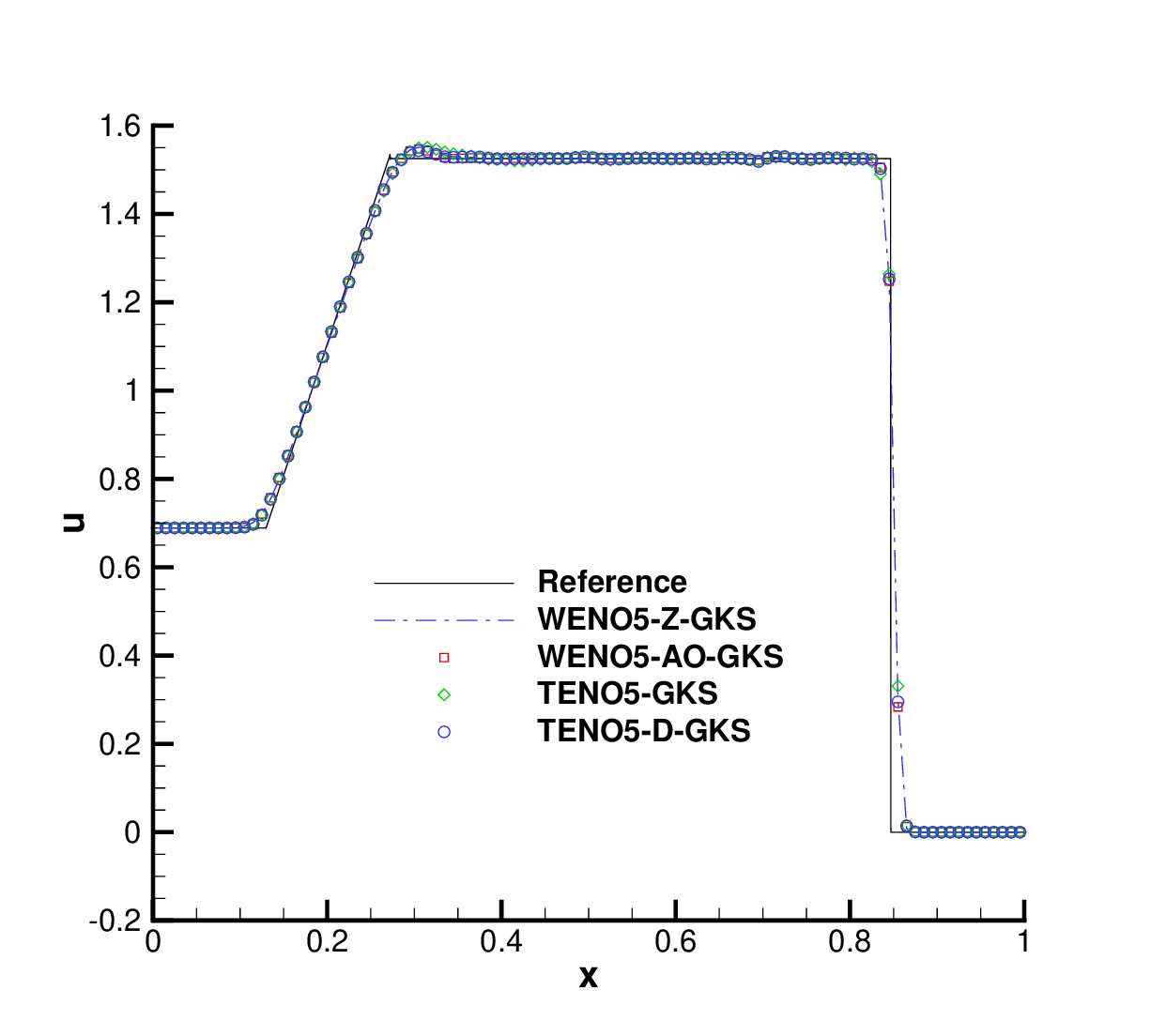}}
	\caption{Lax problem: the density and velocity distributions with 100 cells. $CFL=0.5.$ $T=0.14.$ }
	\label{figLAX}
\end{figure}

The computational domain for the Lax problem~\cite{cpa.3160070112} is $\left[ 0,\ 1 \right]$ with 100 uniform mesh points. The solutions of the Sod problem are presented at $t=0.14$ with non-reflecting boundary condition on both ends. The initial condition is given by
$$\left( \rho ,U,p \right)=\left\{ \begin{aligned}
	& \left( 0.445,0.689,3.528 \right),0<x<0.5, \\ 
	& \left( 0.5,0,0.571 \right),0.5<x<1. \\ 
\end{aligned} \right.$$
In Fig. \ref{figLAX}, both the TENO5 GKS and TENO5-D GKS schemes show sharp shock-capturing property and the computed solutions agree well with the references. They show performance close to that of the GKS with WENO schemes.\\
\emph{(b) Sod problem}\par
The initial condition for the Sod problem~\cite{SOD19781} is 
 $$\left( \rho ,U,p \right)=\left\{ \begin{aligned}
  & \left( 1,0,1 \right),0<x<0.5, \\ 
 & \left( 0.125,0,0.1 \right),0.5<x<1. \\ 
\end{aligned} \right.$$
The solutions of the Sod problem are presented at $t=0.2$ with non-reflecting boundary condition on both ends.\par
\begin{figure}[htbp]
	\centering
	\subfigure{\includegraphics[width=0.4\textwidth]{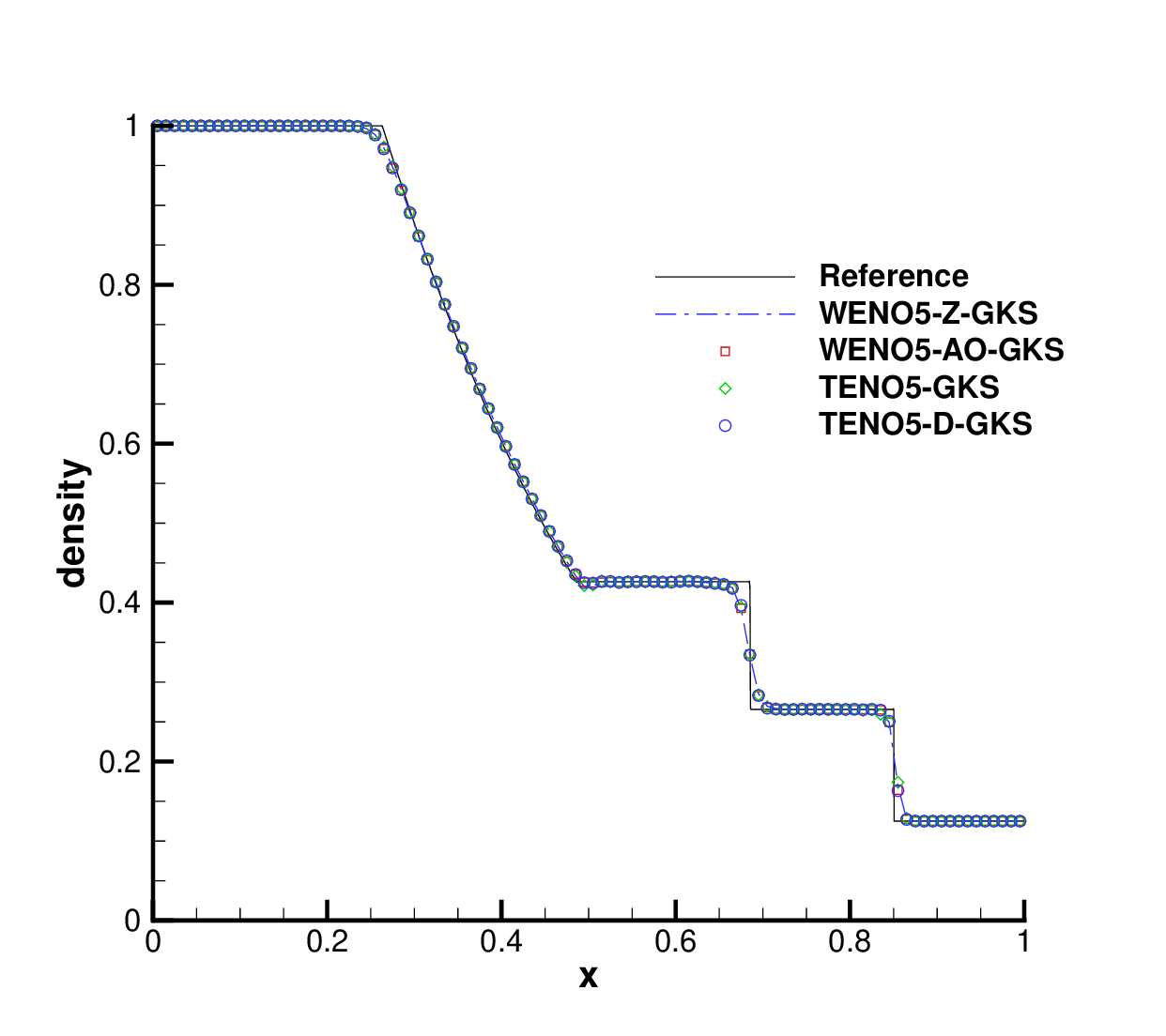}}
	\subfigure{\includegraphics[width=0.4\textwidth]{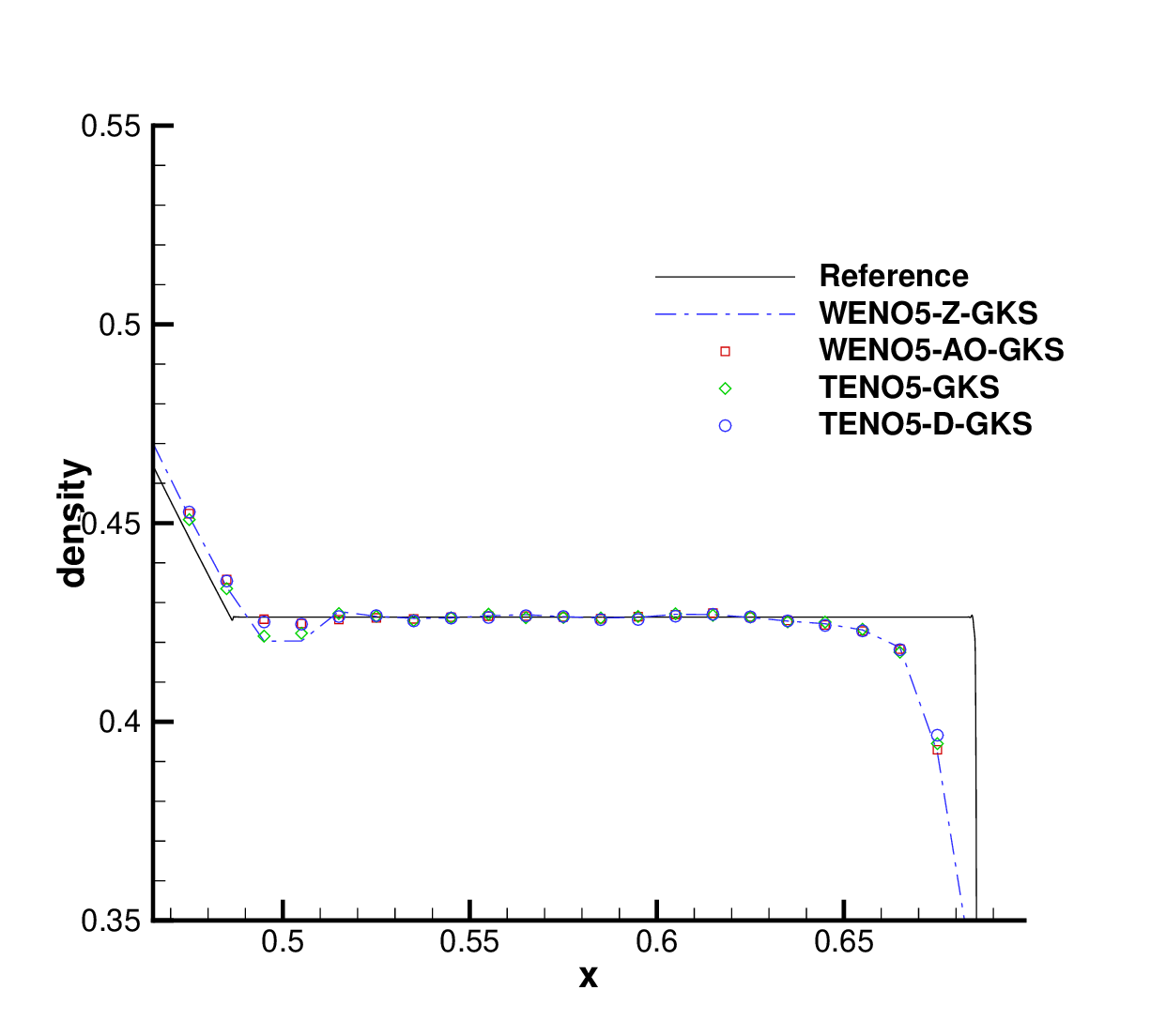}}
	\subfigure{\includegraphics[width=0.4\textwidth]{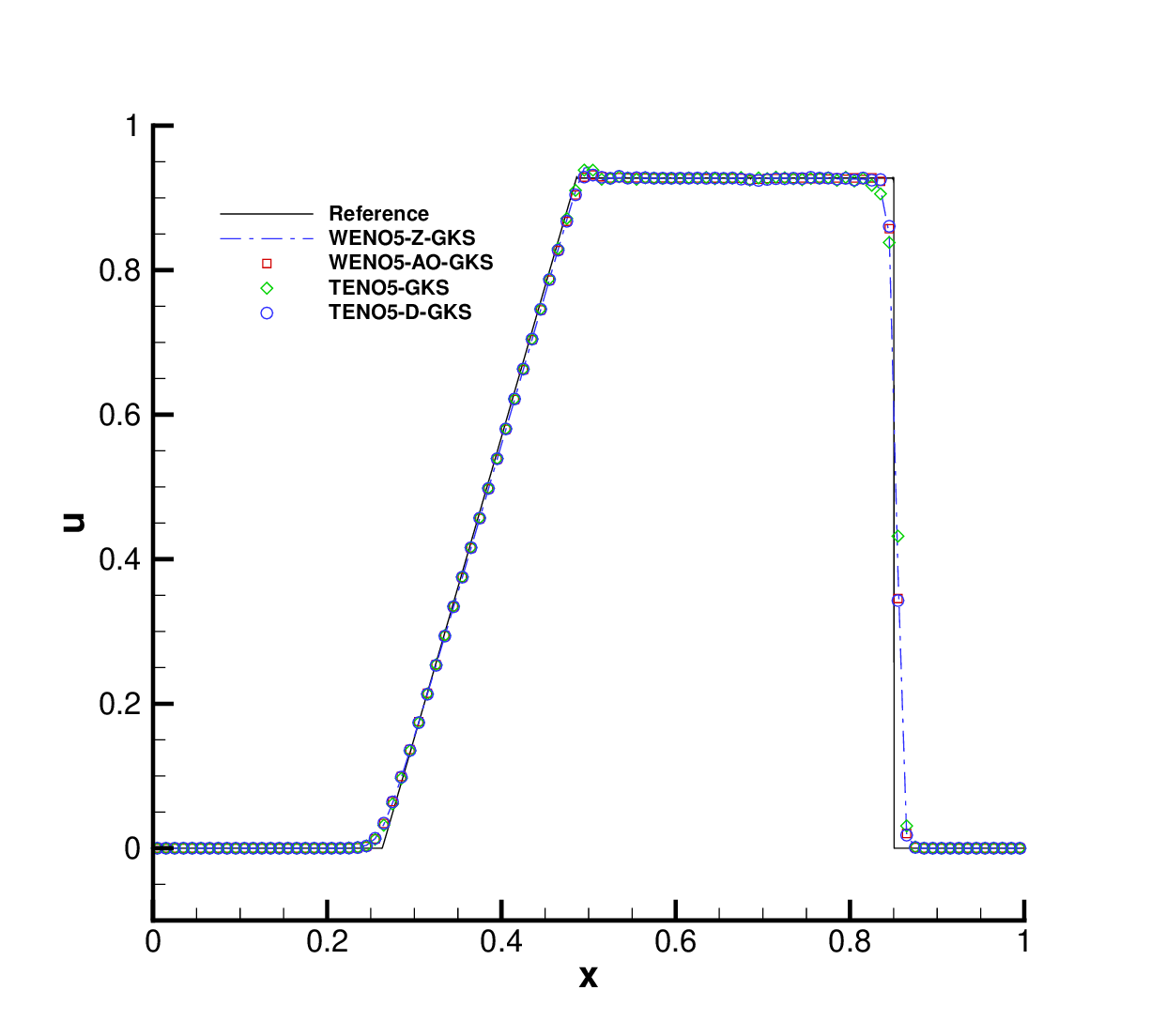}}
	\subfigure{\includegraphics[width=0.4\textwidth]{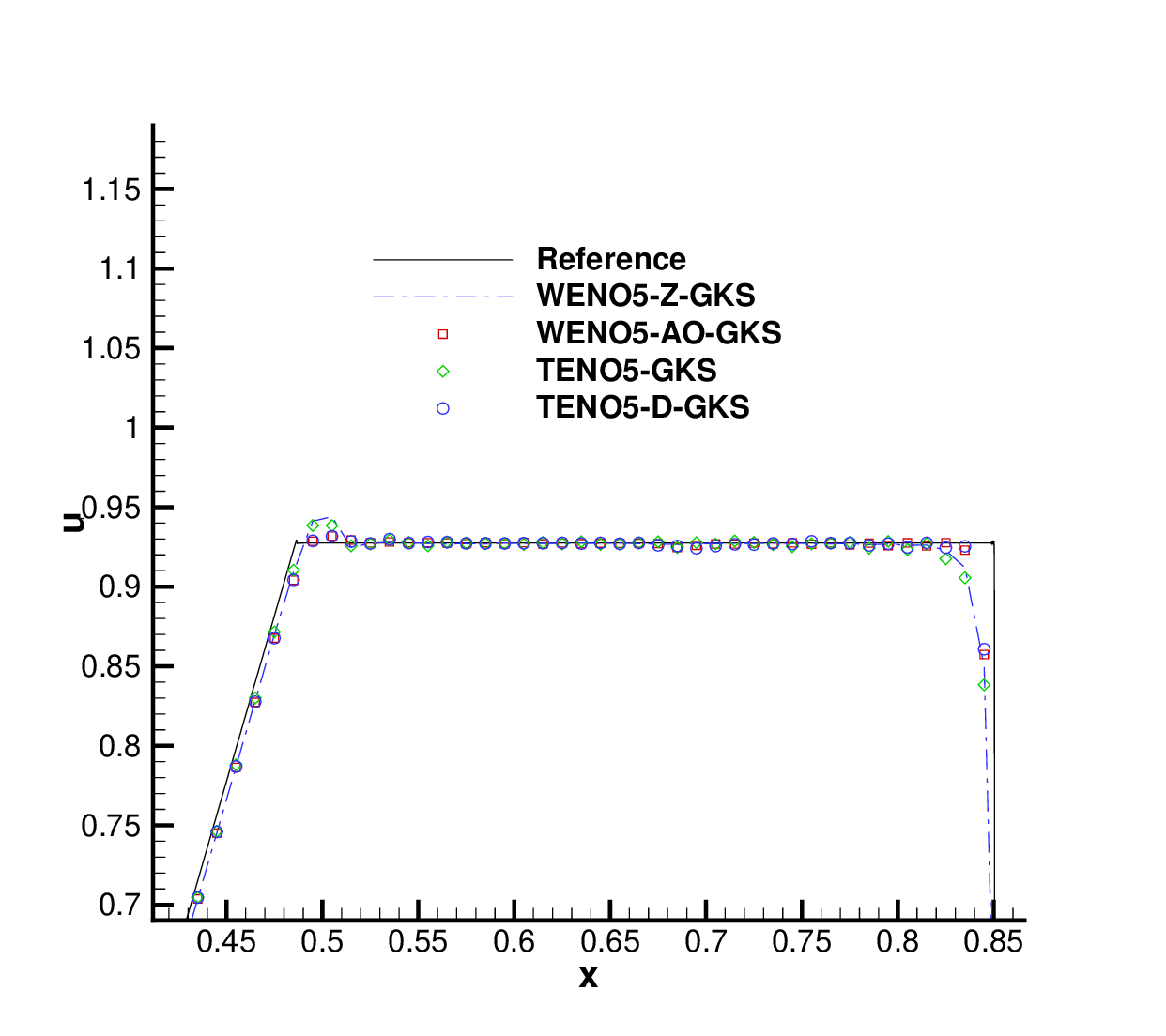}}
	\caption{Sod problem: the density and velocity distributions and local enlargements with 100 cells. $CFL=0.5.$ $T=0.2.$ }
	\label{figSOD}
\end{figure}

 
In Fig. \ref{figSOD}, a comparison is made between the results of the WENO5-Z GKS, WENO5-AO GKS, TENO5 GKS and TENO5-D GKS. Overall, the results of the four reconstruction methods are in good agreement with the reference solutions. From the local enlargements in Fig. \ref{figSOD}, it is observed that the solutions from the WENO5-Z GKS and TENO5 GKS show undershoot or overshoot around the corner of the rarefaction wave. The results obtained using the WENO5-AO GKS and the TENO5-D GKS are almost identical due to the kinetic-weighting method, which is analyzed in ~\cite{CiCP-28-539}.\\
\subsection{ Shock–density wave interaction}
\label{sec5.3} 
\emph{(a) Shu-Osher problem}\par
\begin{figure}[htbp]
	\centering
	\subfigure{\includegraphics[width=0.4\textwidth]{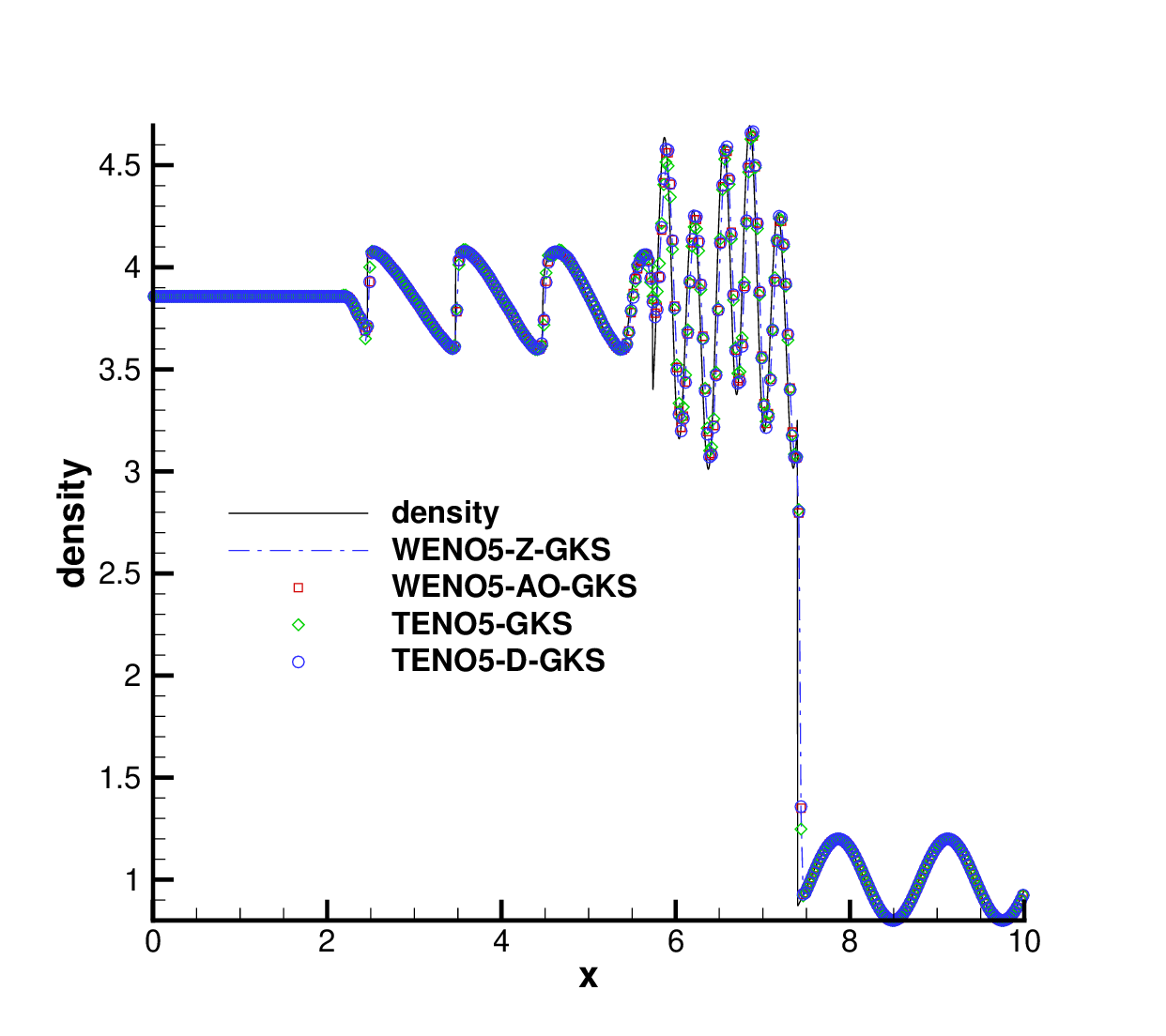}}
	\subfigure{\includegraphics[width=0.4\textwidth]{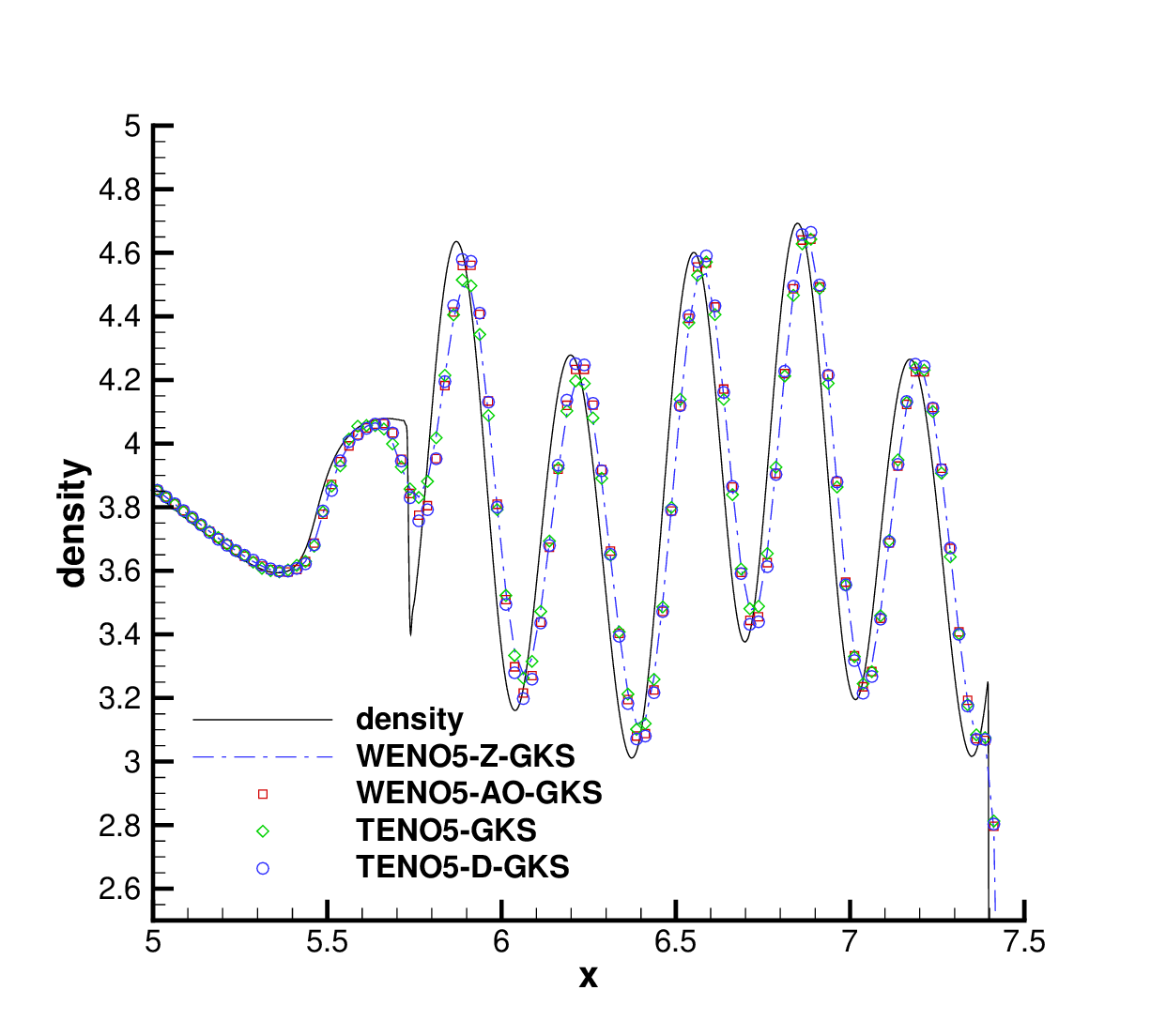}}
	\caption{Shu-Osher problem: the density distributions and local enlargements for different reconstruction schemes with 400 cells. $CFL=0.5.$ $T=1.8.$ }
	\label{figshu}
\end{figure}

The Shu-Osher shock acoustic interaction~\cite{SHU198932} is computed in $[0,\ 10]$ with 400 mesh points. The non-reflecting boundary condition is given on the left, and the fixed wave profile is extended on the right. The initial conditions are
 $$\left( \rho ,U,p \right)=\left\{ \begin{aligned}
  & \left( 3.857134,2.629369,10.33333 \right),0<x\le 1, \\ 
 & \left( 1+0.2\sin \left( 5x \right),0,1 \right),1\le x\le 10. \\ 
\end{aligned} \right.$$
The Fig. \ref{figshu} presents density profiles and enlargements at $t=1.8$. As depicted in Figure  \ref{figshu}, the resolution of sparse waves on the right appears to be similar across all four methods. Upon closer inspection in the enlarged image, the results of TENO5-D GKS show slight improvements compared to the other three methods, albeit not significantly. To further evaluate the performance, we test additional Titarev–Toro problem with high-frequency linear waves as described below. \\
\emph{(b) Titarev–Toro problem}\par
\begin{figure}[htbp]
	\centering
	\subfigure{\includegraphics[width=0.4\textwidth]{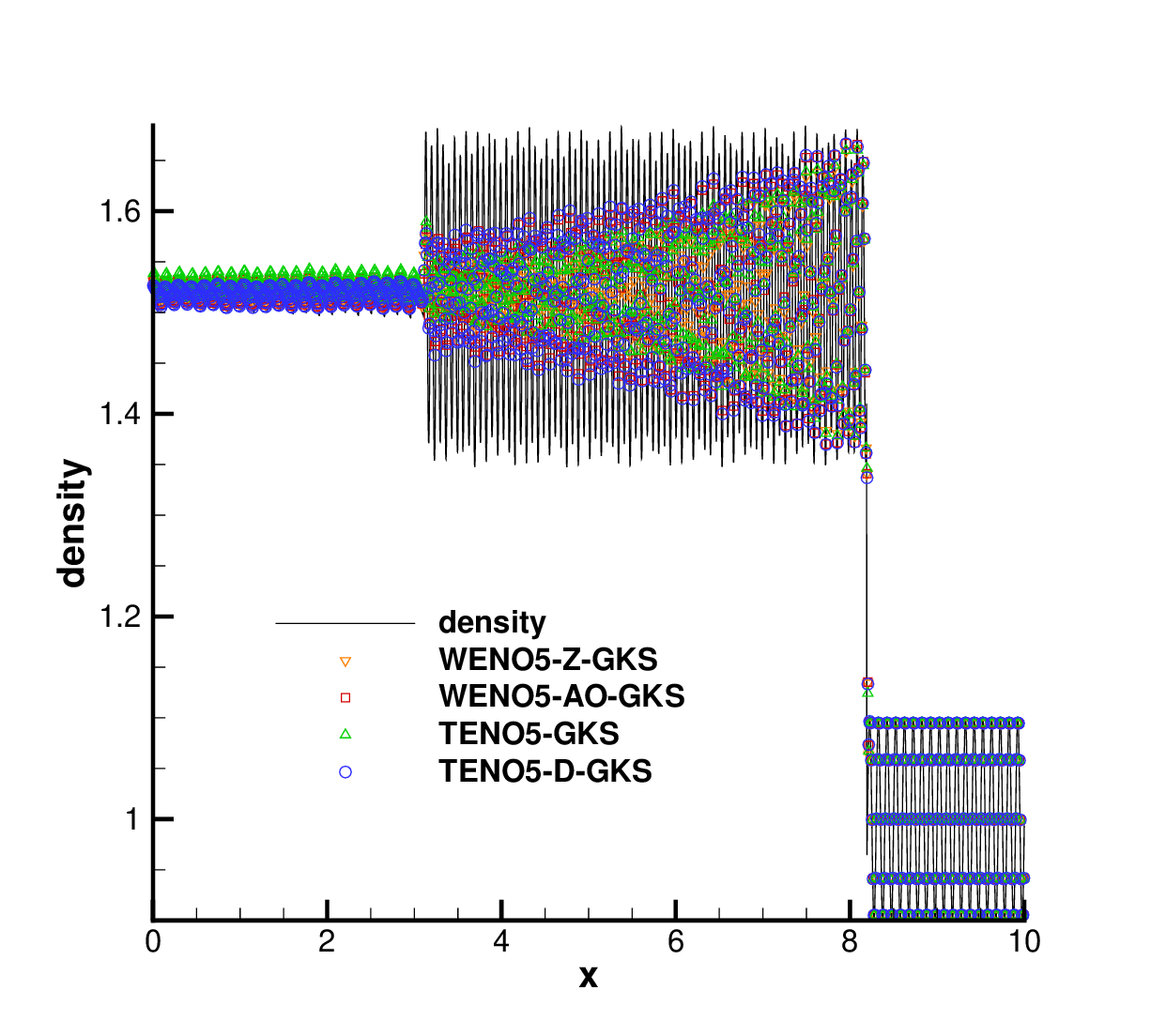}}
	\subfigure{\includegraphics[width=0.4\textwidth]{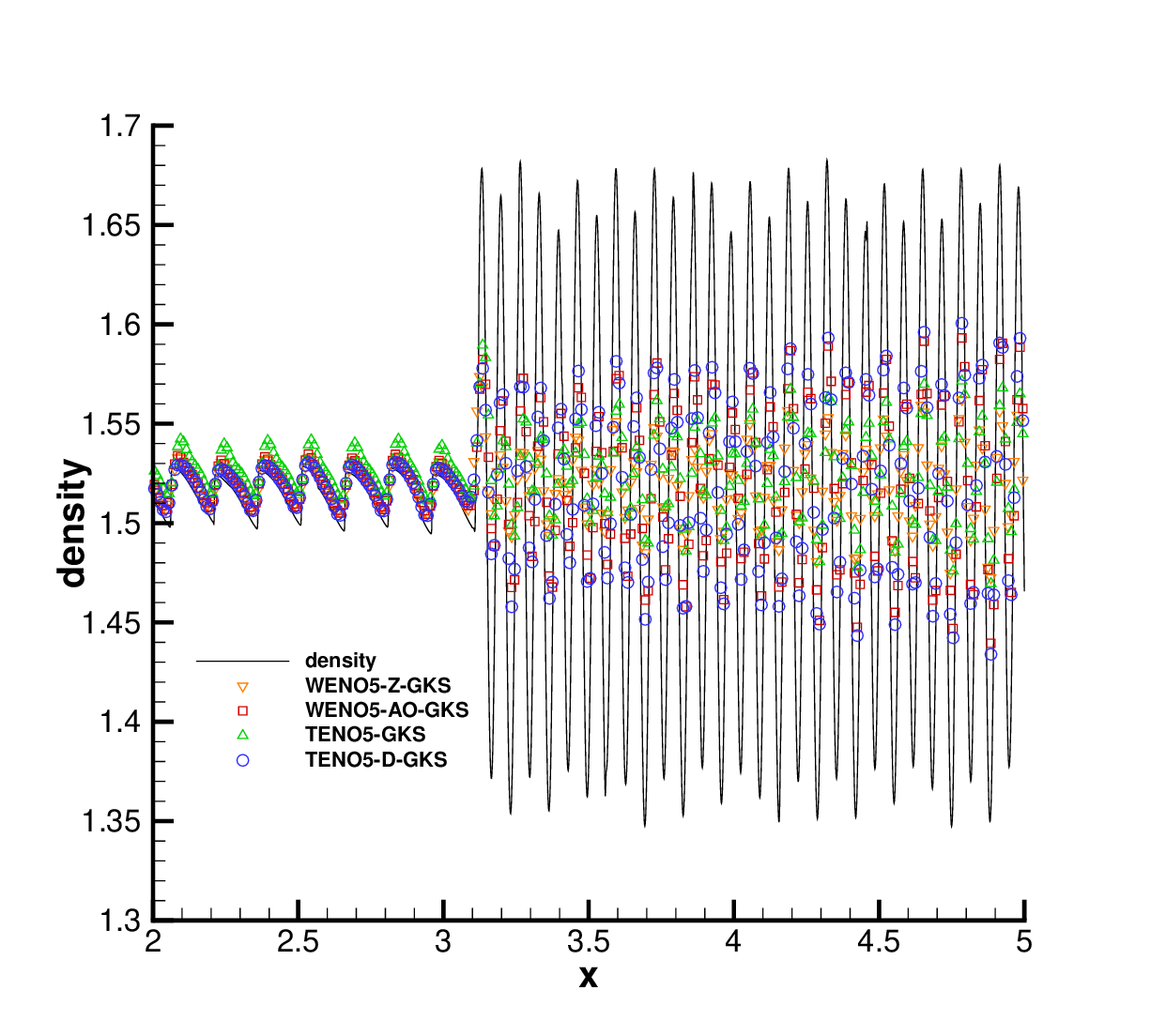}}
	\caption{Titarev-Toro problem: the density distributions and local enlargements for different reconstruction schemes with 1000 cells. $CFL=0.5.$ $T=5.0.$ }
	\label{figtoro}
\end{figure}

As an extension of the Shu-Osher problem, the Titarev-Toro problem ~\cite{TITAREV2004238} is computed in $[0,\ 10]$ with 1000 mesh points. The non-reflecting boundary condition is given on the left, and the fixed wave profile is extended on the right. The initial conditions are
$$\left( \rho ,U,p \right)=\left\{ \begin{aligned}
	& \left( 1.515695, 0.523346, 1.805 \right),0<x\le 0.5, \\ 
	& \left(1 +0.1\sin(20\pi x), 0, 1 \right),0.5\le x\le 10. \\ 
\end{aligned} \right.$$
The Fig. \ref{figtoro} presents density profiles and enlargements at $t=5.0$. As shown in Fig. \ref{figtoro}, all schemes demonstrate their capability to successfully capture acoustic waves with shocklets. When it comes to entropy waves, noteworthy improvements can be observed in both WENO5-AO GKS and TENO5-D GKS. In particular, TENO5-D GKS exhibits the highest resolution in preserving the amplitudes of density waves. The WENO-Z GKS exhibits significant flow feature smearing, displaying the highest level of dissipation among the schemes. In terms of capturing shock waves, both TENO5-D GKS and WENO5-AO GKS yield slightly superior results compared to other schemes. This suggests that WENO5-AO GKS and TENO5-D GKS possess superior wave-resolution properties and shock-capturing capabilities compared to the WENO5-Z GKS and TENO5 GKS schemes.\\
\subsection{Interacting blast waves}
\label{sec5.4} 
\begin{figure}[htbp]
	\centering
	\subfigure{\includegraphics[width=0.4\textwidth]{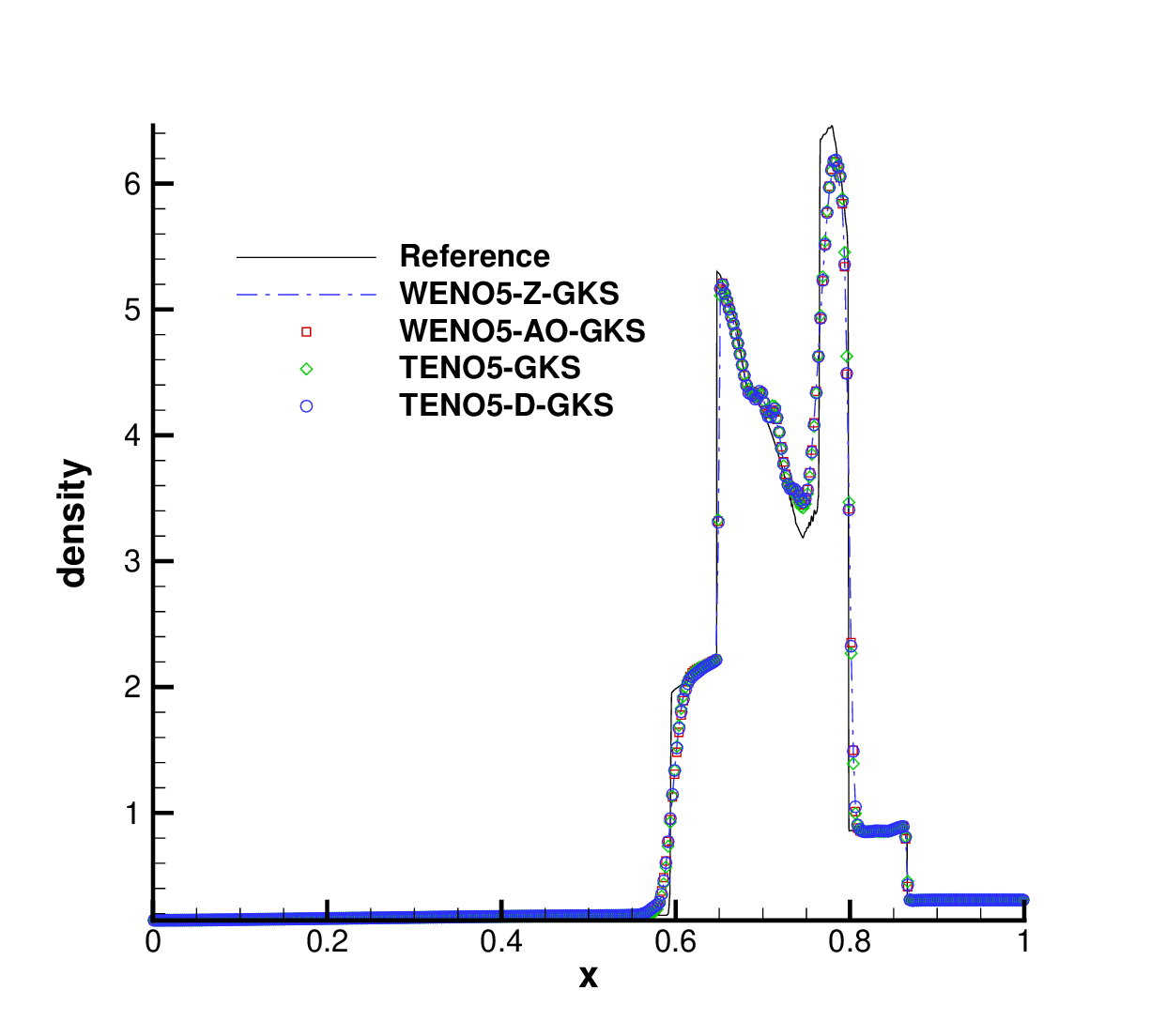}}
	\subfigure{\includegraphics[width=0.4\textwidth]{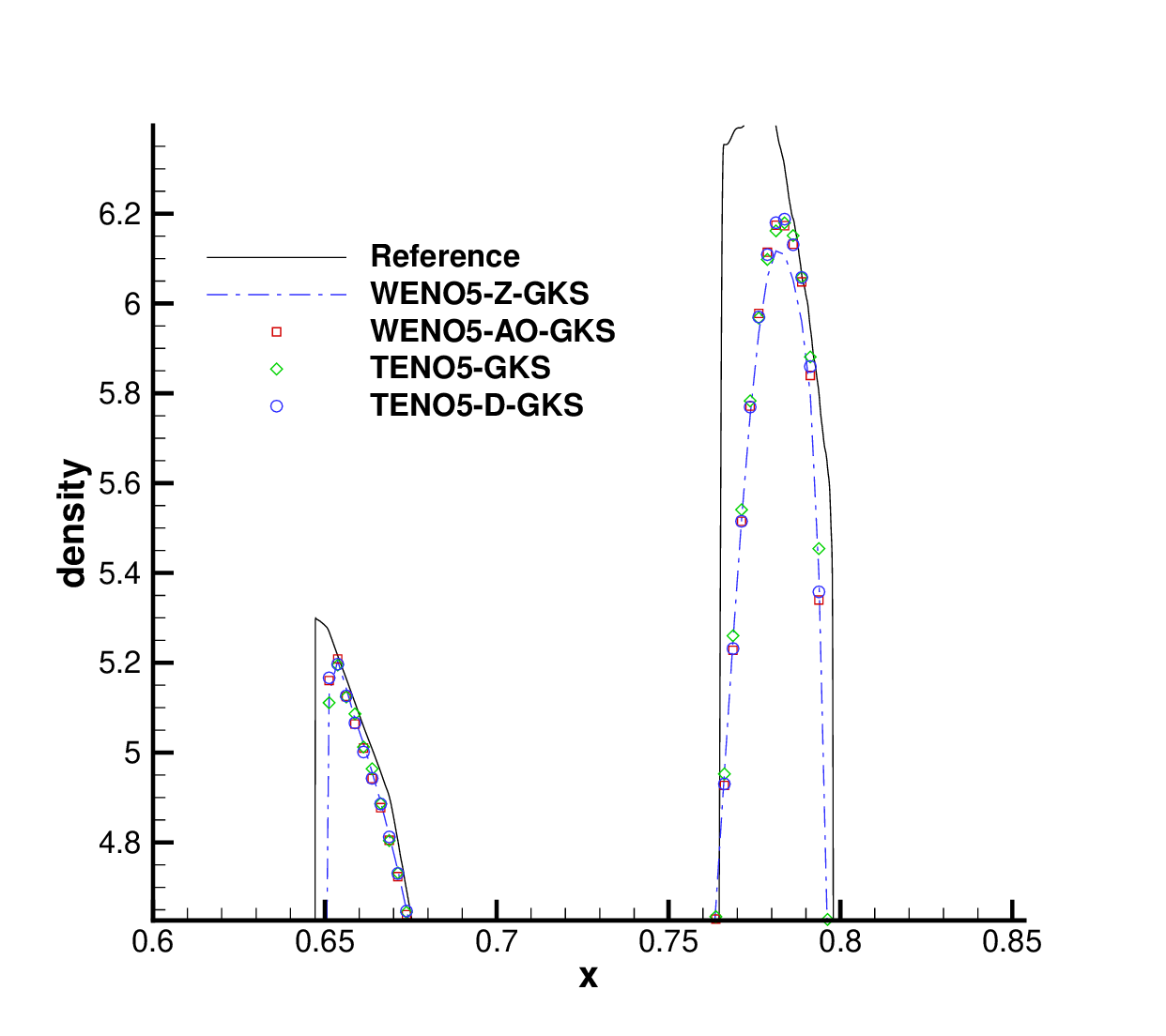}}
	\subfigure{\includegraphics[width=0.4\textwidth]{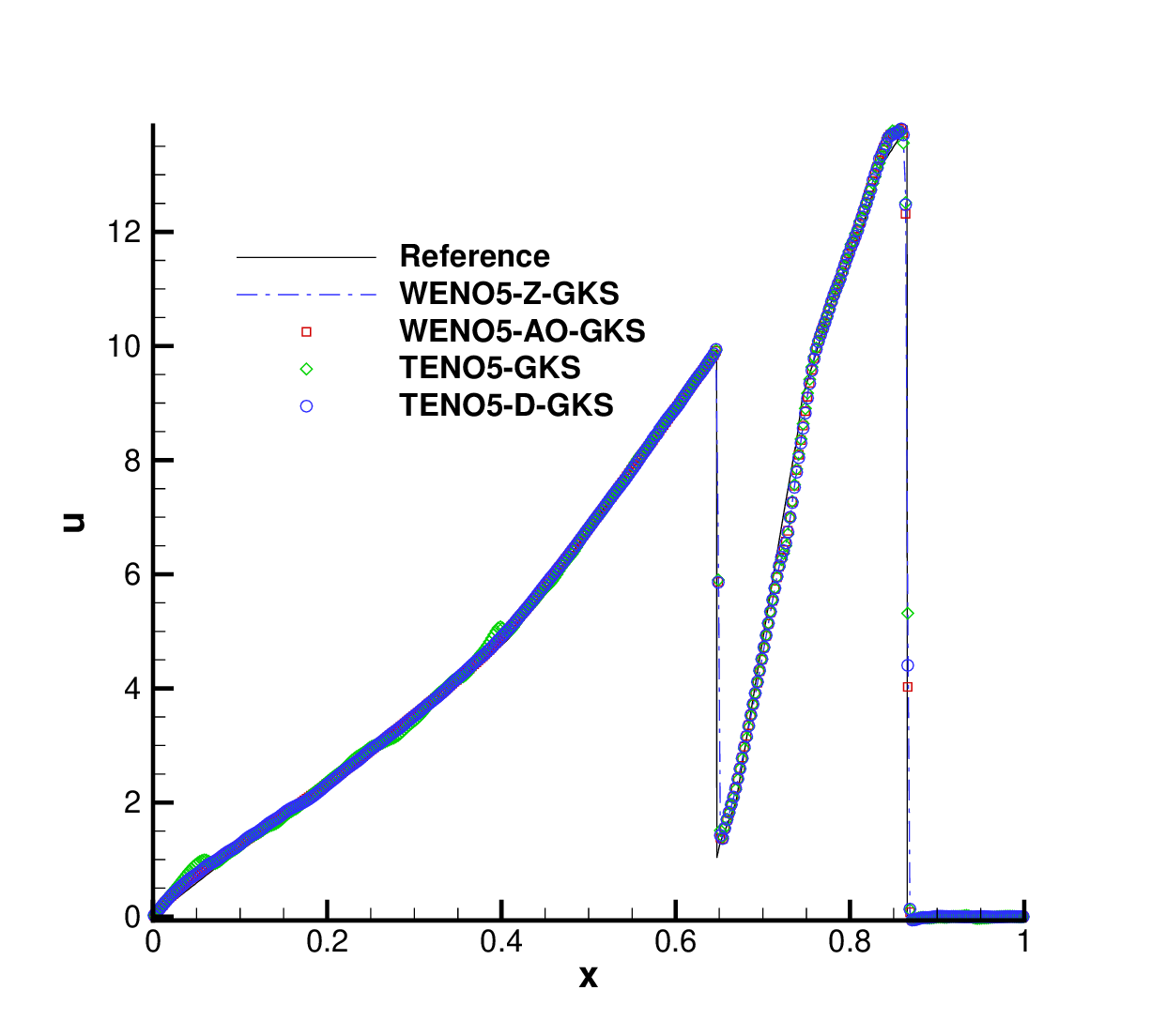}}
	\subfigure{\includegraphics[width=0.4\textwidth]{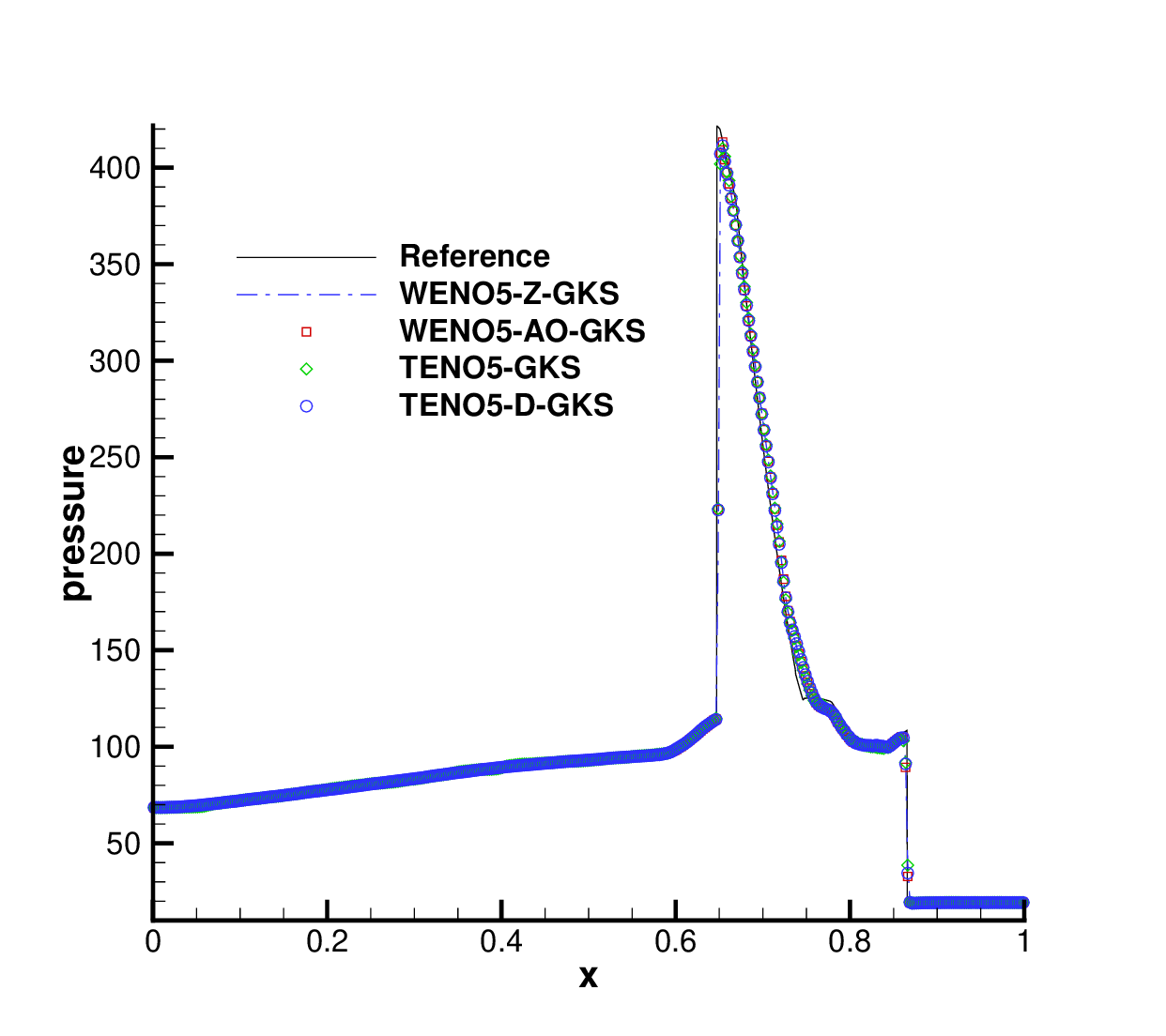}}
	\caption{Blast wave problem: the density, velocity and pressure distributions for different schemes with 400 cells.  $CFL=0.5.$ $T=0.038.$ }
	\label{figblast}
\end{figure}

The blast wave problem is taken from Woodward-Colella blast~\cite{WOODWARD1984115}. The initial conditions for the blast wave problem are given as follows
 $$\left( \rho ,U,p \right)=\left\{ \begin{aligned}
  & \left( 1,0,1000 \right),0\le x<0.1, \\ 
 & \left( 1,0,0.01 \right),0.1\le x<0.9, \\ 
 & \left( 1,0,100 \right),0.9\le x\le 1.0. \\ 
\end{aligned} \right.$$
The computation is conducted with $CFL=0.5$, employing 400 equally spaced grid points while applying reflection boundary conditions at both ends. Fig. \ref{figblast} displays the computed profiles of density, velocity and pressure at t=0.038. The blast wave problem places a high demand on the schemes in terms of both robustness and their ability to capture strong shock waves. The results of all four schemes can be compared favorably with the reference solutions. Specifically, the results of TENO5 GKS, TENO5-D GKS, and WENO5-AO GKS are similar and superior to those of WENO5-Z GKS. Hence, we can conclude that combining HGKS with TENO class reconstruction is also a good choice.\\

\subsection{Accuracy test in 2-D}
\label{sec5.5}
Similar to 1-D case, we choose the advection of density perturbation for accuracy test. The collision time $\tau =0$ is set for this inviscid flow. The initial conditions are 
$$\rho \left( x,y \right)=1+0.2sin\left( \pi \left( x+y \right) \right),U\left( x,y \right)=\left( 1,1 \right),p\left( x,y \right)=1,\left( x,y \right)\in \left[ 0,2 \right]\times \left[ 0,2 \right].$$
The compute domain is $\left[ 0,2 \right]\times \left[ 0,2 \right]$. $N\times N$ uniform mesh is used and the boundary conditions in both directions are periodic. The analytic solution of the 2-D advection of density perturbation is
$$\rho \left( x,y,t \right)=1+0.2sin\left( \pi \left( x+y-t \right) \right),U\left( x,y,t \right)=1,p\left( x,y,t \right)=1,\left( x,y \right)\in \left[ 0,2 \right]\times \left[ 0,2 \right].$$
The CFL number of the time steps are 0.5. HGKS with fifth-order WENO-Z, WENO-AO, TENO and TENO-D reconstruction are tested and presented in the Table \ref{2Daccuracy01} - Table \ref{2Daccuracy04} respectively. The HGKS with TENO class schemes can achieve expected accuracy as the HGKS with WENO class schemes.
\begin{table}
\begin{tabular}{c|cc|cc|cc}
	\hline
	mesh length & ${{L}^{1}}$error & Order & ${{L}^{2}}$error & Order & ${{L}^{\infty }}$error & Order\\
	\hline
	1/5 & 3.4876001e-02 &  & 3.8827381e-02 &  & 5.4193732e-02 & \\ 
	1/10 & 1.7631333e-03 & 4.30 & 1.9210920e-03 & 4.34 & 2.6476711e-03 & 4.36 \\
	1/20 & 4.7700565e-05 & 5.21 & 5.3254793e-05 & 5.17 & 8.1803433e-05 & 5.02\\
	1/40 & 1.3974771e-06 & 5.09 & 1.5674888e-06 & 5.09 & 2.4100934e-06 & 5.09\\
	1/80 & 4.7861422e-08 & 4.87 & 5.3231080e-098 & 4.88 & 7.6561492e-08 & 4.98\\
	\hline
\end{tabular}
\centering 
	\caption{\label{2Daccuracy01}Accuracy test in 2-D for the advection of density perturbation by the WENO-Z reconstruction. $CFL=0.5$}
\end{table}
\begin{table}
\begin{tabular}{c|cc|cc|cc}
	\hline
	mesh length & ${{L}^{1}}$error & Order & ${{L}^{2}}$error & Order & ${{L}^{\infty }}$error & Order\\
	\hline
	1/5 & 3.5140972e-02 &  & 3.8349331e-02 &  & 5.4070852e-02 & \\ 
	1/10 & 1.3599138e-03 & 4.69 & 1.4895633e-03 & 4.69 & 2.1081344e-03 & 4.68 \\
	1/20 & 4.2540389e-05 & 5.00 & 4.7371073e-05 & 4.97 & 6.916346e-05 & 4.93\\
	1/40 & 1.3778268e-06 & 4.95 & 1.5296732e-06 & 4.95 & 2.2380713e-06 & 4.95\\
	1/80 & 4.7722522e-08 & 4.85 & 5.3087765e-08 & 4.85 & 7.6217426e-08 & 4.87\\
	\hline
\end{tabular}
\centering 
	\caption{\label{2Daccuracy02}Accuracy test in 2-D for the advection of density perturbation by the  WENO-AO reconstruction. $CFL=0.5$}
\end{table}
\begin{table}
	\begin{tabular}{c|cc|cc|cc}
		\hline
		mesh length & ${{L}^{1}}$error & Order & ${{L}^{2}}$error & Order & ${{L}^{\infty }}$error & Order\\
		\hline
		1/5 & 3.0748012e-02 &  & 3.4391087e-02 &  & 4.7659306e-02 & \\ 
		1/10 & 1.3206267e-03 & 4.54 & 1.4537748e-03 & 4.56 & 2.0643420e-03 & 4.53 \\
		1/20 & 4.2406682e-05 & 4.96 & 4.7268698e-05 & 4.94 & 6.9003512e-05 & 4.90\\
		1/40 & 1.3771205e-06 & 4.94 & 1.5290724e-06 & 4.95 & 2.2352962e-06 & 4.95\\
		1/80 & 4.7710963e-08 & 4.85 & 5.3077862e-08 & 4.85 & 7.6460775e-08 & 4.87\\
		\hline
	\end{tabular}
	\centering 
	\caption{\label{2Daccuracy03}Accuracy test in 2-D for the advection of density perturbation by the TENO reconstruction. $CFL=0.5$}
\end{table}
\begin{table}
	\begin{tabular}{c|cc|cc|cc}
		\hline
		mesh length & ${{L}^{1}}$error & Order & ${{L}^{2}}$error & Order & ${{L}^{\infty }}$error & Order\\
		\hline
		1/5 & 2.4746434e-02 &  & 2.7497601e-02 &  & 3.9477542e-02 & \\ 
		1/10 & 1.3223778e-03 & 4.23 & 1.4555983e-03 & 4.24 & 2.0744814e-03 & 4.25 \\
		1/20 & 4.2451259e-05 & 4.96 & 4.7293893e-05 & 4.94 & 6.9152436e-05 & 4.91\\
		1/40 & 1.3776848e-06 & 4.95 & 1.5295282e-06 & 4.95 & 2.2379703e-06 & 4.95\\
		1/80 & 4.7722292e-08 & 4.85 & 5.3087505e-08 & 4.85 & 7.6516876e-08 & 4.87\\
		\hline
	\end{tabular}
	\centering 
	\caption{\label{2Daccuracy04}Accuracy test in 2-D for the advection of density perturbation by the TENO-D reconstruction. $CFL=0.5$}
\end{table}
\subsection{Two-dimensional Riemann problems}
\label{sec5.6}
\emph{(a) Configuration 3}\par
\begin{figure}[htbp]
	\centering
	\subfigure{\includegraphics[width=0.4\textwidth]{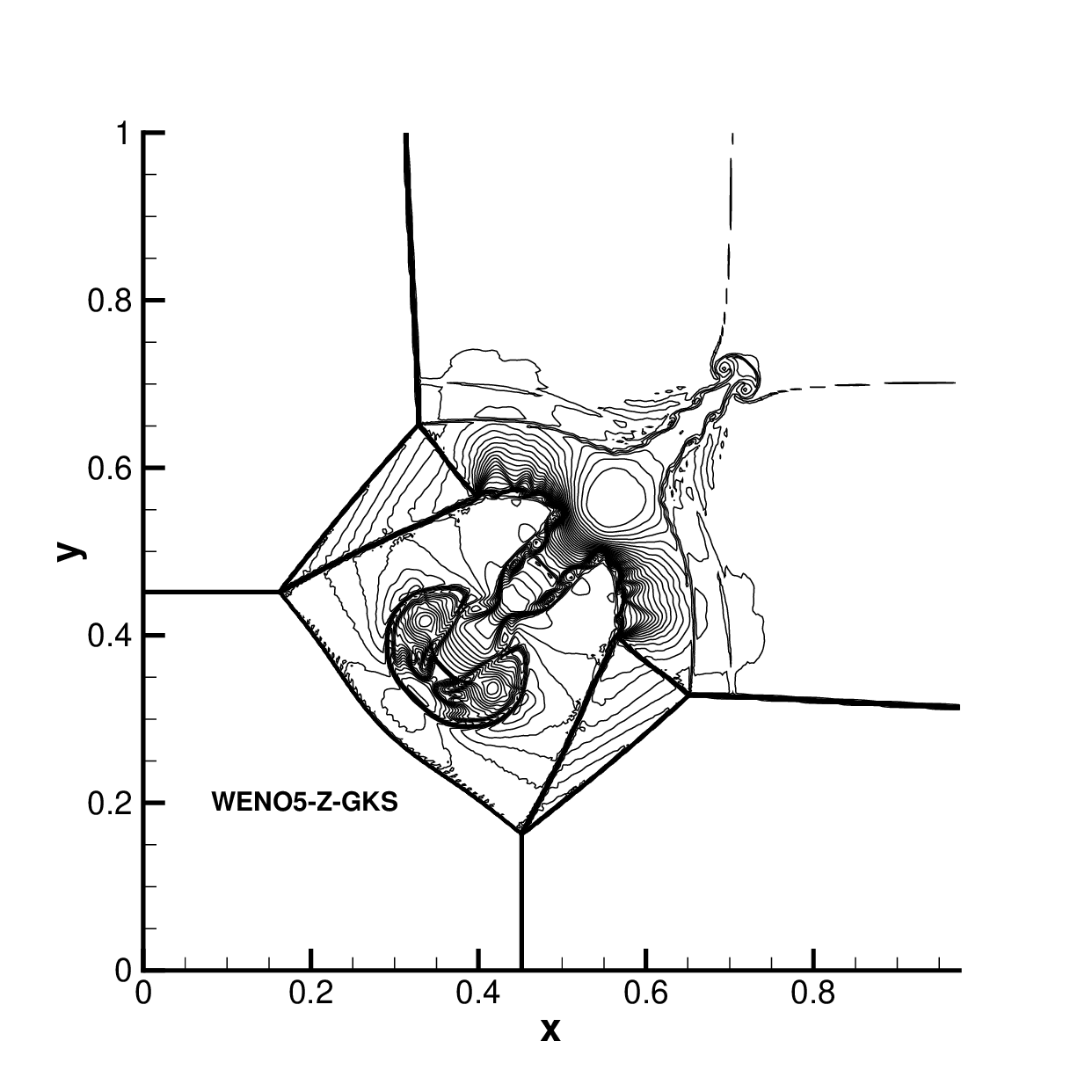}}
	\subfigure{\includegraphics[width=0.4\textwidth]{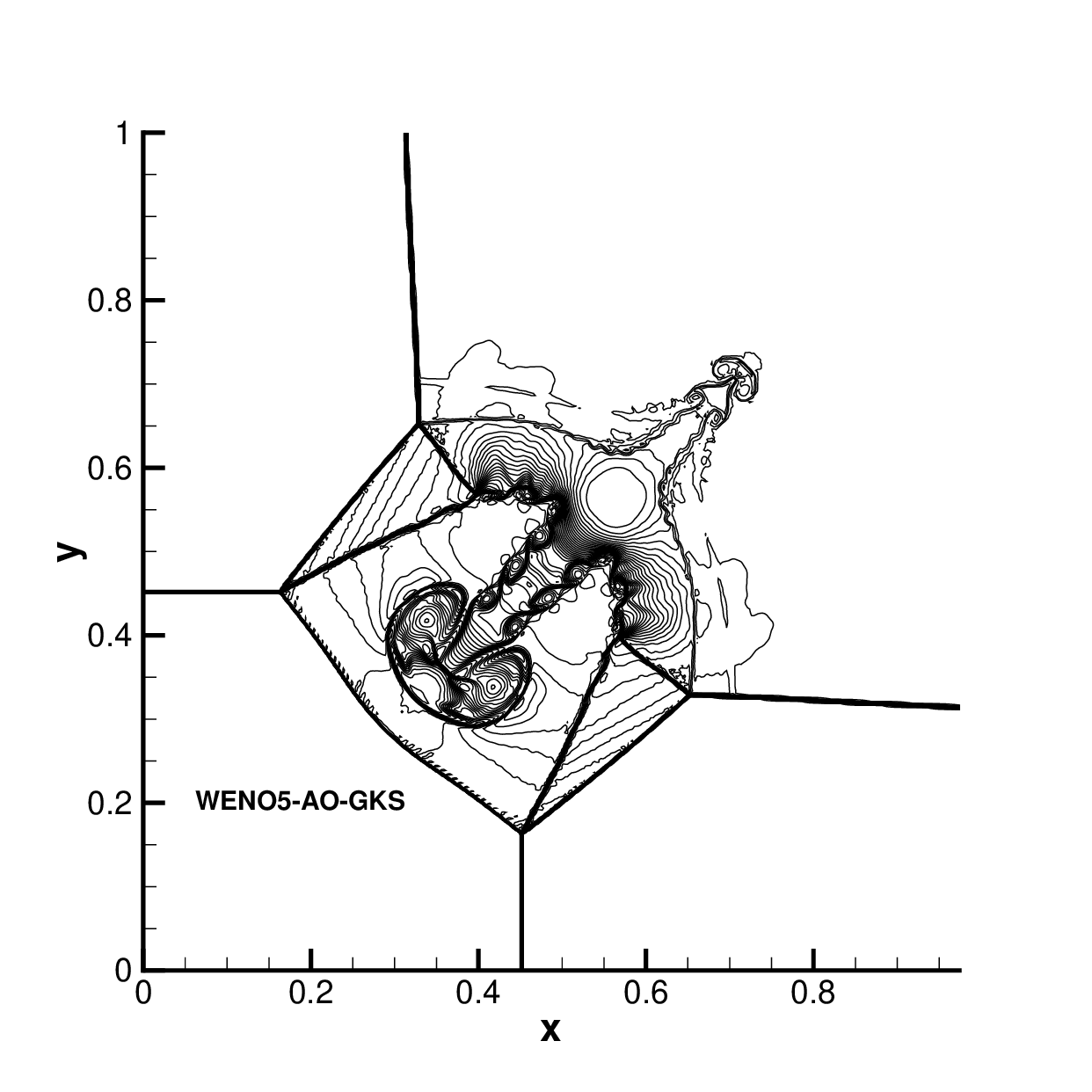}}
	\subfigure{\includegraphics[width=0.4\textwidth]{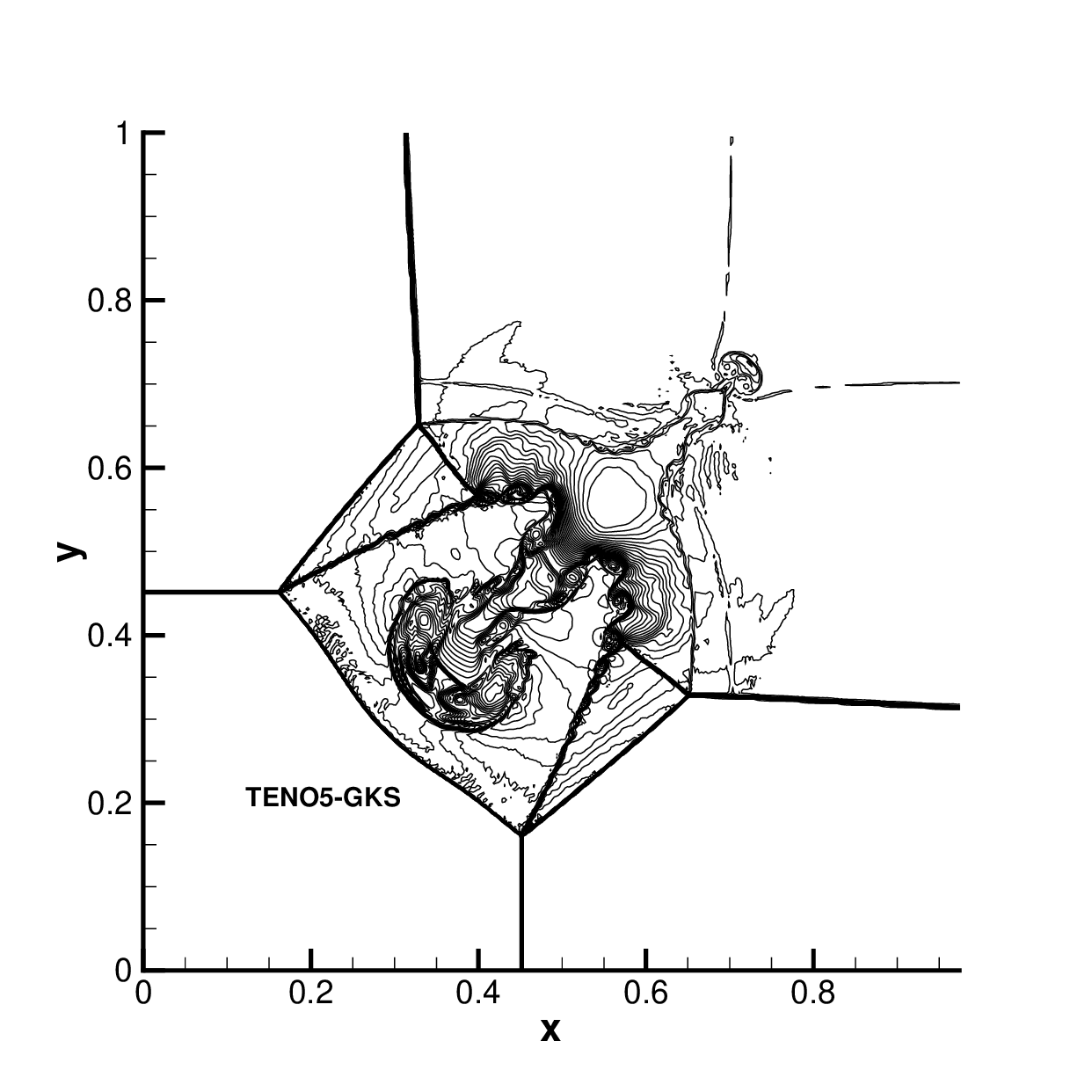}}
	\subfigure{\includegraphics[width=0.4\textwidth]{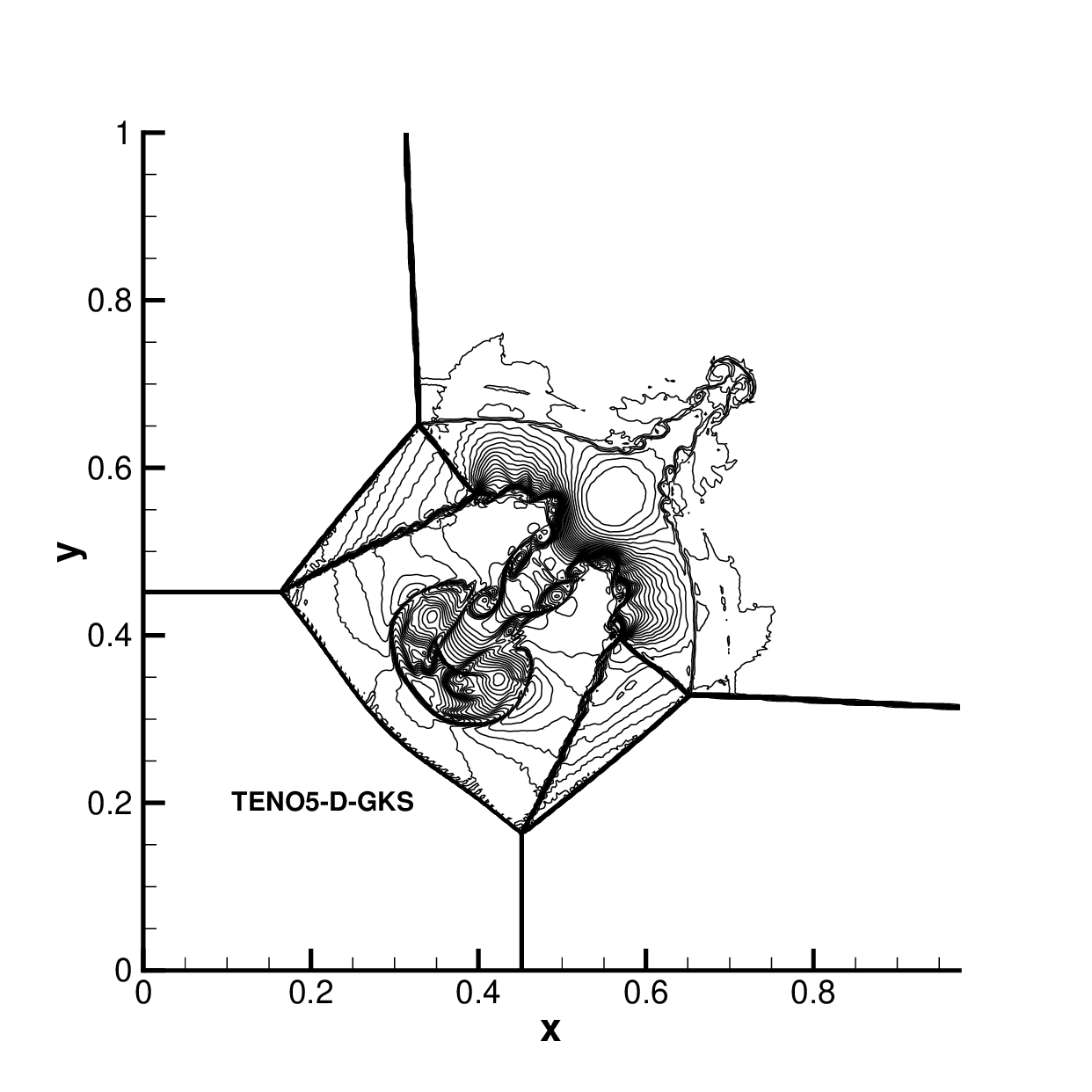}}
	\caption{Two-dimensional Riemann problems: the density distributions for Configuration 3. $CFL=0.5.$ $T=0.6.$ Mesh:$400\times 400.$}
	\label{figliman03}
\end{figure}

The initial conditions of the Configuration 3 are the shock–shock interaction and shock–vortex interaction and given as~\cite{SOTR}
$$\left( \rho ,U,V,p \right)=\left\{ \begin{aligned}
  & \left( 0.138,1.206,1.206,0.029 \right),x<0.7,y<0.7, \\ 
 & \left( 0.5323,0,1.206,0.3 \right),x\ge 0.7,y<0.7, \\ 
 & \left( 1.5,0,0,1.5 \right),x\ge 0.7,y\ge 0.7, \\ 
 & \left( 0.5323,1.206,0,0.3 \right),x<0.7,y\ge 0.7. \\ 
\end{aligned} \right.$$
The results of the Configuration 3 at $t=0.6$ for the WENO5-Z GKS, WENO5-AO GKS, TENO5 GKS and TENO5-D GKS are presented in Fig. \ref{figliman03}. In inviscid flow simulation, higher-order shock capturing schemes with lower numerical dissipation amplitudes tend to produce more detailed fine structures. As depicted in the Fig. \ref{figliman03}, compared to WENO5-Z GKS and WENO5-AO GKS, TENO5 GKS and TENO5-D GKS yield additional vortex structures. Moreover, the low-dissipation properties of TENO5 GKS and TENO5-D GKS lead to induce symmetry breaking of flow field. A more dissipative scheme is more likely to preserve symmetry even when increasing the resolution significantly~\cite{FU2016333}.\\
\emph{(b) Configuration 6}\par
\begin{figure}[htbp]
	\centering
	\subfigure{\includegraphics[width=0.4\textwidth]{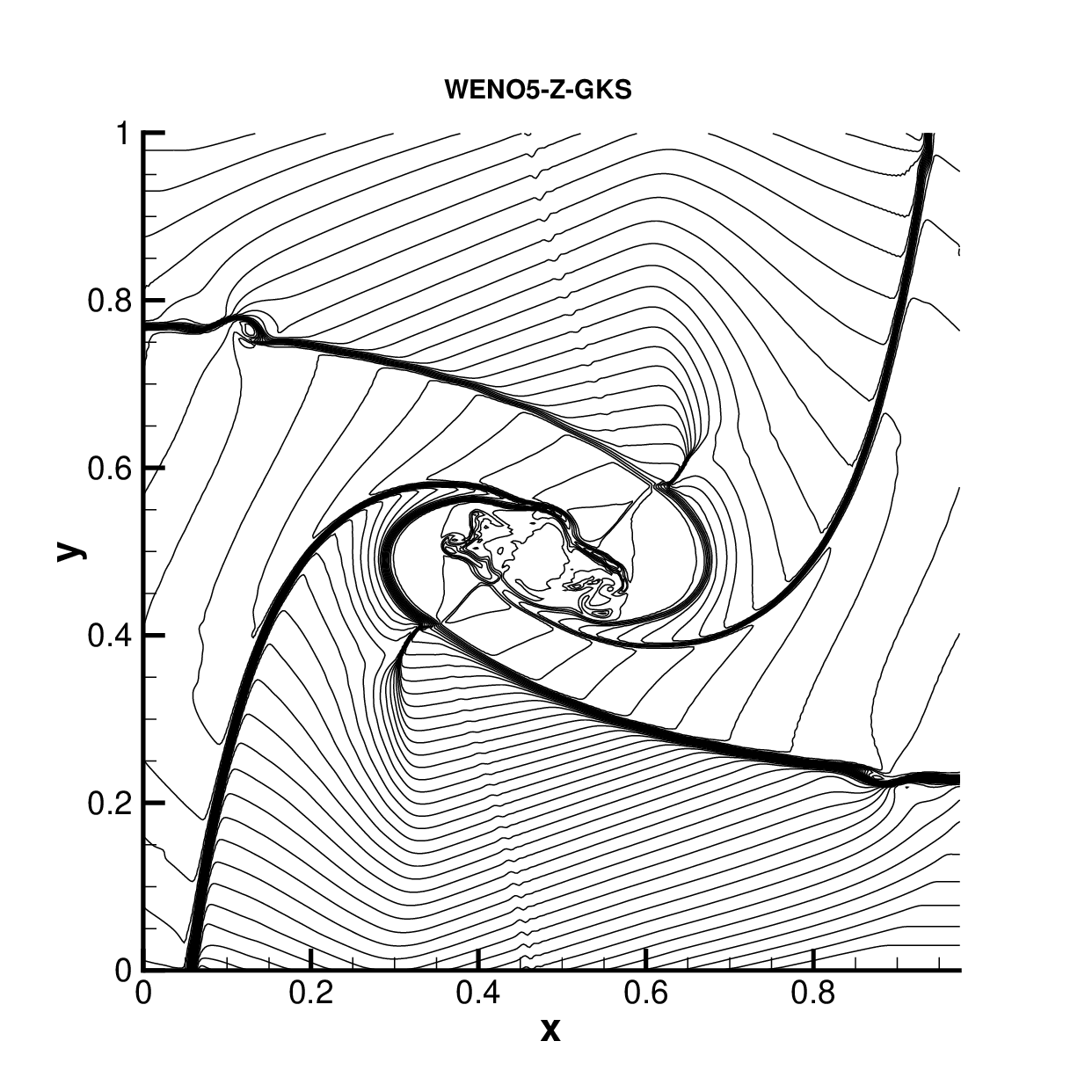}}
	\subfigure{\includegraphics[width=0.4\textwidth]{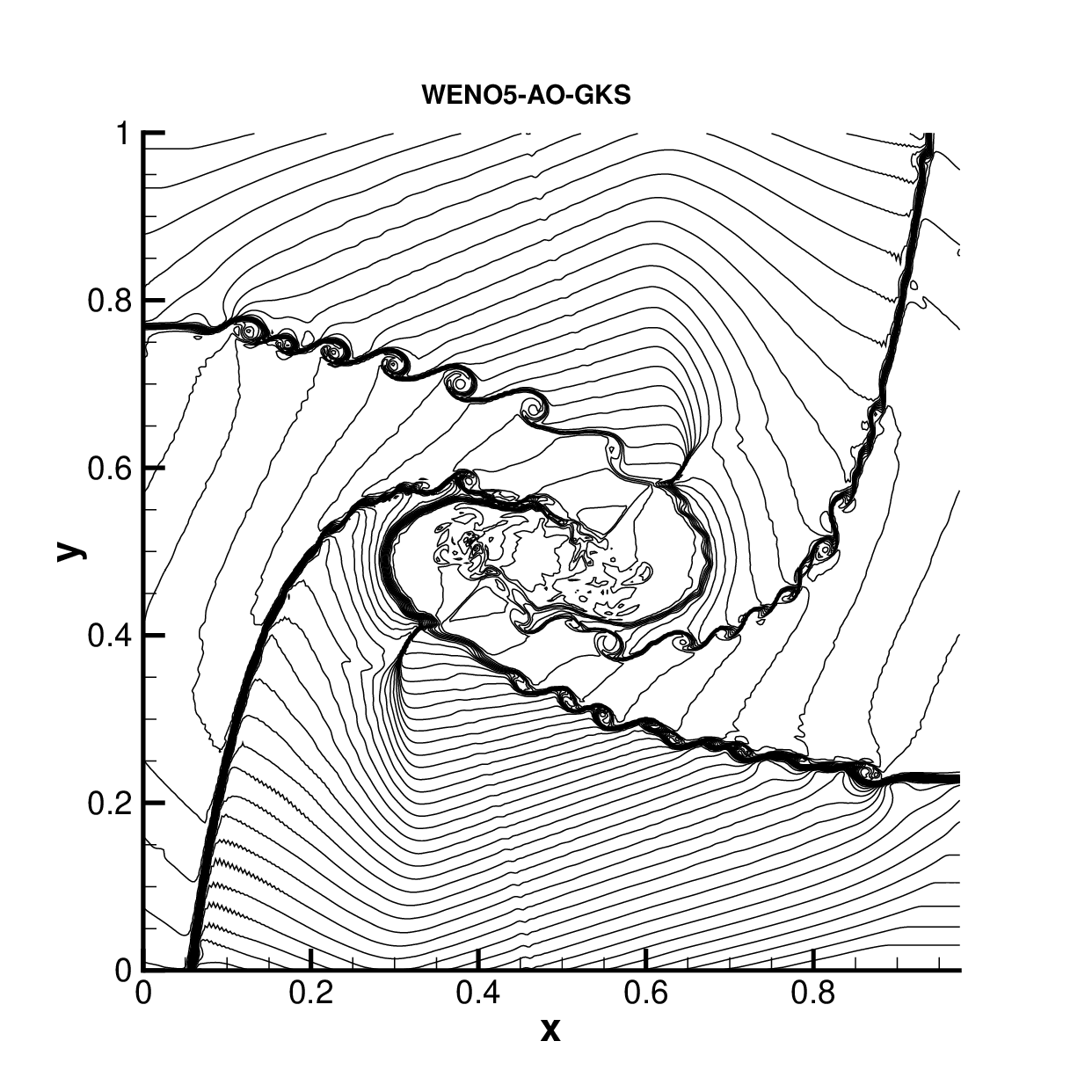}}
	\subfigure{\includegraphics[width=0.4\textwidth]{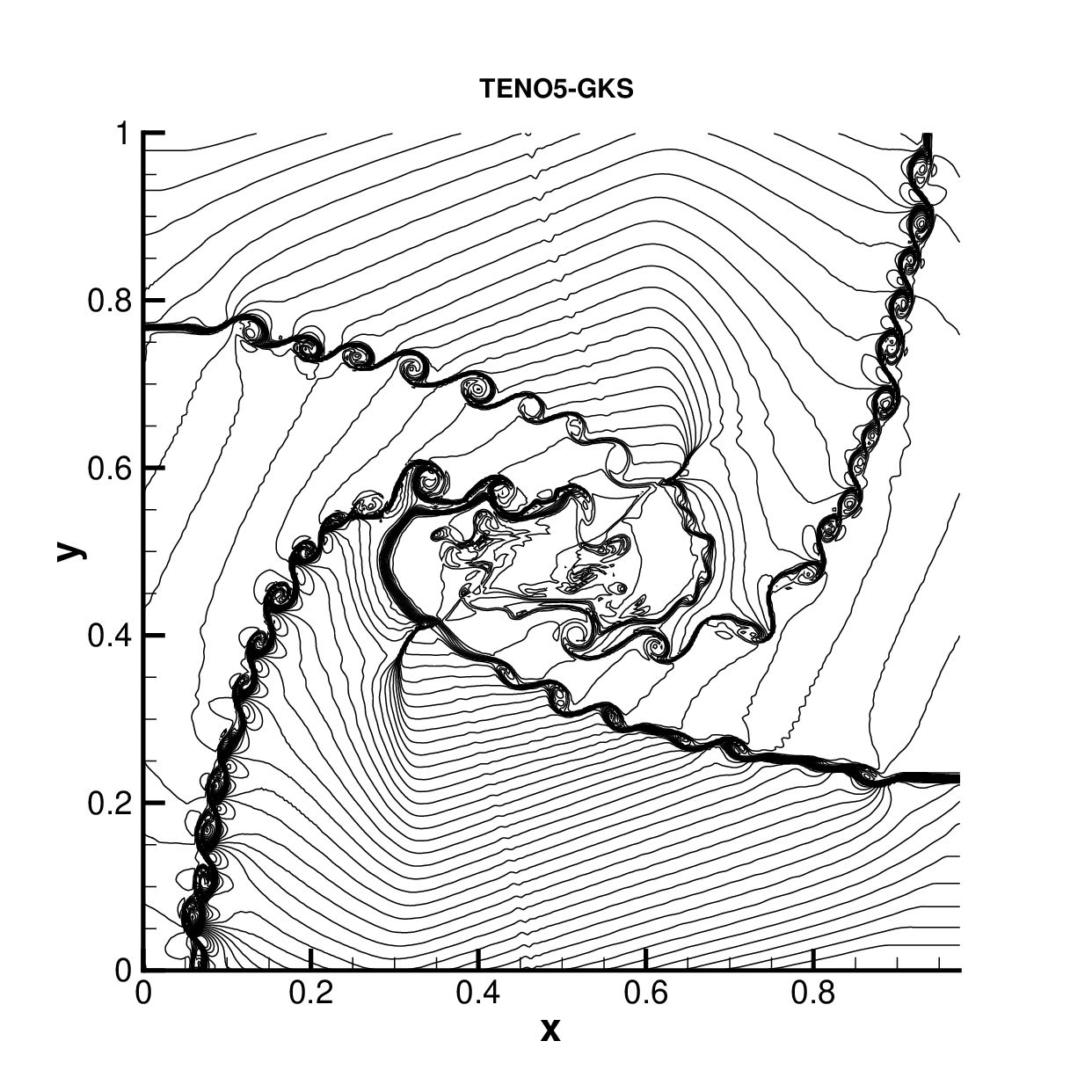}}
	\subfigure{\includegraphics[width=0.4\textwidth]{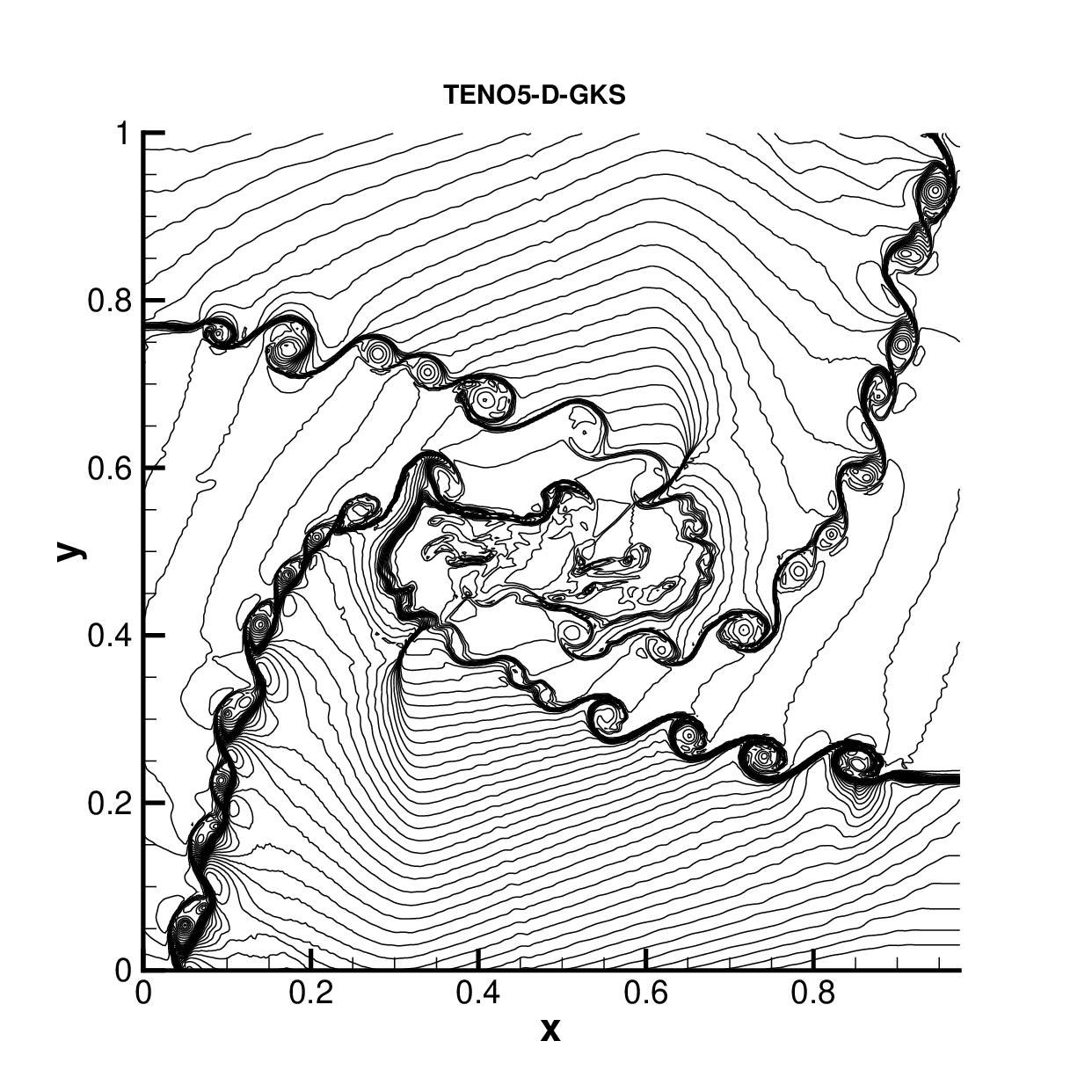}}
	\caption{Two-dimensional Riemann problems: the density distributions for Configuration 6. $CFL=0.8.$ $T=0.6.$ Mesh:$400\times 400.$}
	\label{figliman06}
\end{figure}

For compressible flow, the shear layer flow is one of the most distinguishable flow patterns. For the case of Configuration 6 ~\cite{SOTR}, the initial conditions with four planar contact discontinuities are given as  
 $$\left( \rho ,U,V,p \right)=\left\{ \begin{aligned}
  & \left( 1,-0.75,0.5,1 \right),x<0.5,y<0.5, \\ 
 & \left( 3,-0.75,-0.5,1 \right),x\ge 0.5,y<0.5, \\ 
 & \left( 1,-0.75,-0.5,1 \right),x\ge 0.5,y\ge 0.5, \\ 
 & \left( 2,0.75,0.5,1 \right),x<0.5,y\ge 0.5. \\ 
\end{aligned} \right.$$
These discontinuities trigger the K-H instabilities due to the numerical viscosities. It is commonly believed that the less numerical dissipation corresponds to larger amplitude shear instabilities. It can be observed in Fig. \ref{figliman06} that the HGKS with TENO class schemes predicts less numerical dissipation and more details of vortices than the HGKS with WENO class schemes. Furthermore, the results obtained from WENO5-AO GKS exhibit improvements over those of WENO5-Z GKS, and the results obtained from TENO5-D GKS are superior to those of TENO5 GKS. In both WENO5-AO GKS and TENO5-D GKS, the high-order accuracy of the initial non-equilibrium state helps reduce numerical dissipation, leading to better overall performance.\\

\subsection{Double Mach reflection problem}
\label{sec5.7}
\begin{figure}[htbp]
	\centering
	\subfigure{\includegraphics[width=0.6\textwidth]{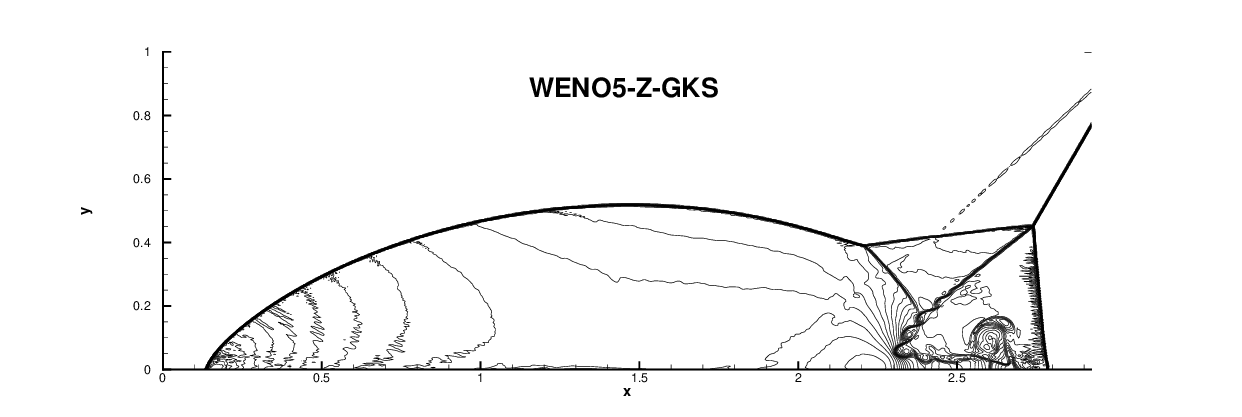}}
	\subfigure{\includegraphics[width=0.2\textwidth]{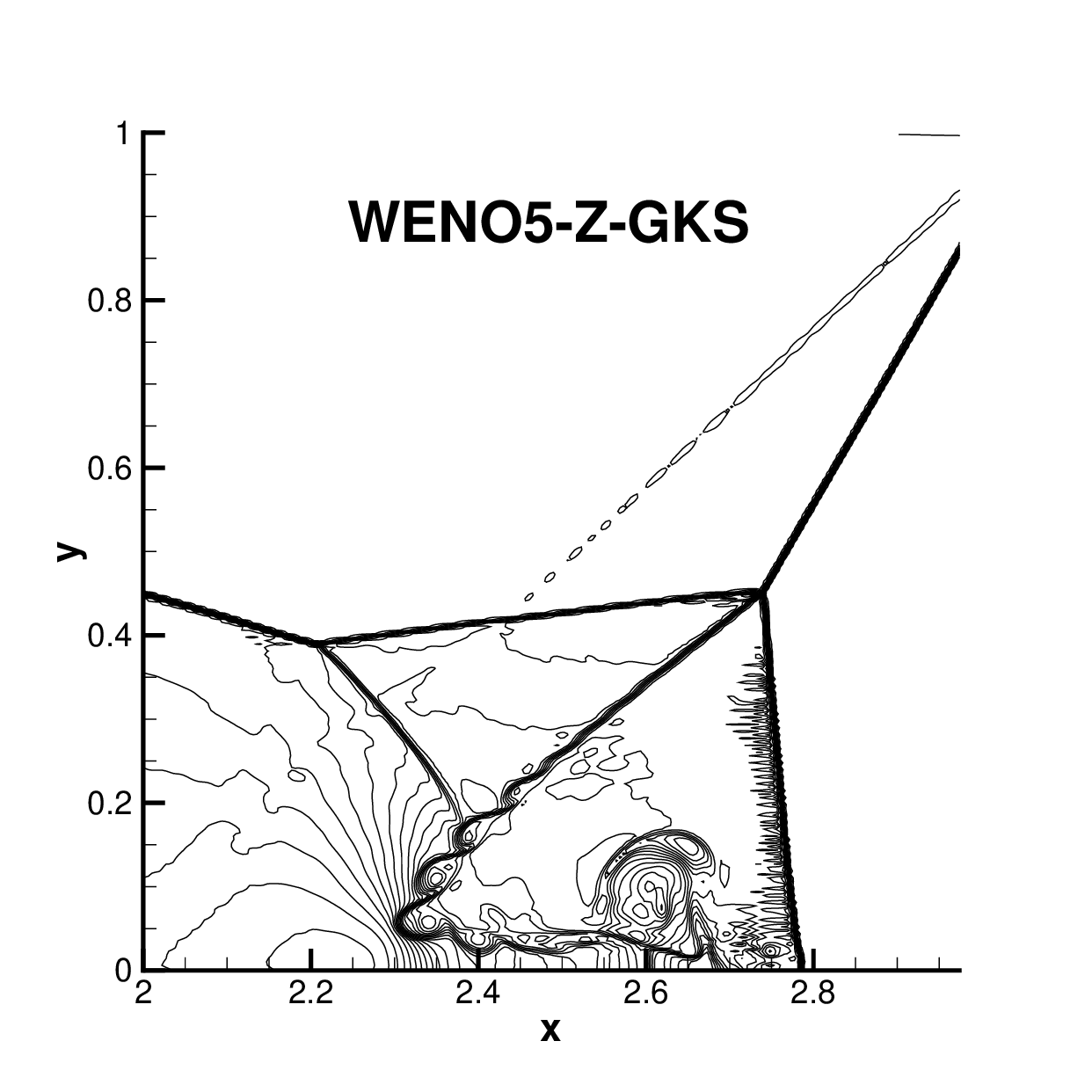}}
	\subfigure{\includegraphics[width=0.6\textwidth]{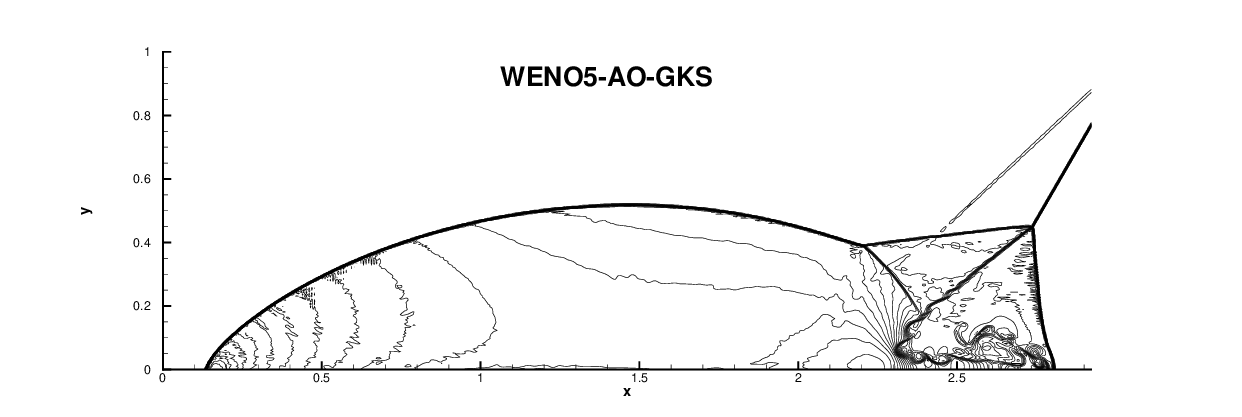}}
	\subfigure{\includegraphics[width=0.2\textwidth]{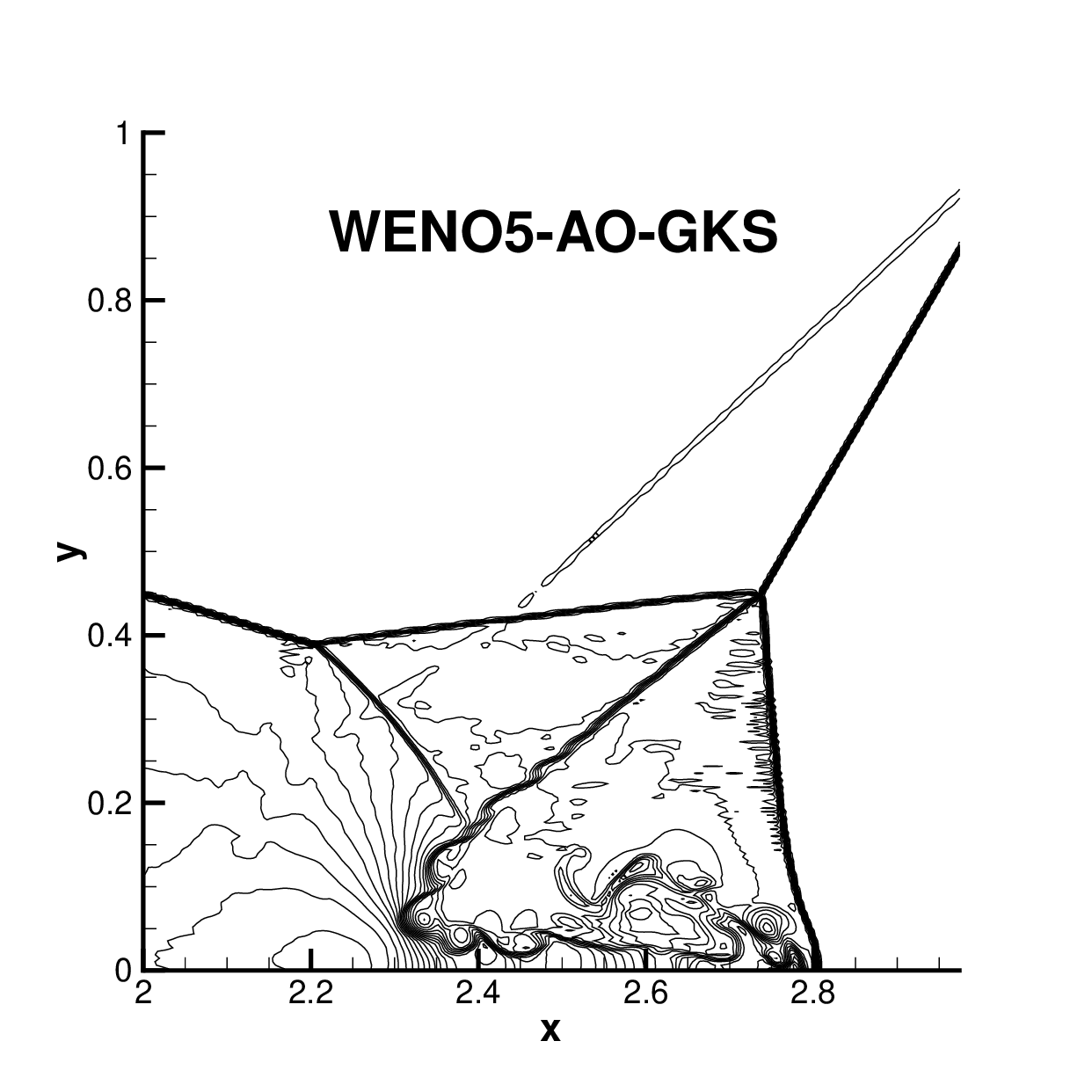}}
	\subfigure{\includegraphics[width=0.6\textwidth]{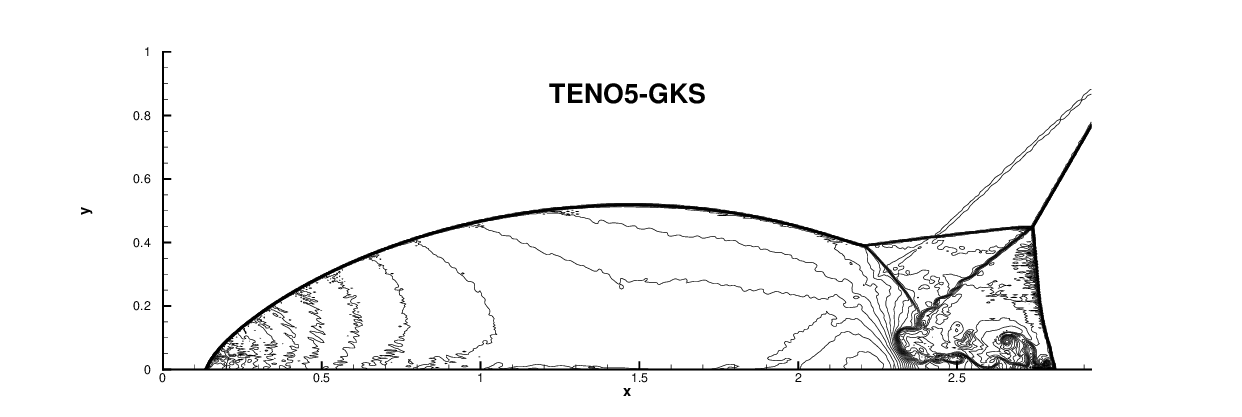}}
	\subfigure{\includegraphics[width=0.2\textwidth]{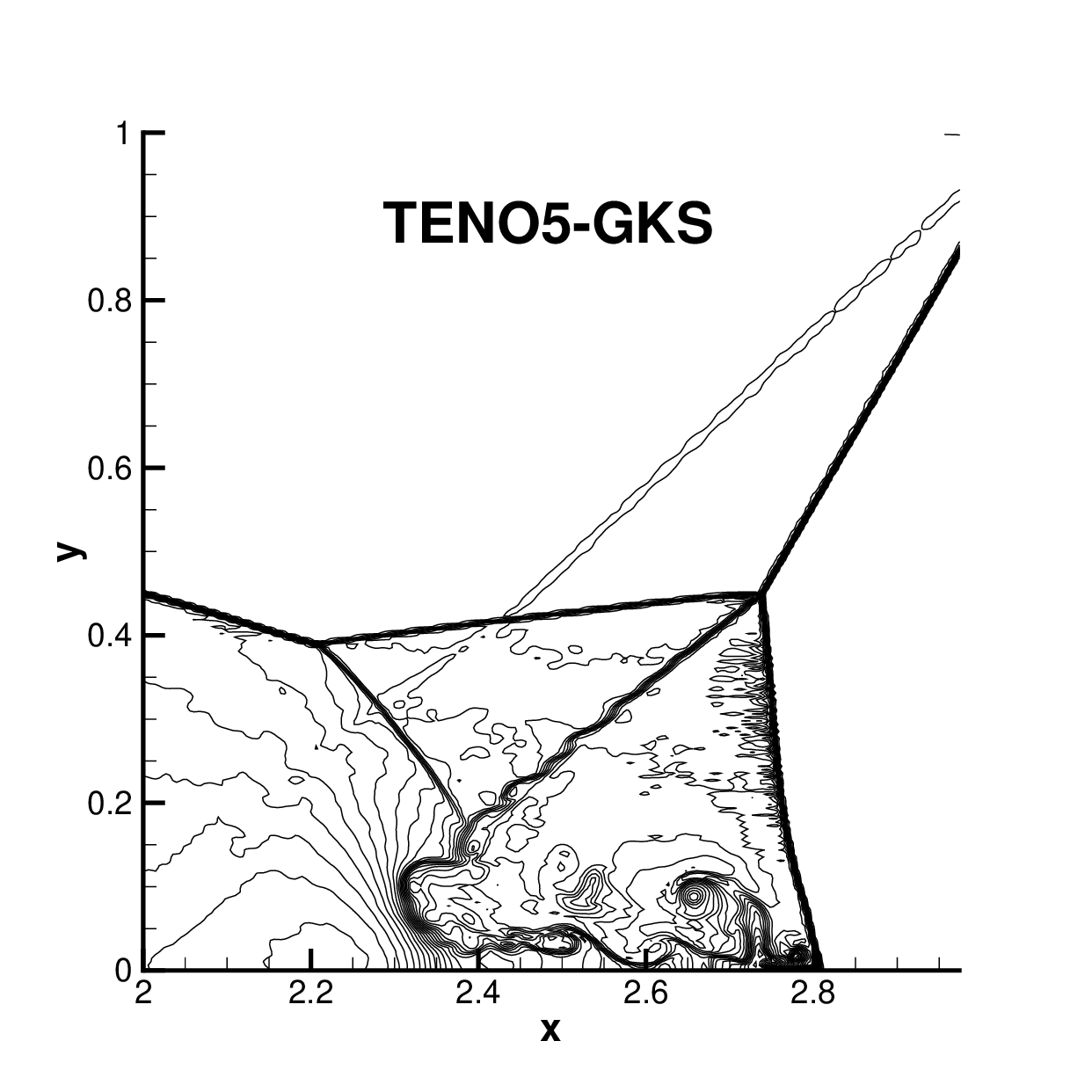}}
	\subfigure{\includegraphics[width=0.6\textwidth]{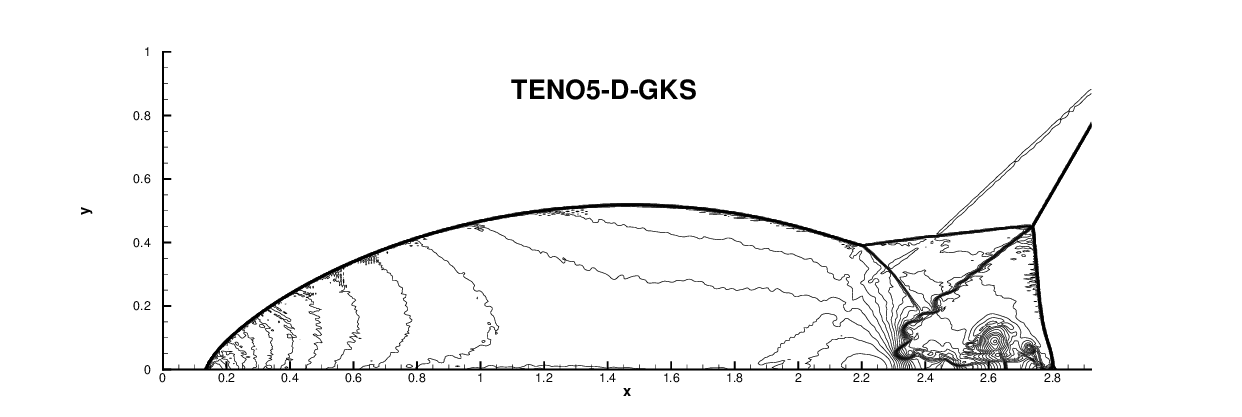}}
	\subfigure{\includegraphics[width=0.2\textwidth]{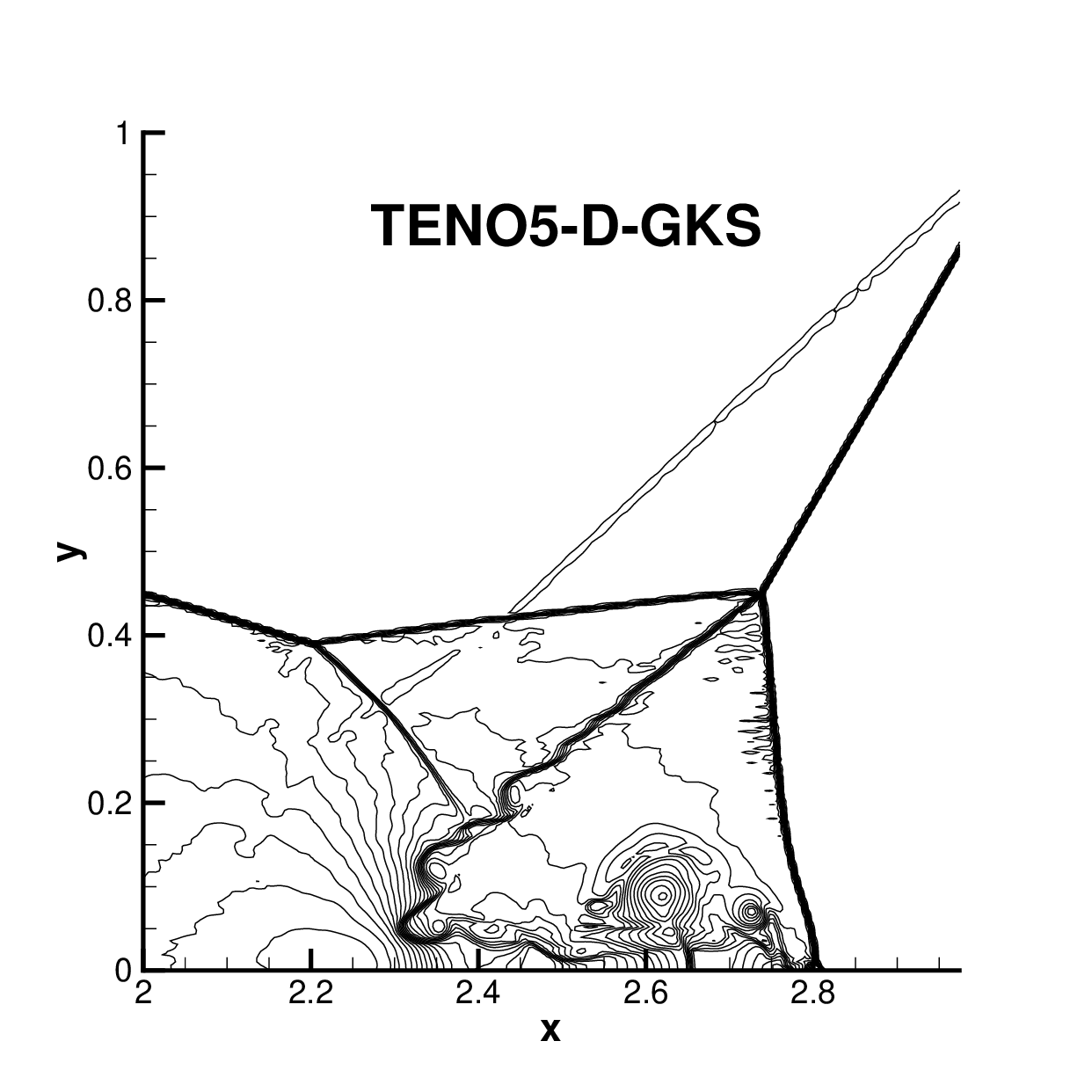}}
	\caption{The density distributions for Double Mach: WENO5-Z-GKS, WENO5-AO-GKS, TENO5-GKS and TENO5-D-GKS scheme. Mesh: $960\times 240.$ $CFL=0.5.$ $T=0.2.$.This figure is drawn with 43
	density contours between 1.887 and 20.9.}
	\label{figdouble}
\end{figure}

The double Mach reflection problem is a inviscid test case designed by Woodward and Colella~\cite{WOODWARD1984115}. The computational domain is $\left[ 0,\ 4 \right]\times \left[ 0,\ 1 \right]$ with a slip boundary condition applied on the bottom of the domain starting from $x=1/6$. The post -shock condition is set for the rest of bottom boundary. At the top boundary, the flow variables are set to describe the exact motion of the Mach 10 shock. The initial pre-shock and post-shock conditions are
$$\begin{aligned}
  & \left( \rho ,U,V,p \right)=\left( 8,4.125\sqrt{3},-4.125,116.5 \right), \\ 
 & \left( \rho ,U,V,p \right)=\left( 1.4,0,0,1 \right). \\ 
\end{aligned}$$
Initially, a right-moving Mach 10 shock with a ${{60}^{\circ }}$ angle against the x-axis is positioned at $\left( x,y \right)=\left( 1/6,0 \right)$.
	The density distributions and local enlargements at $t=0.2$ for the all four schemes are shown in Fig.\ref{figdouble}. \par
	The double Mach reflection problem demonstrates the good robustness of all four schemes. Upon closer examination with local enlargements, it becomes apparent that WENO5-AO GKS and TENO5-D GKS are capable of resolving more rich small-scale structures compared to WENO5-Z GKS and TENO5 GKS because of the higher order accuracy for the initial non-equilibrium states. Compared to WENO5-Z GKS and TENO5 GKS, WENO5-AO GKS and TENO5-D GKS utilize the kinetic-weighting method in Section \ref{sec3.2} to construct the equilibrium state. In this method, the upwind mechanics are introduced in the determination of $g_{x}^{0}$. As a result, these schemes effectively reduce oscillations. This improvement in handling oscillations can be attributed to the use of the kinetic-weighting method and the incorporation of upwind mechanics in determining $g_{x}^{0}$ for WENO5-AO GKS and TENO5-D GKS. In this problem, TENO5 GKS method must satisfy ${{C} _ {T}} > {{10} ^ {-5}} $ in normal reconstruction and ${{C} _ {T}} > {{10} ^ {-3}} $ in tangential reconstruction. The parameter only satisfies ${{C}_{T}}>{{10}^{-5}}$ in tangential reconstruction and no limitation in normal reconstruction for TENO5-D GKS. TENO schemes also provide an efficient and direct means to control the spectral properties of the underlying
	scheme for specific problems. 
\\

\subsection{Viscous shock tubes problem}
\label{sec5.8}
This viscous shock tube problem~\cite{KIM2005570} was investigated to valid the capability of the present four HGKS schemes for low Reynolds number viscous flow with strong shocks. In a two-dimensional unit box $\left[ 0,\ 1 \right]\times \left[ 0,\ 1 \right]$, a membrane located at $x=0.5$ separates two different states of the gas and the dimensionless initial states are 
$$\left( \rho ,U,p \right)=\left\{ \begin{aligned}
  & \left( 120,0,120/\gamma  \right),0<x<0.5, \\ 
 & \left( 1.2,0,1.2/\gamma  \right),0.5<x<1, \\ 
\end{aligned} \right.$$
where $\gamma =1.4$, Prandtl number $\Pr =0.73$ and Reynolds number $\operatorname{Re} =200$ and $\operatorname{Re} =1000$. The computational domain is $\left[ 0,\ 1 \right]\times \left[ 0,\ 0.5 \right]$ with a symmetric boundary condition imposed on the top $x\in \left[ 0,\ 1 \right]$, $y=0.5$ and non-slip adiabatic conditions applied on the other three solid wall boundaries. The output time is $t=1.0$. 
At $t=0$, the membrane is removed and a wave interaction occurs. A shock wave with Mach number $Ma=2.37$ moves to the right side. Then the shock wave followed by a contact discontinuity reflects at the right wall. After the reflection, the shock interacts with the contact discontinuity. During their propagation, both of them interact with the horizontal wall and create a thin boundary layer. The solution will develop complex two-dimensional shock/shear/boundary-layer interactions and the dramatic changes for velocities above the bottom wall introduce strong shear stress.\par
\begin{figure}
	\centering
	\includegraphics[width=0.4\textwidth]{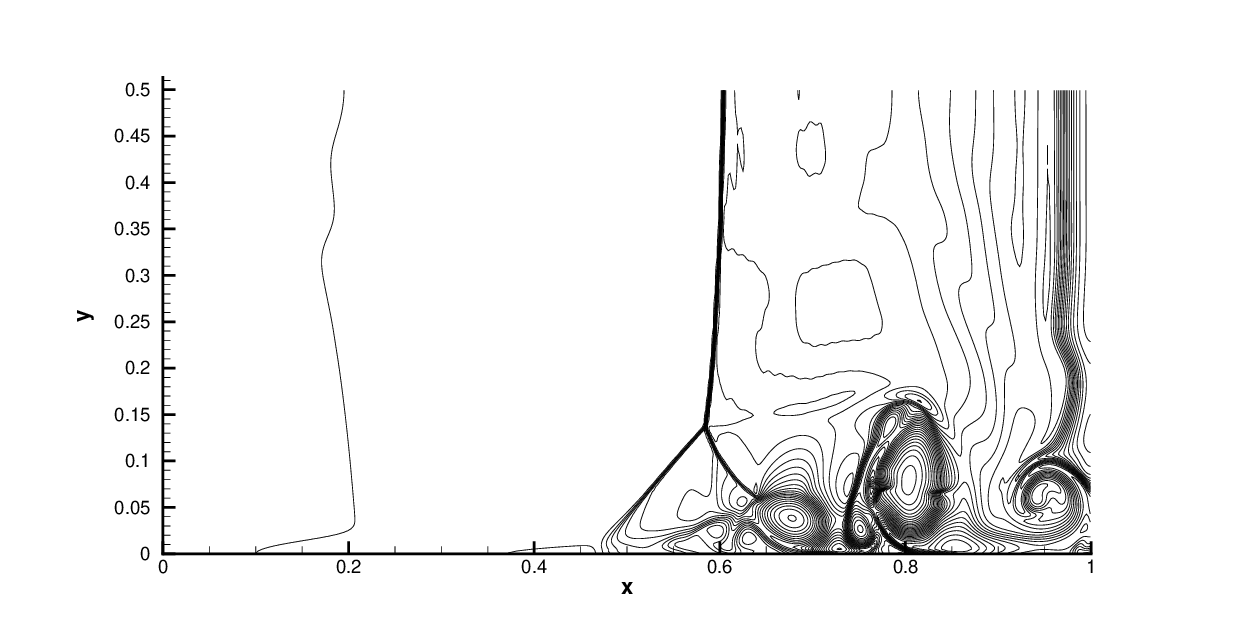}
	\includegraphics[width=0.4\textwidth]{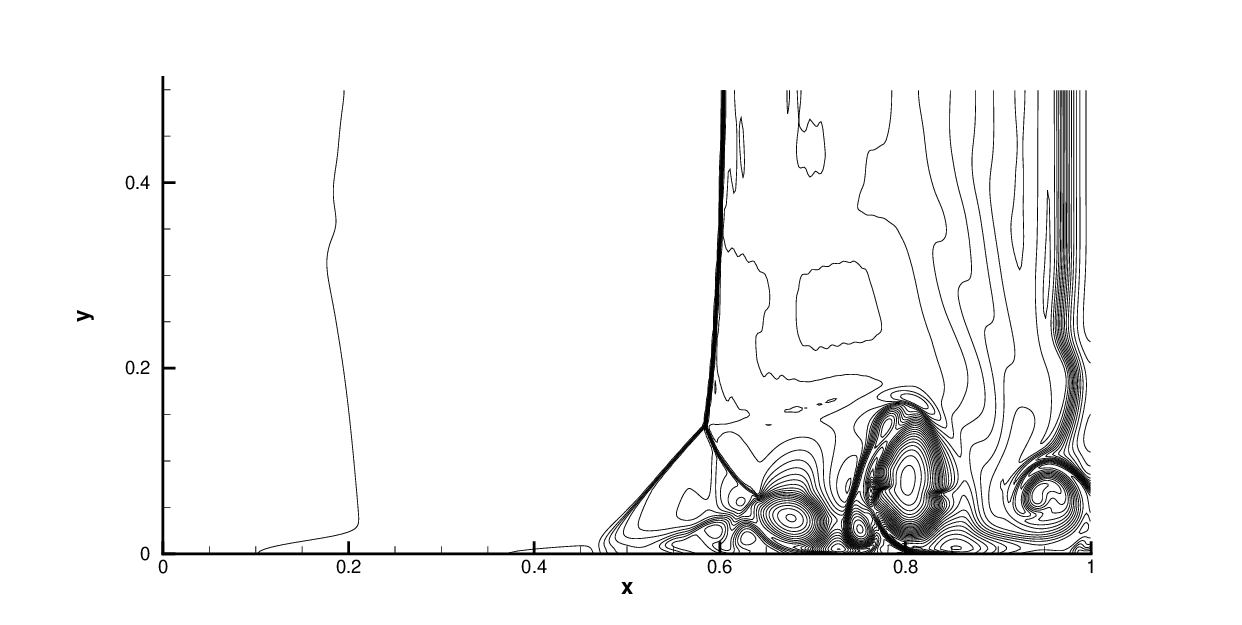}
	\includegraphics[width=0.4\textwidth]{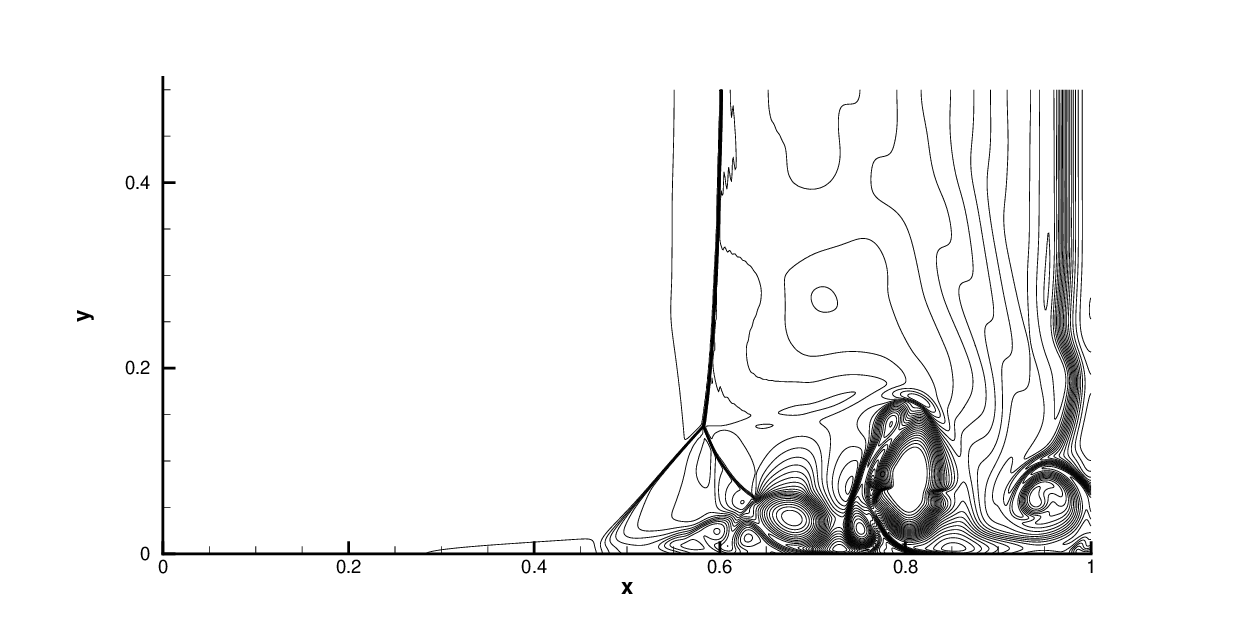}
	\includegraphics[width=0.4\textwidth]{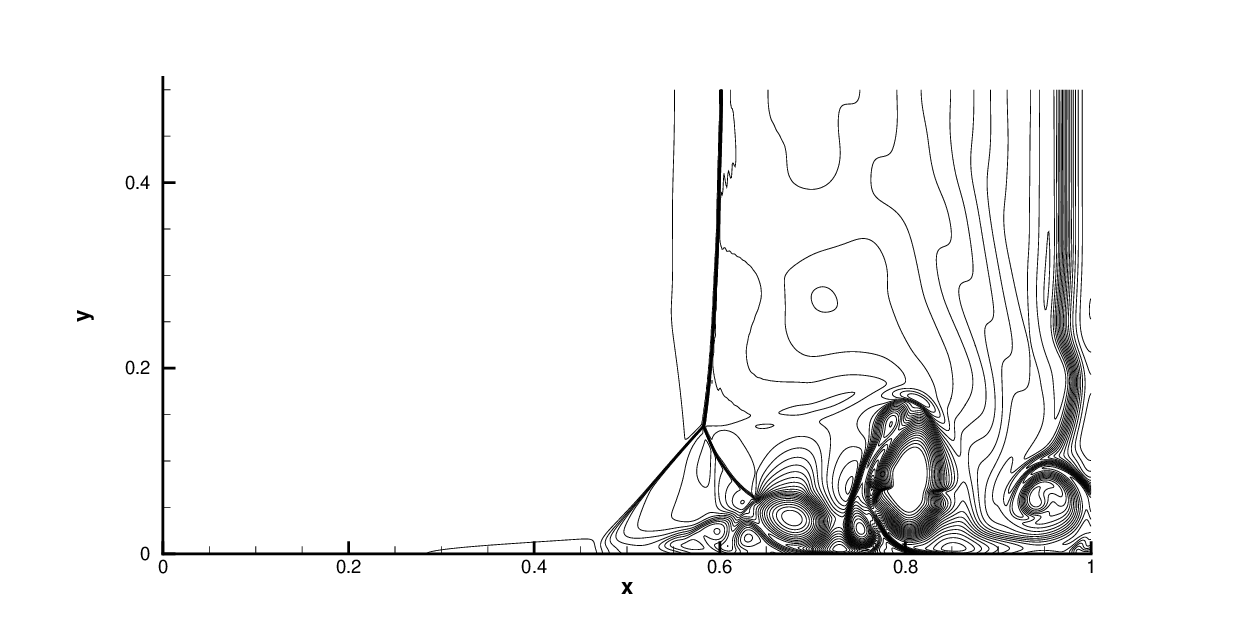}
	\caption{\label{figviscous200}Viscous shock tube problem with $Re =200$ by WENO5-AO-GKS scheme (left) and TENO5-D-GKS scheme (right): density distribution. For all cases, the CFL number is 0.2. For the top two figures, the mesh number is $500\times250$; and for the bottom two figures, the mesh number is $1000 \times500$.}
\end{figure}

\begin{figure}
	\centering
	\includegraphics[width=1.0\textwidth]{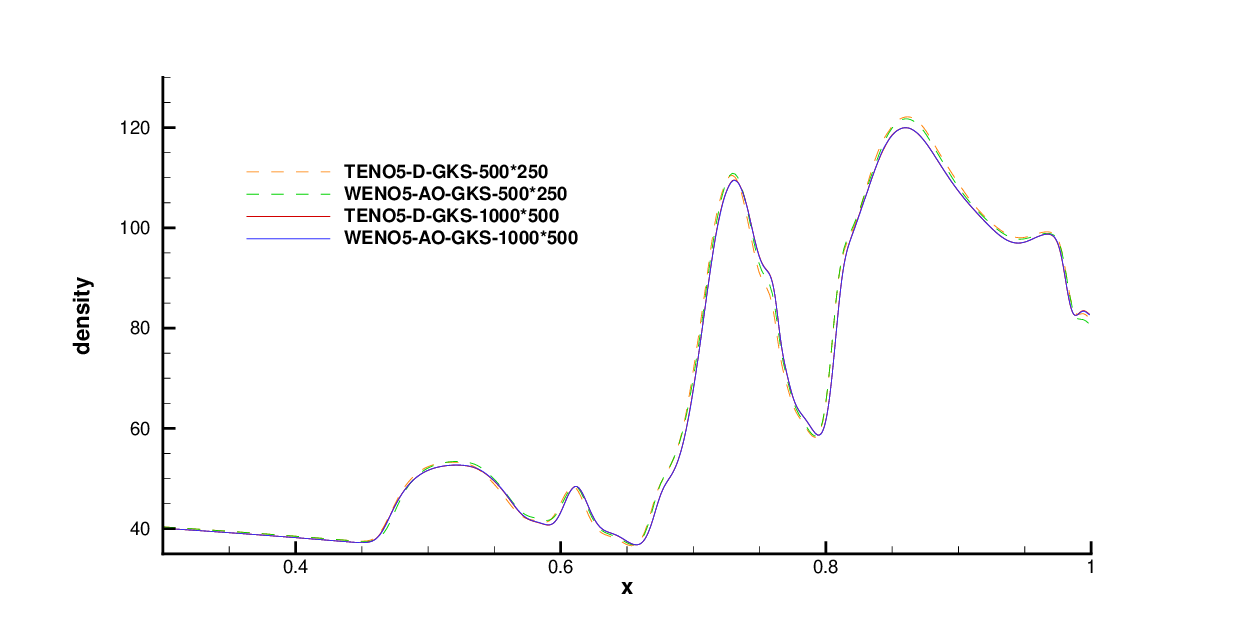}
	\caption{\label{figviscous200rho}Viscous shock tube problem of $Re =200$ by WENO5-AO-GKS scheme and TENO5-D-GKS scheme: density profile along the bottom wall $(y =0)$. For all cases, the CFL number is 0.2.}
\end{figure}

\begin{figure}
	\centering
	\includegraphics[width=0.4\textwidth]{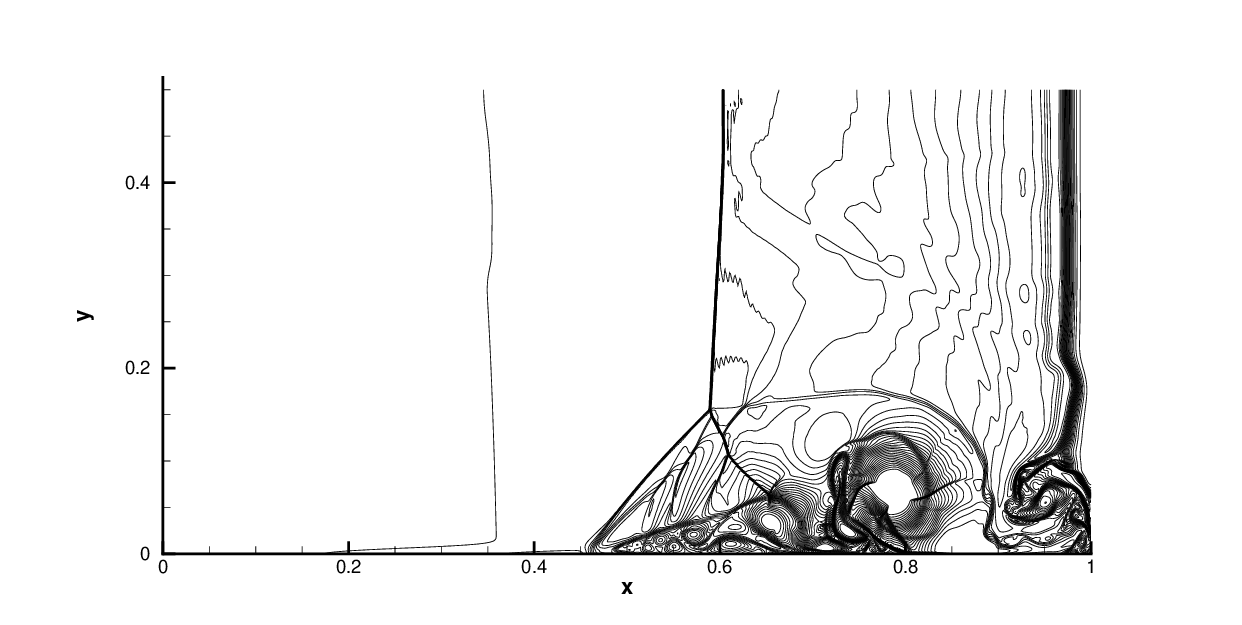}
	\includegraphics[width=0.4\textwidth]{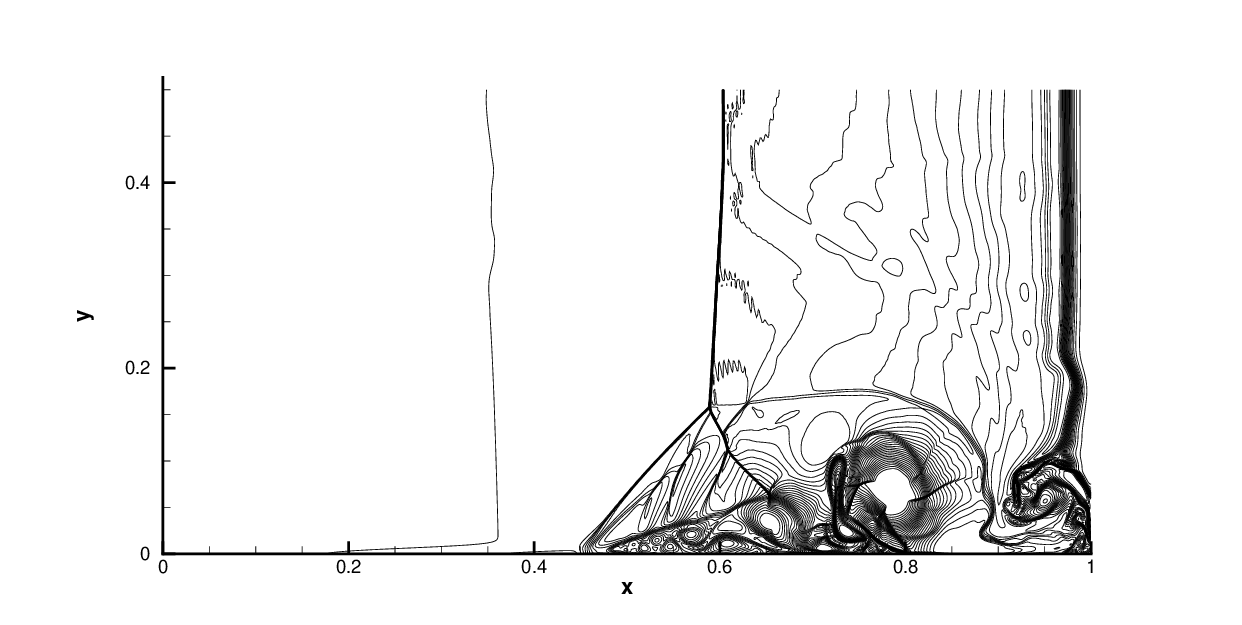}
	\includegraphics[width=0.4\textwidth]{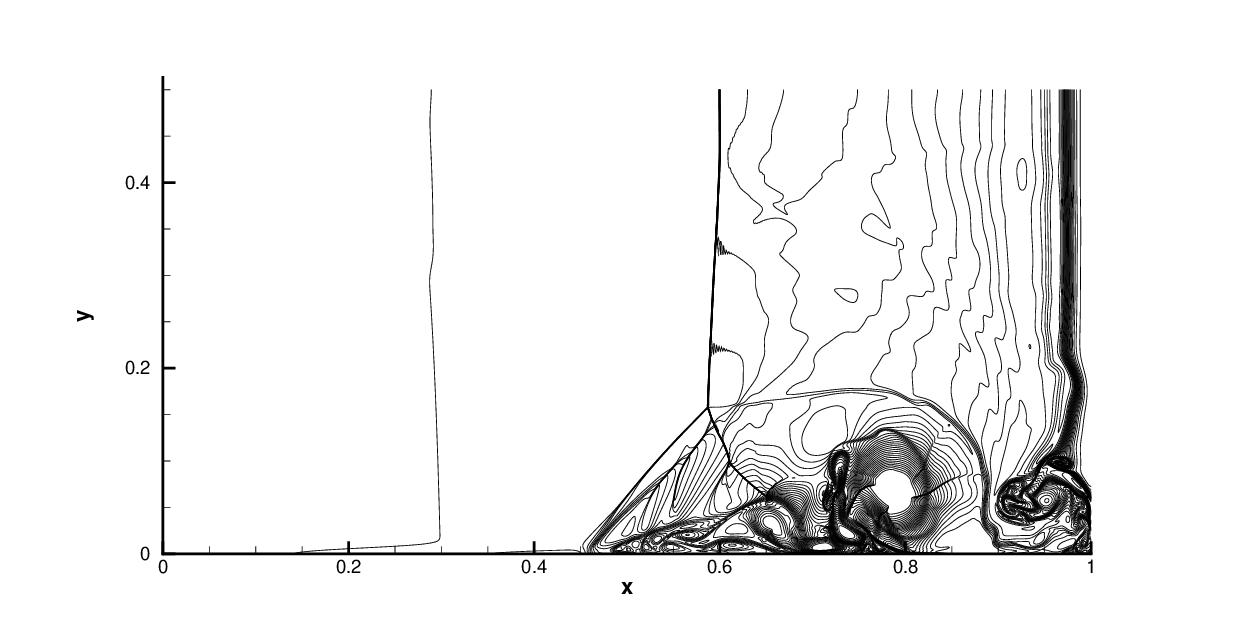}
	\includegraphics[width=0.4\textwidth]{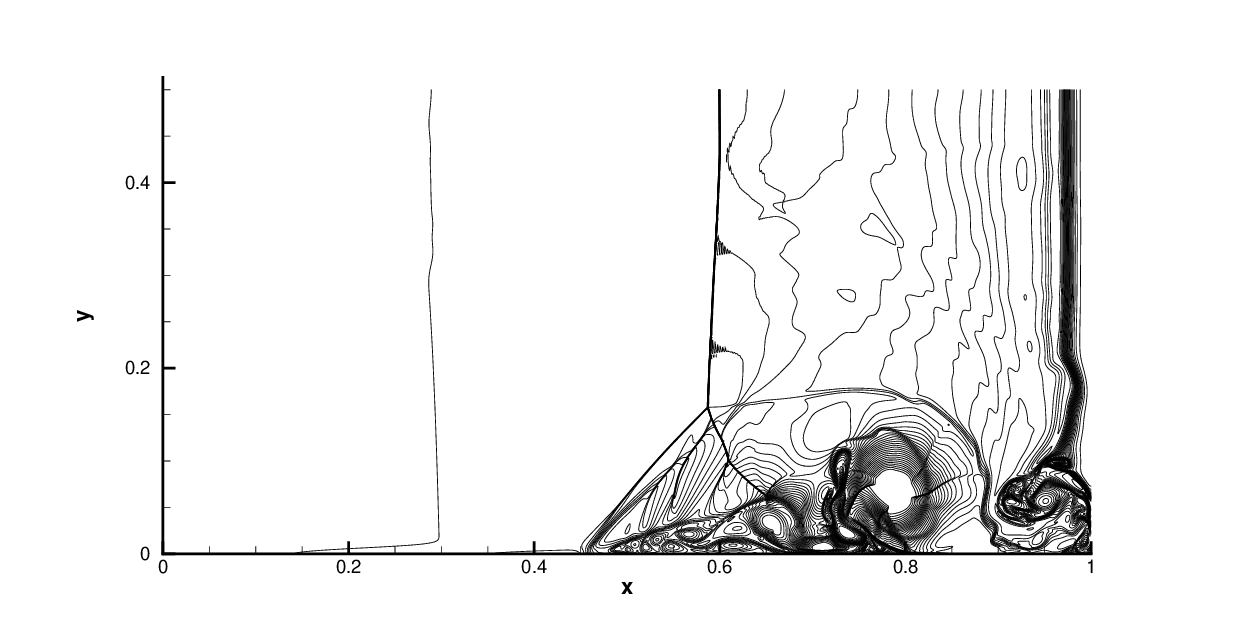}
	\caption{\label{figviscous1000}Viscous shock tube problem with $Re =1000$ by WENO5-AO-GKS scheme (left) and TENO5-D-GKS scheme (right): density distribution. For all cases, the CFL number is 0.2. For the top two figures, the mesh number is $1000\times500$; and for the bottom two figures, the mesh number is $2000 \times1000$.}
\end{figure}

\begin{figure}
	\centering
	\includegraphics[width=1.0\textwidth]{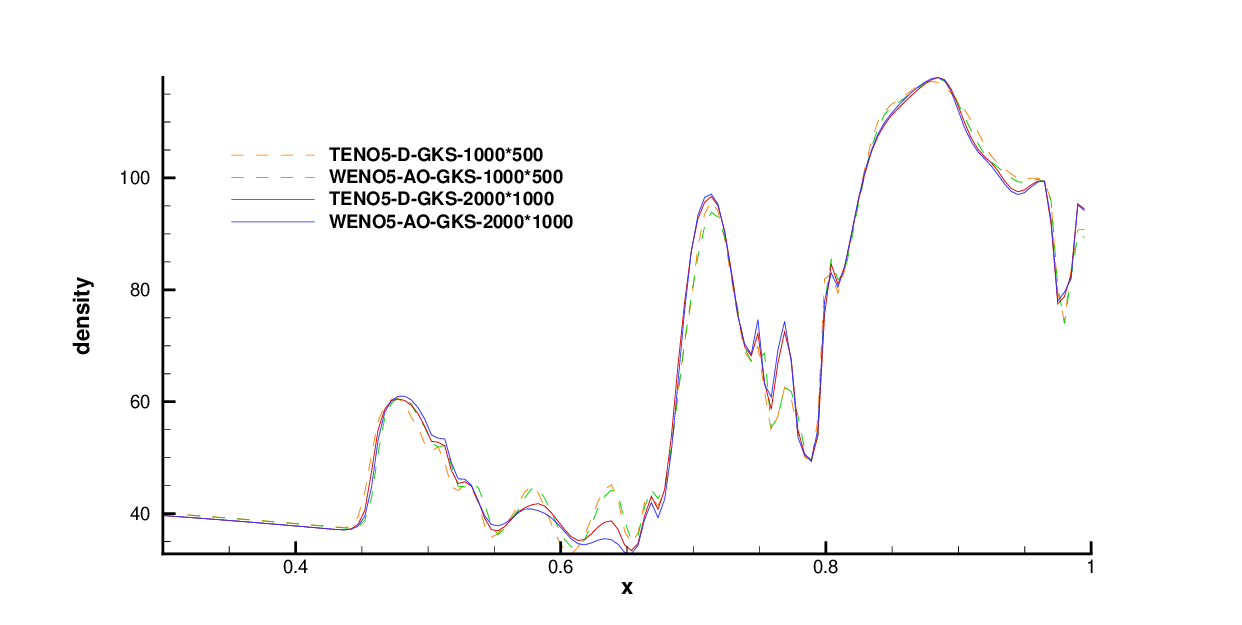}
	\caption{\label{figviscous1000rho}Viscous shock tube problem of $Re =1000$ by WENO5-AO-GKS scheme and TENO5-D-GKS scheme: density profile along the bottom wall $(y =0)$. For all cases, the CFL number is 0.2.}
\end{figure}

 
For the viscous shock tube with $Re=200$, the density distributions for the WENO5-AO-GKS and TENO5-D-GKS are plotted in Fig. \ref{figviscous200}. The density profiles along the bottom wall are shown in Fig. \ref{figviscous200rho}. And for the viscous shock tube with $Re=1000$ the density distributions for both schemes are plotted in Fig. \ref{figviscous1000}. Their density profiles along the bottom wall are shown in Fig. \ref{figviscous1000rho}. \par
The WENO5-Z GKS replaces WENO-Z type weights with WENO-JS weights to pass this test case. Even if the normal reconstruction ${{C} _ {T}} > {{10} ^ {-3}}$ and the tangential reconstruction ${{C} _ {T}} > {{10} ^ {-3}}$, TENO5-GKS still cannot pass this test. Fig. \ref{figviscous200} and Fig. \ref{figviscous1000} show that the WENO5-AO GKS and TENO5-D GKS can survive for this problem and give a reasonable resolution. In Fig. \ref{figviscous200rho} and Fig. \ref{figviscous1000rho}, the density profiles along the bottom wall of the two schemes are nearly identical in a numerical simulation conducted using the same grid points. It suggests that both schemes are producing consistent results in capturing the density distribution at the lower region of the flow. This similarity indicates that both schemes are capable of accurately representing the density profiles near the bottom and are likely performing well in resolving the corresponding flow behavior. Besides, WENO5-AO GKS and TENO5-D GKS are more robust than WENO5-Z GKS and TENO5 GKS for such shock boundary layer interaction problems.

\section{Conclusion}
\label{sec6}
In this paper, we propose two types of HGKS schemes with TENO reconstruction methods: TENO5 GKS and TENO5-D GKS. \par
TENO5 GKS and TENO5-D GKS as two TENO class HGKS schemes compared with the previous two WENO class HGKS schemes, the performance improvement is mainly due to the TENO class reconstruction method. In the TENO scheme, controlling the dissipation of smooth and non-smooth regions is achieved through stencil selection similar to ENO. Unlike traditional WENO schemes, which tend to select smoother stencils through convex combinations, the TENO scheme completely suppresses the stencil when non-smoothness is detected based on a predetermined threshold. This unique characteristic of the TENO scheme enables precise control over the dissipation behavior in both smooth and non-smooth regions. Furthermore, the TENO class methods consistently contribute to the final reconstruction process with their standard weight. Compared to traditional WENO schemes, the TENO class methods exhibit lower numerical dissipation and a more pronounced ability to capture discontinuities. The results presented in this paper demonstrate that, in most of the conducted tests, the TENO class HGKS schemes outperform the WENO class GKS schemes in terms of resolution and accuracy.\par
The GKS flux solver necessitates the use of point values and derivatives of both non-equilibrium and equilibrium states at the cell interface. In the spatial reconstruction process, both WENO5-Z GKS and TENO5 GKS methods reconstruct the non-equilibrium states and equilibrium states separately. In contrast, WENO5-AO GKS and TENO5-D GKS schemes avoid the need for separate reconstruction of the equilibrium state. This is achieved by utilizing a reconstruction process that is sufficiently accurate for large stencils, allowing for high-order non-equilibrium derivatives to be obtained. The non-equilibrium distribution functions, such as those related to particle collisions, are obtained through dynamic modeling. The utilization of this approach in WENO5-AO GKS and TENO5-D GKS schemes helps to effectively reduce spurious oscillations. Besides, both TENO5-D GKS and WENO5-AO GKS demonstrate significantly improved robustness compared to WENO5-Z GKS and TENO5 GKS. WENO5-Z GKS fails in the viscous shock tube test, while TENO5 GKS is unable to pass the test even with ${{C}_{T}}>{{10}^{-3}}$. But both WENO5-AO GKS and TENO5-D GKS are able to easily pass the test with satisfactory results. This improved robustness highlights the effectiveness of WENO5-AO GKS and TENO5-D GKS in accurately capturing and resolving complex flow.\par
In a comprehensive comparison of the four schemes, TENO5-D GKS stands out for several reasons. Firstly, due to the inherent characteristics of the TENO method, it exhibits lower dissipation and better shock capturing capability compared to the WENO method. This enables TENO5-D GKS to provide the highest resolution of small-scale flow field structures in various tests. Furthermore, both TENO5-D GKS and WENO5-AO GKS share the same equilibrium state reconstruction process. This means that TENO5-D GKS inherits the advantages of WENO5-AO GKS. These advantages include robustness, low numerical viscosity, reduced spurious oscillations at weak discontinuities, and excellent resolution of shear instability. Overall, TENO5-D GKS combines the strengths of the TENO method in shock capture and dissipation with the advantages of the WENO-AO reconstruction process. This results in a scheme that excels in resolving complex flow, provides high-resolution solutions for small-scale features, and maintains stability and accuracy in various simulations. Indeed, the WENO-AO-type HGKS scheme with arbitrary linear weights is known to be effective for unstructured mesh due to its ease in reconstructing values at Gaussian points. The TENO5-D GKS scheme also shares this advantage by utilizing simplified weights, making it suitable for developing TENO-D class HGKS scheme on unstructured mesh as well. This characteristic enables accurate and efficient computation on unstructured grids, further enhancing the applicability and versatility of the TENO5-D GKS scheme in various numerical simulation for compressible flow.

\section*{Acknowledgments}
The authors would like to thank Yining Yang, Peiyuan Geng and Boxiao Zou for helpful discussion.
This work is sponsored by the National Natural Science Foundation of China [grant numbers 11902264, 12072283, 12172301], the National Numerical Wind Tunnel Project and the 111 Project of China (B17037).


\end{document}